\documentclass{ws-ijmpa-mod}

\def\lt{\left}
\def\rt{\right}
\def\p{\partial}
\def\th{{\hat{t}}}
\def\o{\omega}
\def\g{\gamma}

\def\O{\Omega}
\def\N{{\cal N}}
\def\F{{\cal F}}
\def\ih{{\hat{i}}}
\def\jh{{\hat{j}}}
\def\kh{{\hat{k}}}
\def\lh{{\hat{l}}}

\def\d{\delta}
\def\et{\tilde{e}}
\def\nt{\tilde{\nabla}}
\def\veps{\varepsilon}
\def\ben{\begin{equation}}
\def\een{\end{equation}}
\def\bea{\begin{eqnarray}}
\def\eea{\end{eqnarray}}
\def\nn{\nonumber}
\def\plh{{\hat{+}}}
\def\mih{{\hat{-}}}
\def\half{{\frac12}}

\def\omt{\tilde{\omega}}
\def\Rt{\tilde{R}}
\def\Jh{\hat{J}}
\def\Mh{\hat{M}}
\def\Nc{{\cal N}}
\def\omh{\hat{\omega}}
\def\qb{\overline{q}}
\def\Yb{\overline{Y}}
\def\Fb{\overline{F}}

\begin{document}

\markboth{Alejandra Castro, Joshua L. Davis, Per Kraus, Finn Larsen}
{String Theory Effects on Five-Dimensional Black Hole Physics}

%
%

\title{String Theory Effects on Five-Dimensional Black Hole Physics}

\author{Alejandra Castro${}^\spadesuit$\footnote{aycastro@umich.edu} , Joshua L. Davis${}^\clubsuit$\footnote{davis@physics.ucla.edu} , Per Kraus${}^\clubsuit$\footnote{pkraus@physics.ucla.edu} , Finn Larsen${}^\spadesuit{}^\diamondsuit$\footnote{larsenf@umich.edu}\medskip}


\address{${}^\spadesuit$Michigan Center for Theoretical Physics, University of Michigan, 450 Church St.\\
Ann Arbor, Michigan 48109-1120,
United States}

\address{${}^\clubsuit$Department of Physics and Astronomy, UCLA, Box 951547\\
Los Angeles, California 90095-1547, United States }

\address{${}^\diamondsuit$ Theory Division \\
CERN, CH-1211, Geneva 23, Switzerland }

\maketitle


\begin{abstract}
We review recent developments in understanding quantum/string corrections to BPS black holes and strings in five-dimensional supergravity. These objects are solutions to  the effective action obtained from M-theory compactified on a Calabi-Yau threefold, including the one-loop corrections determined by anomaly cancellation and supersymmetry. We introduce the off-shell formulation of this theory obtained through the conformal supergravity method and review the methods for investigating supersymmetric solutions.  This leads to quantum/string corrected attractor geometries, as well as asymptotically flat black strings and spinning black holes. With these solutions in hand,  we compare our results with analogous studies in four-dimensional string-corrected supergravity, emphasizing the distinctions between the
four and five dimensional theories.

\keywords{Attractors; Black holes; String theory.}
\end{abstract}

\ccode{PACS numbers: 04.50.-h, 04.50.Gh, 04.65.+e, 04.70.Dy}

\pagebreak

\section{Introduction}

One of the great successes of string theory  has been in elucidating the microphysics of black holes. As is expected for any candidate theory
of quantum gravity, string theory has provided an accounting of the Bekenstein-Hawking entropy of many black holes in terms of a microscopic
counting of states
\ben\label{a}
S_{BH} = {A \over 4 G}=  \log \O_{string}~.
\een
However, it must be emphasized that the Bekenstein-Hawking entropy is only a leading order result, derived from the classical Einstein-Hilbert action.
Any  theory of quantum gravity will in general contain higher dimension operators in the low energy effective Lagrangian, \textit{i.e.} terms which contain more than two derivatives of the fundamental fields.   The area law for the black hole entropy is therefore
only valid in the limit that the black hole is much larger than the Planck and string scales.   Analogously, the explicit counting
of states is usually done in the limit of large mass and charge, where powerful formulas for the asymptotic degeneracies
are available.
Thus one expects corrections on both sides of (\ref{a}), and matching these corrections leads to an even more detailed understanding
of string theory and black holes.   An ambitious long term goal is to verify (\ref{a}) {\it exactly}, as
this would surely signal that we have achieved a fundamental understanding of quantum gravity.

Recent years have seen much progress in  analyzing the effects of string and quantum gravity corrections to black holes.
The Bekenstein-Hawking area law formula has been generalized to the Bekenstein-Hawking-Wald entropy.\cite{Wald}
The Wald formula applies to any diffeomorphism and gauge invariant local effective action,
and so can be  applied to effective Lagrangians arising  in string theory or otherwise.   The Wald formula  greatly simplifies for an
extremal black hole with a near horizon AdS$_2$ or AdS$_3$ factor, as is often the case.    For AdS$_2$ the entropy function
formalism\cite{Sen:2005wa} is appropriate,  while for AdS$_3$ $c$-extremization\cite{Kraus:2005vz} is most efficient.
Supersymmetric black hole solutions in four dimensional supergravity with $R^2$ corrections were found in Refs.~\refcite{Behrndt:1996jn}--\refcite{LopesCardoso:2000qm}, and their entropies successfully matched with a microscopic counting, including subleading
corrections.\cite{Maldacena:1997de}
Much recent work has also been stimulated by the connection with topological strings, starting with the OSV conjecture.\cite{Ooguri:2004zv}

An especially interesting application of these ideas is to so-called
\textit{small black holes},  those whose Bekenstein-Hawking entropy
vanishes at the leading order.  In two-derivative supergravity these
correspond to solutions with a naked singularity, but higher-order
corrections provide a string/Planck scale horizon to cloak the
singularity.\cite{Sen:1995in}\cdash\cite{Hubeny:2004ji} This
provides a beautiful example of how quantum and string effects can
smooth out a classical singularity.    In some cases, the entropies
of these small black holes can be reproduced from the microscopic
side, even though they are far from the regime of classical general
relativity.

The simplest examples of small black holes are 5D extended string solutions, which can be identified in one duality frame
as fundamental heterotic strings.    When these carry momentum they have an event horizon and a corresponding entropy
that matches that of the heterotic string.    We can alternatively consider the solution without any momentum, in which
case we obtain a completely smooth and horizon free solution representing an unexcited heterotic string.\cite{Castro:2007hc}

\subsection{Why five dimensions?}

In this review we focus on stationary solutions which preserve some fraction of supersymmetry.
Four and five dimensional black holes have a privileged status, for it is only these that can be supersymmetric.\footnote{By a D-dimensional black hole we mean one whose geometry  is asymptotic to D-dimensional Minkowski space, possibly times a compact space. In six dimensions there are supersymmetric black strings but no black holes (at least at the two-derivative level.)}       Five dimensions is especially interesting for a number of
reasons.   In four dimensions,
single-center black hole solutions  preserving supersymmetry can carry any number of electric and magnetic charges, but no angular momentum,  and the horizon is restricted to be an $S^2$.\footnote{That 4D black holes have horizons with $S^2$ topology was proven by Hawking\cite{Hawking:1971vc} in the context of general relativity. However it is possible that $S^1 \times S^1$ horizons are allowed in
a higher derivative theory; see Ref.~\refcite{Iizuka:2007sk} for a recent discussion.}   The space of solutions is much richer in
five dimensions.    First of all,  there are supersymmetric, electrically charged,  rotating  (BMPV) black holes.\cite{Strominger:1996sh}\cdash\cite{Sabra:1997yd}   Second,
there are
supersymmetric solutions carrying magnetic charge; such objects are \textit{black strings}  extended in one spatial direction.\cite{Gibbons:1994vm,Chamseddine:1999qs} These can carry momentum and traveling waves.
Finally, there are  \textit{black rings}\cite{Elvang:2004rt,Bena:2004de} which have horizons with $S^1 \times S^2$ topology. These
objects carry electric charges and magnetic dipoles, the latter usually referred to as \textit{dipole charges}. Additionally, they carry  angular momenta in two independent planes.    As a further motivation for focussing on five dimensions, we note that  four dimensional black holes
can be recovered by the Kaluza-Klein reduction of five dimensional solutions along a circle isometry.

Although the full effective action of string theory is expected to contain an infinite series of higher derivative terms, we confine our
attention to the leading order corrections, which are four-derivative terms.   On the one hand, this is a product of
necessity, since terms with more derivatives are not fully known, especially in the off-shell formalism that is needed
in order for explicit calculations to be feasible.    Fortunately, as we will see, most of the physics that we wish
to uncover, such as singularity resolution, is  captured already at the four-derivative level.   In fact, for black holes
containing a near horizon AdS$_3$ factor, a non-renormalization theorem\cite{Kraus:2005vz,Kraus:2005zm} can be proven, stating that the entropy
gets no corrections from terms with more than four-derivatives.     This theorem allows us to systematically compute the
entropy of small black holes, even though we would, {\it a priori}, expect to need to use the entire series of higher derivative
terms since there is no small expansion parameter to control the derivative expansion.  This argument also applies
to non-BPS and near extremal black holes, and explains the fact that we have excellent control over the microscopic and
macroscopic entropies of these objects even though they are non-supersymmetric.      These powerful consequences  of AdS$_3$ underlie all successful entropy matchings in string theory, and provide yet another
motivation for working in five dimensions.   Conversely,   the full symmetries of these solutions  are not manifest in the description of these
black holes in terms of four-dimensional supergravity.

\subsection{Finding solutions to $D=5$ $R^2$-corrected supergravity}

It is useful to think of five-dimensional supergravity as arising from the dimensional reduction of M-theory on a
Calabi-Yau threefold.    The theory contains some number of vector multiplets, determined by the Hodge numbers of
the Calabi-Yau.    It then turns out that the action up to four derivatives is completely determined in terms of two additional
pieces of topological data, namely  the triple intersection numbers and second Chern class of the Calabi-Yau.
The task of finding solutions to this theory is greatly simplified by working in a fully off-shell formalism.\cite{Bergshoeff:2001hc}\cdash\cite{Hanaki:2006pj} This means that
enough auxiliary fields are introduced so that the supersymmetry transformations are independent of the
action.\footnote{A familiar example of this is  $N=1$  supersymmetric field theory in four dimensions, where the superspace
construction ensures that the supersymmetry algebra closes without having to use the equations of motion.}    This is a
great advantage, because the supersymmetry transformation laws are very simple, while the explicit action is quite
complicated.   In looking for BPS solutions we first exhaust the conditions implied by unbroken supersymmetry; in the
off-shell formalism it follows that this part of the analysis proceeds the same whether one considers the two, four, or even
higher derivative solutions.    Much of the solution thereby can  be determined without great effort.   Only at the very end do we need to
consider some of the equations of motion in order to complete the solutions.    In general, we find that  the full
solution can be expressed algebraically in terms of a single function, which obeys a nonlinear ordinary differential equation.
This equation is straightforward to solve numerically.

\subsection{Overview of results}

We now give an overview of the results contained in this review.    We begin with a review of standard two-derivative
$D=5$ supergravity coupled to an arbitrary number of vector multiplets,\cite{Gunaydin:1983bi} and discuss the relevant solutions of this theory:
black holes, black strings, and black rings.   These are the solutions that we want to generalize to the higher derivative
context.   We next turn to the formalism of $D=5$ $R^2$-corrected supergravity, obtained via the gauge fixing of a superconformal
theory.    The technical virtue of introducing the physically extraneous superconformal symmetry is that it facilitates
the construction of fully off-shell multiplets.   Our goal is to highlight the main conceptual steps of the superconformal
program, while leaving the technical details to the original literature.\cite{Bergshoeff:2001hc}\cdash\cite{Hanaki:2006pj}   The four derivative part of the action is to be
thought of as the supersymmetric completion of a certain mixed gauge-gravitational Chern-Simons term.  This term
is related to gauge and gravitational anomalies, and it is this relation that leads to the non-renormalization theorem
mentioned above.

All of the solutions that we consider have the property of having a near horizon region with enhanced supersymmetry.
This fact implies that the near horizon geometries are much simpler to obtain than the full asymptotically flat solutions,
since some of the equations of motion can be traded for the simpler conditions following from enhanced supersymmetry.
Therefore, in the interests of pedagogy, we first show how to obtain the near horizon solutions directly, postponing
the full solutions until later.     The near horizon geometries fall into two classes, depending on whether an AdS$_2$ or
AdS$_3$ factor is present, and we analyze each in turn.  We also exhibit the higher derivative version of the attractor
mechanism, which fixes the moduli in terms of the electric and magnetic fluxes in the near horizon region.

Our discussion of the asymptotically flat solutions naturally divides into two parts.   Half BPS solutions have a distinguished
Killing vector formed out of the Killing spinors, and the analysis hinges on whether this Killing vector is timelike or null.\cite{Gauntlett:2002nw}
The timelike case gives rise to 5D black holes\cite{Strominger:1996sh}\cdash\cite{Sabra:1997yd} and black rings,\cite{Elvang:2004rt,Bena:2004de} while the null case corresponds to 5D black strings.\cite{Gibbons:1994vm,Chamseddine:1999qs}
We show how to systematically construct these solutions, starting by applying the conditions of unbroken supersymmetry,
and then imposing the equations of motion for the Maxwell and auxiliary fields.

After constructing the full solutions and observing that they indeed contain the near horizon regions with enhanced
supersymmetry, we turn to evaluating the black hole entropy.    This is not completely straightforward, since Wald's
formula does not directly apply, as the action contains  non-gauge invariant Chern-Simons terms.  We apply two
different strategies, depending on whether an AdS$_3$ or AdS$_2$ factor is present.     In the AdS$_3$ case,
finding the black hole entropy can be reduced to finding the generalized Brown-Henneaux central charges\cite{Brown:1986nw}
of the underlying Virasoro algebras.\footnote{To be precise, the central charges take into account the
contributions from all {\it local} terms in the effective action.  Additional nonlocal terms are also present due to the
fact that the black hole has a different topology than Minkowski space.  These contributions come from the worldlines
of particles winding around the horizon.\cite{Dijkgraaf:2000fq}\cdash\cite{Kraus:2006nb}}    For a general Lagrangian, an efficient $c$-extremization formula is available,\cite{Kraus:2005vz}
which reduces the computation of the central charges to solving a set of algebraic equations.    In the supersymmetric
context the procedure is even simpler, since the central charges can be read off from the coefficients of the Chern-Simons
terms.\cite{Harvey:1998bx,Kraus:2005vz,Kraus:2005zm}  We carry out both procedures and show that they agree.    For AdS$_2$, there is a similar extremization
recipe based on the so-called entropy function.\cite{Sen:2005wa}    Applying the entropy function here requires a bit of extra work, since
the Chern-Simons terms need to be rewritten in a gauge invariant form in order for Wald's formula to apply.    We
carry this out and obtain the explicit entropy formulas for our black hole solutions.   The results turn out to be remarkably
simple.  In the case of non-rotating 5D black holes the effects of the higher derivative terms manifest themselves simply
as a shift in the charges.

We are also able to shed light on the connection between four and five dimensional black holes.   One way to
interpolate between these solutions is by placing the 5D black object at or near the tip of a Taub-NUT geometry, as in
Refs.~\refcite{Bena:2005ay}--\refcite{Bena:2005ni}.
The Taub-NUT contains a circle of freely adjustable size, which is to be thought of as the Kaluza-Klein circle that takes
us from five to four dimensions.   When the circle is large the solution is effectively five dimensional, while in the other
limit we have a four dimensional black hole.   In the case of BPS black holes, the entropy is independent of moduli
(except possibly for isolated jumps at walls of marginal stability),  and so this interpolation lets us relate the entropies
of four and five dimensional black holes.\cite{Gaiotto:2005gf}     At the two-derivative level this works out very simply.
In Section \ref{sec:fourdfived} we also review how the full solutions can be mapped back and forth, by demonstrating the
equivalence of the BPS equations in the two cases.\cite{BLM}    On the other hand, higher derivative corrections bring in some
new complications.  As we shall see explicitly, due to the existence of curvature induced charge densities, the relations between
the four and five dimensional charges gets corrected in a nontrivial way.   Also, for reasons that we discuss,  we find that there is apparently no simple  relation between the two sets of BPS equations.

\subsection{Additional references}

This review is mainly based on Refs.~\refcite{Castro:2007sd,Castro:2007hc}, and \refcite{Castro:2007ci}.  However, we have also
taken the opportunity to include some new results, in particular extending the higher derivative BPS equations
to the general case.

The literature on higher derivative corrections to supersymmetric black holes is by now quite large.
There are a number of reviews available which overlap with some of the topics discussed here, for example Refs.~\refcite{Mohaupt:2000mj}--\refcite{Sen:2007qy}.   Further references
on supergravity solutions in the presence of higher derivatives include Refs.~\refcite{Saida:1999ec}--\refcite{Cvitan:2007hu}.    Some references relating
higher derivative corrections to the topological string and the OSV conjecture include Refs.~\refcite{Lopes Cardoso:2004xf}--\refcite{Denef:2007vg}.   Works discussing small black
holes in the context of AdS/CFT include Refs.~\refcite{Dabholkar:2005qs} and \refcite{Iizuka:2005uv}--\refcite{Dabholkar:2006tb}.   For more on supergravity and the attractor mechanism in five dimensions see Refs.~\refcite{Sabra:1997yd,Chamseddine:1999qs} and \refcite{Ferrara:1996dd}--\refcite{Kraus:2005gh}.
Fundamental string solutions
are further discussed in Refs.~\refcite{Giveon:2006pr}--\refcite{Kraus:2007vu}.

\section{On-shell Formalism for $D=5$ Supergravity}\label{on-shell}

In this section we briefly review the two-derivative supergravity theory in which we are interested, $N =2$ supergravity in five dimensions coupled to an arbitrary number of vector multiplets. We review how this theory is embedded in string theory as a dimensional reduction of eleven-dimensional supergravity on a Calabi-Yau threefold.\cite{Cadavid:1995bk,Antoniadis:1995vz} We also summarize the panoply of supersymmetric solutions to this theory, including their origin as wrapped M-branes in the higher-dimensional theory. This is all done in the familiar ``on-shell'' formalism (in contrast to the formalism to be introduced subsequently in this review).
\subsection{M-theory on a Calabi-Yau threefold}

We begin with the action
\ben
S_{11} = -{1\over2\kappa_{11}^2} \int ~d^{11}x~ \sqrt{-G} \lt(R + \half |{\cal F}_4|^2\rt) +\frac{1}{12\kappa_{11}^2} \int~ {\cal A}_3 \wedge {\cal F}_4 \wedge {\cal F}_4~,
\een
which is the bosonic part of the low energy eleven-dimensional supergravity theory. The
perturbative spectrum contains the graviton $G_{MN}$, the gravitino $\Psi_M$, and the three-form potential ${\cal
A}_{3}$ with field strength ${\cal F}_4=d{\cal A}_3$. This theory is maximally supersymmetric, possessing 32 independent supersymmetries. There are spatially extended half-BPS solitons, the $M2$ and $M5$-branes, which carry the electric and magnetic charges, respectively, of the flux
${\cal F}_{4}$.

The five-dimensional theory of interest is obtained via compactification on a Calabi-Yau threefold $CY_3$ and depends only on topological data of the compactification manifold. Let $J_I$ be a basis of closed $(1,1)$-forms spanning
the Dolbeault cohomology group $H^{(1,1)}\lt(CY_3\rt)$ and let $h^{(1,1)}=dim\lt(H^{(1,1)}\lt(CY_3\rt)\rt)$. We can then expand the K\"ahler form $J$
on $CY_3$ as
\ben
J=M^I J_I ~,\quad\quad I=1\ldots h^{(1,1)}~.
\een
By de Rham's theorem\footnote{This theorem asserts the duality between the homology group $H_2 \lt(\mathcal{M}\rt)$ and the \textit{de Rham} cohomology $H^2\lt(\mathcal{M}\rt)$ for a manifold $\mathcal{M}$. For a Calabi-Yau threefold there are no $(0,2)$ or $(2,0)$ forms in the cohomology so we have a duality between $H_2$ and the \textit{Dolbeault} cohomology $H^{(1,1)}$.} we can choose a basis of two-cycles $\o^K$ for the homology group $H_2 \lt(CY_3\rt)$ such that
\ben
\int_{\o^K} J_I = \d^K_I~.
\een
Thus the real-valued expansion coefficients $M^I$ can be understood as the volumes of the two-cycles $\o^I$
\ben
M^I = \int_{\o^I} J~.
\een
The $M^I$ are known as \textit{K\"ahler moduli} and they act as scalar fields in the effective five-dimensional theory. We will often refer to the $M^I$ simply as the \textit{moduli} since
the other Calabi-Yau moduli, the \textit{complex structure moduli}, lie in $D=5$ hypermultiplets and are decoupled for the purposes of investigating
stationary solutions.\footnote{By \textit{decoupled}, we mean that they can be set to constant values in a way consistent with the BPS conditions and equations of motion of the theory. } We will therefore largely ignore the hypermultiplets in the following.

The eleven-dimensional three-form potential can be decomposed after compactification as
\ben
{\cal A}_3 = A^I \wedge J_I~,
\een
where $A^I$ is a one-form living in $D=5$. This results in $h^{(1,1)}$ vector fields in the five-dimensional effective theory.  Since the $J_I$ are closed, the field strengths are given by
\ben
{\cal F}_4 = F^I \wedge J_I~,
\een
where $F^I =dA^I$. The eleven-dimensional Chern-Simons term reduces to
\bea
\int_{{\cal M}_{11}} {\cal A}_3 \wedge {\cal F}_4 \wedge {\cal F}_4 &=& \int_{CY_3} J_I \wedge J_J \wedge J_K ~\int_{{\cal M}_5} A^I \wedge F^J \wedge F^K~\cr
&=& c_{IJK} \int_{{\cal M}_5} A^I \wedge F^J \wedge F^K~,
\eea
where in the last line we have used the definition of the Calabi-Yau manifold \textit{intersection numbers}
\ben
c_{IJK} = \int_{CY_3} J_I \wedge J_J \wedge J_K~.
\een
The nomenclature arises since $c_{IJK}$ can be regarded as counting the number of triple intersections of the four-cycles $\o_I$, $\o_J$, and
$\o_K$, which are basis elements of the homology group $H_4\lt(CY_3\rt)$. This basis has been chosen to be dual to the previously introduced basis of
$H_2 \lt(CY_3\rt)$, \textit{i.e.} with normalized inner product
\ben
\lt(\o^I , \o_J \rt) = \d^I_J~,
\een
where this inner product counts the number of intersections of the cycles $\o^I$ and $\o_J$.

The above is almost sufficient to write down the $D=5$ Lagrangian, but there is an important constraint that must be considered separately. To fill up the
five-dimensional supersymmetry multiplets one linear combination of the aforementioned vectors must reside in the gravity multiplet. This vector is
called the \textit{graviphoton} and is given by
\ben
A^{grav}_\mu = M_I A^I_\mu~,
\een
where the $M_I$ are the volumes of the basis four-cycles $\o_I$
\ben
M_I = \half \int_{\o_I} J \wedge J = \half \int_{CY_3} J \wedge J \wedge J_I = \half c_{IJK}M^J M^K~.
\een
Since one combination of the vectors arising from compactification does not live in a vector multiplet, the same must be true of the scalars. It turns
out that the total Calabi-Yau volume, which we call $\N$, sits in a hypermultiplet. Due to the decoupling of hypermultiplets we can
simply fix the value of the volume, and so we arrive at the {\it very special geometry} constraint
\ben\label{constraint}
\N \equiv \frac{1}{3!} \int_{CY_3} J\wedge J\wedge J = \frac16 c_{IJK} M^IM^JM^K =1~.
\een
Due to the above considerations the index $I$ runs over $1\ldots (n_V+1)$, where $n_V$ is the number of independent vector multiplets in the effective
theory.

Choosing units\footnote{Equivalently, our units are such that the five-dimensional Newton's constant is $G_5 = \frac{\pi}{4}$.} such that $\kappa^2_{11}=\kappa^2_5 {\cal N}=2\pi^2$, the action for the theory outlined above is
\ben\label{onshellact}
S=\frac{1}{4\pi^2}\int_{{\cal M}_5} d^5 x \sqrt{|g|} {\cal L}~,
\een
with Lagrangian
\ben\label{onshelllag}
{\cal L}=-R-G_{IJ}\partial_aM^I\partial^aM^J-\frac12 G_{IJ}F^I_{ab}F^{Jab}+\frac{1}{24}c_{IJK} A^I_{a}F^{J}_{bc}F^{K}_{de}\epsilon^{abcde}~.
\een
The metric on the scalar moduli space is\cite{Strominger:1985ks}
\ben
G_{IJ} =\half \int_{CY_3} J_I \wedge * J_J =-\half\lt.\partial_I\partial_J(\ln{\cal
N})\rt|_{{\cal N}=1}=\half\left( {\cal N}_I{\cal N}_J-{\cal
N}_{IJ}\right)~,
\een
where the $*$ denotes Hodge duality within the Calabi-Yau and $\N_I$ and $\N_{IJ}$ denote derivatives of $\N$ with respect to the moduli
\ben
\N_I \equiv \p_I \N = \frac12 c_{IJK} M^JM^K=M_I~, \quad \N_{IJ} \equiv \p_I \p_J \N = c_{IJK}M^K~.
\een

As previously stated, the eleven-dimensional theory is maximally supersymmetric with 32 independent supersymmetries. A generic Calabi-Yau manifold has $SU(3)$ holonomy, reduced from $SU(4)(\cong SO(6))$ for a generic six-dimensional manifold; thus it preserves 1/4 supersymmetry, or 8 independent supersymmetries. More precisely, this is the number of explicit
supersymmetries for general $c_{IJK}$; for special values of the $c_{IJK}$ there are more supersymmetries which are implicit in our formalism. These
values correspond to compactification on a manifold ${\cal M}$ of further restricted holonomy; for ${\cal M}=T^2\times K3$ there are 16 supersymmetries, while for ${\cal M}=T^6$
there are 32.

\subsection{$M$-branes and $D=5$ solutions}

Eleven-dimensional supergravity has asymptotically flat solutions with non-trivial four-form flux. In the full quantum description, these solutions are
understood to be sourced by certain solitonic objects, the $M$-branes. Specifically, the \textit{$M2$-brane} is an extended object with a
$(2+1)$-dimensional worldvolume which carries the unit electric charge associated with ${\cal A}_3$. Conversely, the \textit{$M5$-brane} carries the unit
magnetic charge of ${\cal A}_3$ and has a $(5+1)$-dimensional worldvolume. The worldvolumes of these objects can be wrapped around various cycles in a
Calabi-Yau and so lead to sources in the effective five-dimensional theory.

The five-dimensional theory has a wealth of interesting supersymmetric solutions including black holes, black strings
and black rings. As indicated above, these each can be embedded into M-theory as a bound state of $M$-branes. In particular, wrapping
an $M2$-brane around one of the basis two-cycles $\o^I$ leads to a five-dimensional solution carrying electric charges
\ben
q_I \equiv - \int_{S^3} {\d S \over \d F^I}= {1\over2\pi^2}\int_{S^3} G_{IJ}  \star F^J~,
\een
where the integral is taken over the asymptotic three-sphere surrounding the black hole. Wrapping an $M5$ brane around one of the basis four-cycles $\o_I$ gives an infinitely extended one-dimensional string,\footnote{This is
not to be confused with a fundamental string, although special configurations are dual to an infinite heterotic string.} carrying the magnetic charges
\ben
p^I = -{1\over 2\pi}\int_{S^2} F^I~,
\een
where one integrates over the asymptotic two-sphere surrounding the string. Further, there are dyonic solutions constructed from both $M2$ and $M5$-branes. These can take the form of either infinite strings with an
extended electric charge density along their volume, or a black ring, where the $M5$-branes contribute non-conserved magnetic dipole
moments.

In the next section, we briefly describe the various supersymmetric solutions of the $D=5$ on-shell supergravity.

\subsection{Solutions of the  two-derivative theory}

In this section we collect the standard black hole, black string, and black ring solutions in the
two-derivative theory.    The remainder of this review consists of finding the analogous solutions
in the presence of higher derivatives.

When the distinguished Killing vector is timelike, the 5D metric and gauge field strengths take the form\footnote{We derive these results is section \ref{sec:timelikesusy}.}
\bea\label{sa} ds^2 &=& e^{4U(x)}(dt+\omega)^2 -e^{-2U(x)} ds_B^2~, \\
F^I &=& d [M^I e^{2U} (dt+\omega)] +\Theta^I~. \label{saa}
\eea
The 4D base metric $ds_B^2$, with coordinates $x$, is required by
supersymmetry to be hyperK\"ahler.\footnote{In general, the base
metric can have indefinite
signature,\cite{Giusto:2004kj}\cdash\cite{Bena:2007kg} but here we
restrict to Euclidean signature.}     $U$, $\omega$, and $\Theta^I$
are, respectively, a function, a one-form, and a two-form on the
base.  The  Bianchi identity implies that $\Theta^I$ is closed.  The
moduli obey the {\it very special geometry} constraint \ben
\label{spec} {1 \over 6}c_{IJK}M^I M^J M^K=1~. \een

Half BPS solutions are those obeying the following set of equations\cite{Gauntlett:2002nw,Gutowski:2004ez}
\bea \Theta^I &=& - \star_4 \Theta^I~,  \label{sb} \\
\nabla^2   (M_I e^{-2U} )& =&{1\over 2} c_{IJK} \Theta^J \cdot \Theta^K~,  \label{sc} \\
d\omega - \star_4 d\omega &=& -e^{-2U} M_I \Theta^I~, \label{sd}
\eea
where the above equations should all be understood as tensor equations on the 4D base space.     When  higher derivatives are added,  the metric and gauge fields
will still take the form (\ref{sa},\ref{saa}),  $\Theta^I$ will remain closed and anti-self-dual,  but the remaining two equations (\ref{sc},\ref{sd})
will be modified.

\subsubsection{Solutions with Gibbons-Hawking base}
\label{sec:twodghsol}
If we take the base space to be a Gibbons-Hawking space\cite{Gibbons:1979zt} then the above equations admit
a general solution in terms of locally harmonic functions.\cite{Gauntlett:2002nw,Gauntlett:2004qy}    The Gibbons-Hawking space is
\ben ds_B^2 = (H^0)^{-1}(dx^5+\chi)^2+ H^0 dx^m dx^m~,
\een
where $H^0$ is harmonic on  $\mathbb{R}^3$, up to isolated singularities, and the 1-form $\chi$ obeys
\ben \vec{\nabla} \times \vec{\chi} = \vec{\nabla} H^0~, \een
and $\vec{\nabla}$ denotes the gradient on $\mathbb{R}^3$. The
compact coordinate $x^5$ has period $4\pi$.

The closed, anti-self-dual 2-form $\Theta^I$ can now be expressed as
\ben\label{am}  \Theta^I = {1 \over 2} (dx^5 +\chi) \wedge \Lambda^I-{1 \over 4} H^0\epsilon_{mnp} \Lambda^I_m dx^n \wedge dx^p~,
\een
with
\ben\label{al}
\Lambda^I = d\left( {H^I \over H^0}\right)~,
\een
for some set of functions $H^I$ harmonic on $\mathbb{R}^3$. We can then solve (\ref{sc}) as
\ben
\label{Msoltwo}
M_I -{1 \over 8} { c_{IJK} H^J H^K \over H^0} = H_I~,
\een
with $H_I$ harmonic on $\mathbb{R}^3$.   Imposing the special geometry constraint (\ref{spec}) allows us to
solve for $e^{-2U}$ in terms of $M^I$ and the harmonic functions.

We next decompose $\omega$ as
\ben\label{ana} \omega = \omega_5(dx^5+ \chi) +\omh~, \een
with
\ben \omh =\omh_mdx^m~.\een
Equation (\ref{sd}) then becomes
\ben\label{apd} \vec{\nabla}\times \vec{\omh} = H^0 \vec{\nabla}\omega_5 -\omega_5 \vec{\nabla} H^0 -{1 \over 2}\Big(H^0 H_I +{1 \over 8}c_{IJK} H^J H^K \Big)\vec{\nabla} \left({H^I \over H^0}\right)~.
\een
Taking the gradient we get
\ben\label{ape}\nabla^2 \omega_5= \nabla^2 \Big[{1 \over 4}{H_I H^I \over H^0} +{1 \over 48}{c_{IJK} H^I H^J H^K\over (H^0)^2}\Big]~,
\een
and we then solve for $\omega_5 $ in terms of another harmonic function $H_0$,
\ben\label{apf} \omega_5={1 \over 4}{H_I H^I \over H^0} +{1 \over 48}{c_{IJK} H^I H^J H^K\over (H^0)^2} +H_0~.
\een
Finally, substituting back into (\ref{apd}) gives
\ben\label{apg} \vec{\nabla}\times \vec{\omh} =  H^0 \vec{\nabla}H_0 -H_0 \vec{\nabla} H^0 -{1 \over 4}[ H_I \vec{\nabla}H^I - H^I \vec{\nabla}H_I] ~.
\een

To summarize, the full solution is  given by choosing harmonic functions  $(H^0, H^I; H_0 ,H_I)$, in terms of which
we have
\bea
M_Ie^{-2U}  &=& {1 \over 8} { c_{IJK} H^J H^K \over H^0}+ H_I~,   \\
e^{-2U} &=& {1 \over 3}\left(H_I M^I +{1 \over 8} { c_{IJK} M^I H^J H^K\over H^0} \right)~,  \\
\Theta^I &=& {1 \over 2} (dx^5 +\chi)\wedge d \left({H^I \over H^0}\right) -{1 \over 4} H^0 \epsilon_{mnp} \p_m  \left({H^I \over H^0}\right) dx^n \wedge dx^p ~,\\
\omega &=& \omega_5 (dx^5 +\chi) +\omh ~,\\
\omega_5 &=& {1 \over 4}{H_I H^I \over H^0} +{1 \over 48} {c_{IJK} H^I H^J H^K \over (H^0)^2} +H_0~,  \\
\vec{\nabla} \times \vec{\omh} &=&  H^0 \vec{\nabla} H_0-H_0 \vec{\nabla} H^0  +{1 \over 4} ( H^I \vec{\nabla} H_I - H_I  \vec{\nabla} H^I )~. \label{se}
\eea
We usually take the harmonic functions to have isolated singularities where $\vec{\nabla}^2 H  \propto
\delta^{(3)}( \vec{x}- \vec{x}_i)$,     in which case (\ref{se}) implies a nontrivial integrability constraint
on the harmonic functions, obtained by taking the divergence of both sides.        We now consider various examples.

\subsubsection{5D static black hole}

This  corresponds to the choice,
\ben\label{sga} H^0 ={1 \over |\vec{x}|}~, ~~H^I=0~;~~ H_0 =0~,~~
H_I = H_I^\infty + {q_I \over 4|\vec{x}|}~.\een
In this case $\omega = \Theta^I =0$, and we have a spherically
symmetric 5D black hole carrying the electric charges $q_I$.
Note that  the Gibbons-Hawking space becomes simply flat
$\mathbb{R}^4$ in nonstandard coordinates.  The near horizon
geometry is AdS$_2 \times S^3$ with scale sizes $\ell_A = \half
\ell_S$.  The near horizon moduli are $M_I = {q_I \over 4
\ell_A^2}$, from which $\ell_A$ can be computed from the special
geometry constraint (\ref{constraint}).   In particular, if we
define the dual charges $q^I$ through
\ben\label{sg}  q_I ={1 \over 2} c_{IJK} q^J q^K~,
\een
then  $\ell_A = \half Q^{1/ 2}$ with\footnote{Note that $Q$ has the same dimension as the physical electric charges $q_I$.}
\ben\label{sh} Q^{3/ 2}  = {1 \over 6} c_{IJK} q^I q^J q^K~.
\een
The entropy is
$S = 2\pi \sqrt{Q^3}$.   Finding explicit expressions in terms of the charges $q_I$ requires that the
equation (\ref{sg}) can be inverted to find $q^I$, but this is only
possible for special choices of $c_{IJK}$.

\subsubsection{5D spinning (BMPV) black hole}

To add rotation we take  $H_0 =  {q_0 \over 16 |\vec{x}|}$  and keep
the other harmonic functions as in (\ref{sga}).   The integer
normalized eigenvalues of the $SU(2)_L \times SU(2)_R$ rotation
group are $(J_L= q_0,~ J_R=0)$. A nontrivial fact is that the
functions $M^I$ and $U$ are unaffected by the inclusion of angular
momentum.\footnote{This will no longer be true with higher
derivatives.}    The near horizon geometry is now described by a
circle fibered over an AdS$_2\times S^2$ base, and the entropy is $S
= 2\pi \sqrt{ Q^3 -{1 \over 4} q_0^2}$, with $\ell_A$ the same as in
the static case.

\subsubsection{5D spinning  black hole on Taub-NUT $=$ 4D black hole}

We now take the base to be a charge $p^0$ Taub-NUT by changing $H^0$ and $H_0$  to
\ben H^0 = 1+{p^0 \over |\vec{x}|}~, \quad H_0 = H_0^\infty+
{q_0\over 16 |\vec{x}|}~.\een
The integrability condition fixes $H_0^\infty = {q_0 \over 16 p^0}$.
Taub-NUT is asymptotically $\mathbb{R}^3 \times S^1$, and so this
gives rise to a 4D black hole after reduction on the $S^1$, with
magnetic charge $(p^0,~ p^I=0)$ and electric charges $(q_0,~ q_I)$.
For $p^0 \neq 1$ the Taub-NUT has a $\mathbb{Z}_{p^0}$ singularity
at the origin; the full 5D metric is smooth, however.    The near
horizon geometry is again a circle fibered over an AdS$_2\times S^2$
base. The near horizon moduli and $\ell_A$ are unchanged, and the
entropy is $S=2\pi \sqrt{p^0 Q^3 -{1\over 4} (p^0 q_0)^2}$.

\subsubsection{5D black ring}

We now choose the harmonic functions
\ben H^0 = {1 \over |\vec{x}|},~  H^I =  {p^I \over  |\vec{x} + R \hat{n}|},~
H_0 = {q_0 \over 16} \left(  {1  \over |\vec{x} + R \hat{n}|}-{1 \over R}\right)  ,~ H_I=H_I^\infty + {\qb_I \over  4|\vec{x} + R \hat{n}|} ~,
\een
where $\hat{n}$ is an arbitrary unit vector in $\mathbb{R}^3$.   The parameter $R$ is
interpreted as the ring radius, and is fixed by the integrability condition to be  $R= { -q_0 \over
4 p^I H^\infty_I}$.      The conserved electric charges measured at infinity are
$q_I = \qb_I +{1 \over 2} c_{IJK} p^J p^K$, and there are also non-conserved magnetic dipole
charges $p^I$.     The near horizon geometry is   AdS$_3 \times S^2$ with
$\ell_A = 2\ell_S = ({1 \over 6} c_{IJK}p^I p^J p^K)^{1\over 3}$.    The near horizon moduli are
$M^I = {p^I \over \ell_A}$.   The black ring entropy is given by setting
$p^0=1$ in formula (\ref{Sring}) below.

\subsubsection{5D black ring on Taub-NUT $=$ 4D  two-center black hole}

This is the most general  case that we'll consider, with harmonic functions
\ben H^0 = 1+{p^0 \over |\vec{x}|},~  H^I =  {p^I \over  |\vec{x} + R \hat{n}|},~
H_0 = {q_0 \over 16} \left(  {1  \over |\vec{x} + R \hat{n}|}-{1 \over R}\right)  ,~ H_I=H_I^\infty + {\qb_I \over 4 |\vec{x} + R \hat{n}|} ~,
\een
and the radius is now determined by $1+{p^0 \over R} = {4 p^I H_I^\infty \over -q_0}$.    From the 4D
perspective this solution is a two-centered geometry, with one center
being a magnetic charge
$p^0$ and the other being a 4D dyonic black hole with charges
proportional to $(p^I,~q_0,~ \qb_I)$.    The entropy of the black hole is\cite{Shmakova:1996nz}
\ben\label{Sring} S =2\pi \sqrt{ p^0 Q^3 - (p^0)^2 J^2}~,
\een
with
\begin{eqnarray} &&Q^{3\over 2} = {1 \over 6} c_{IJK} y^I y^J y^K~, \cr
&& c_{IJK}y^J y^K = 2 \qb_I + {c_{IJK} p^J p^K \over p^0}~, \cr && J =
{q^0 \over 2} +{{1 \over 6} c_{IJK}  p^I p^J p^K \over (p^0)^2}+{p^I
\qb_I \over 2p^0}~.
\end{eqnarray}

\subsubsection{5D black string}

Starting with the black ring but taking $H^0=1$, so that the base becomes $\mathbb{R}^3 \times S^1$,
 yields a 5D black string carrying electric and
magnetic charges.\cite{Bena:2004wv}  An even simpler black string is obtained by taking the limit $H^0 \rightarrow 0$.
In this limit the metric on the base degenerates, but the full 5D solution is well behaved:
\bea\label{si} ds^2 &=& {4 \over ({1 \over 6} c_{IJK} H^I H^J H^K)^{1/3}}\left(dt dx^5 +\hat{H}_0 (dx^5)^2 \right)
-{1 \over 4}  ({1 \over 6} c_{IJK} H^I H^J H^K)^{2/3} d\vec{x}^2~, \cr F^I &=& d \left[ {{1 \over 3} H^I H_I \over   ({1 \over 6} c_{IJK} H^I H^J H^K)^{2/3}} \right] \wedge dx^5 -{1 \over 4} \epsilon_{mnp} \p_m  H^I dx^n \wedge dx^p ~,\cr
 M^I &=& {H^I \over  ({1 \over 6} c_{IJK} H^I H^J H^K)^{1/3}}~,
\eea
with
\ben\label{sj} \hat{H}_0  = H_0 - \half c^{IJ}H_I H_J~,
\een
and $c^{IJ}$ is the inverse of $c_{IJ} = c_{IJK} H^K$.   The near horizon geometry is locally
AdS$_3 \times S^2$ with $\ell_A =2\ell_S = ({1 \over 6}c_{IJK}p^I p^J p^K)^{1\over 3}$, and the
near horizon moduli are $M^I =  {p^I \over \ell_A}$.   Note that these are the same as for the
black ring.   The entropy is
\ben\label{Sbr}
S = 2\pi \sqrt{ {1 \over 6}c_{IJK}p^I p^J p^K \hat{q}_0}~,
\een
with $\hat{q}_0 = q_0 -{1\over 32} C^{IJ}\qb_I \qb_J$, and $C^{IJ}$ is the inverse of $C_{IJ}= c_{IJK}p^K$.
Here we continued to identify $x^5\cong x^5 +4\pi$, but we are to free to drop the identification
and obtain an infinite 5D black string.

Note that by taking the $H^0\rightarrow 0$ limit the formerly timelike Killing vector ${\partial \over \partial t}$ has become null.    The solution (\ref{si}) therefore does not appear directly in the classification based
on timelike supersymmetry, but rather lies in the domain of null supersymmetry.

There are also more general solutions without translation invariance along the string\cite{Gauntlett:2002nw,Gutowski:2005id} but we will refrain from discussing them in detail here.

\section{Conformal Supergravity}\label{confsect}

The low energy limit of a supersymmetric compactification of string
theory is a supergravity theory.  While the Lagrangians of these
theories can in principle be extracted from string S-matrix
computations, in practice a more efficient method is to work
directly in field theory, by demanding invariance under local
supersymmetry.     This approach typically  uses the so-called
Noether method.   In this procedure one starts with an action
invariant under global supersymmetry, and then attempts to
incorporate local invariance iteratively.   For a two-derivative
Lagrangian the possible  matter couplings are usually known, and the
process of constructing the action and transformation rules for the
fields only involves a finite number of steps.
 The incorporation of
higher derivative terms increases enormously the possible terms in
the action and transformation rules. For example, one might start by
including a specific four-derivative interaction and using the two-derivative
transformation rules. This will generate additional four-derivative
terms that will necessitate  modifications to the  supersymmetry
transformations. Now, these modified transformations will generate
six-derivative terms in the Lagrangian and so forth. In general, it may take many steps, if not an
infinite number,
for this  iterative procedure to terminate,
making the construction extremely difficult and tedious.

A more systematic approach to obtaining an invariant action is by constructing off-shell
representations of the supersymmetry algebra. The advantage of this
formalism is that the construction of invariants is well-defined
since the transformation rules are fixed.    Now, the theory we are aiming for is an off-shell
version of Poincar\'e supergravity.   However, it turns out that the construction of off-shell
multiplets is greatly simplified by first considering a theory with a larger gauge invariance,
and then at the end gauge fixing down to Poincar\'e supergravity.
In five dimensions, it
turns out that extending conformal supergravity to a gauge theory
described by the superalgebra $F(4)$ gives an irreducible off-shell
realization of the gravity and matter multiplets. The cost of this
procedure is the inclusion of additional symmetries and compensating
fields, which have no physical degrees of freedom. The construction
of a supergravity theory from the gauge theory is first done by
imposing constraints that identify the gauge theory as a
gravity theory. Then, gauge fixing appropriately the values of
certain compensating fields, one reduces the superconformal theory
to Poincar\'e supergravity. This has been extensively studied for
$d\leqslant6$ superconformal theories; for more details we refer the
reader to
Refs.~\refcite{Bergshoeff:2001hc}--\refcite{Hanaki:2006pj}, \refcite{Mohaupt:2000mj}, and \refcite{Kugo:2000hn}--\refcite{Bergshoeff:1985mz}.

One of the fruitful applications of this formalism is the construction of higher-derivative
Lagrangians. Specifically, we will consider the four-derivative
corrections to $N=2$ supergravity which arise from string
theory.  In five dimensional theories, there is a special mixed
gauge-gravitational Chern-Simons term given by
\begin{equation}\label{lcs}
{\cal L}_{\rm CS} = {c_{2I}\over 24\cdot 16} \epsilon_{abcde}
A^{Ia} R^{bcfg}R^{de}_{~~fg}~.
\end{equation}
The coefficient of this term is precisely determined in string/M-theory by $M5$-brane anomaly cancellation via anomaly inflow.\cite{Duff:1995wd} The constants $c_{2I}$ are understood as the
expansion coefficients of the second Chern class of the Calabi-Yau
threefold on which the eleven-dimensional M-theory is compactified. In
Ref.~\refcite{Hanaki:2006pj} all terms related by supersymmetry to
(\ref{lcs}) were derived using the superconformal formalism.

Our present goal is to simply introduce the main concepts of the
superconformal formalism and the specific results we will use
to study black holes and other such objects in the next sections. In
the following subsections we outline this construction by first
describing the full superconformal algebra and the necessary
identifications to obtain the gravity theory. The field content and
transformation rules for the gravity and matter multiplets are given,
and we briefly explain how to obtain invariant actions for these
multiplets. At the end of this section, after gauge fixing the
superconformal theory, we present the $R^2$ supersymmetric
completion of $N=2$ supergravity.

\subsection{Superconformal formalism}

The five dimensional theory is obtained by first constructing a
gauge theory with gauge symmetry given by the supergroup $F(4)$. The
generators ${\bf X}_A$ and corresponding gauge fields $h_\mu^A$ for this theory are
\begin{eqnarray}\label{ab}\nonumber
{\bf X}_A:~\mathbf{P}_a~,&~ {\bf M}_{ab}~,&~ {\bf D}~,~{\bf
K}_a~,~{\bf U}_{ij}~,~{\bf Q}_i~,~{\bf S}^i~
\\
h_\mu^A:~e^a_{\mu}~,&~\omega^{ab}_{\mu}~,&~b_\mu~,~f^a_{\mu}~,~V^{ij}_{\mu}~,~\psi^i_\mu~,~\phi^i_\mu
\end{eqnarray}
where $a,b=0,\dots,4$ are tangent space indices, $\mu,\nu=0,\dots,4$ are
(curved) spacetime indices and $i,j=1,2$ are $SU(2)$ indices. The generators
of the Poincar\'e algebra are translations ${\bf P}_a$ and Lorentz
transformations ${\bf M}_{ab}$. The special conformal transformations and
dilatations are generated by ${\bf D}$ and ${\bf K}_a$,
respectively. ${\bf U}_{ij}$ is the generator of $SU(2)$ and the
fermionic generators for supersymmetry and conformal supersymmetry
are the symplectic Majorana spinors ${\bf Q}_i$ and ${\bf S}^i$.

The next step is to construct from the superconformal gauge theory a
conformal supergravity theory, {\it i.e.} our symmetries have to be
realized as space-time symmetries rather than internal symmetries.
In order to make this a theory of diffeomorphisms of spacetime one
needs to identify the non-compact translations ${\bf P}_a$ with
general coordinate transformations generated by ${\cal D}_a$. This
procedure is well known\cite{VanProeyen:1999ni} and it is achieved
by applying torsion-less constraints\footnote{In the literature,
(\ref{ca}) are often called the {\it conventional constraints}.}
over the curvatures, which are
\begin{equation}\label{ca}
\hat{R}_{\mu\nu}^a(P)=0~,\quad
\gamma^\mu\hat{R}_{\mu\nu}^i(Q)=0~,\quad \hat{R}_{\mu}^{~a}(M)=0~.
\end{equation}
Here the curvatures are defined as commutators of the conformal supercovariant
derivatives, that is
\begin{equation}
[\hat{\cal D}_\mu,\hat{\cal D}_\nu]=-\hat{R}^A_{\mu\nu}{\bf X}_A~,
\end{equation}
with
\begin{equation}\label{ba}
\hat{\cal D}_\mu=\partial_\mu-h^A_\mu {\bf X}_A~,
\end{equation}
where we are summing over ${\bf X}_A=\big\{{\bf M}_{ab},~ {\bf D},~ {\bf
K}_a,~ {\bf U}^{ij},~{\bf Q}_i, ~{\bf S}^i\big\}$. By solving (\ref{ca}),
some of the gauge fields will become dependent fields. Assuming that
the vielbein $e^a_{\mu}$ is invertible, the first constraint will
determine the connection $\omega^{ab}_\mu$. The second and third
constraints fix $\phi^i_\mu$ and $f^a_\mu$, respectively, making
them dependent fields as well.\cite{Bergshoeff:2001hc,Fujita:2001kv}

The final step in constructing the off-shell gravity multiplet is
adding auxiliary fields. In order to understand this, it is useful
to track the number of independent bosonic and fermionic components.
Before imposing (\ref{ca}) the gauge fields are composed of 96
bosonic and 64 fermionic gauge fields. Explicitly, the number of
independent gauge fields is the total number of components $h_\mu^A$
minus the number of generators ${\bf X}_A$:
\begin{eqnarray}\nonumber
h_\mu^A:&~e^a_{\mu}~,~\omega^{ab}_{\mu}~,&~b_\mu~,~f^a_{\mu}~,~V^{ij}_{\mu}~,~\psi^i_\mu~,~\phi^i_\mu~.\\
\# :& ~20~,~40~,&~4~,~~20~,~~12~,~~32~,~32~.
\end{eqnarray}
The curvature constraints fix the connections $\o^{ab}_\mu$, $\phi^i_\mu$ and $f^a_\mu$, eliminating their degrees of freedom. The new number of degrees of freedom is then the total number of components of the remaining gauge fields minus the total number of the generators ${\bf X}_A$. This counting results in 21+24 degrees of freedom. Adding auxiliary fields, which will include
extra transformation rules and modifications to the supersymmetry algebra, solves this final mismatch in the number of bosonic and fermionic degrees of freedom. The procedure has been outlined in
Ref.~\refcite{Bergshoeff:1985mz}.

\subsubsection{Weyl multiplet}

The construction sketched above gives the irreducible Weyl multiplet (denoted by {\bf W})
describing $32+32$ degrees of freedom. The multiplet contains the
following fields,
\begin{equation}
e^{a}_\mu~,~V^{ij}_{\mu}~,~b_\mu~,~v_{ab}~,~D~,~\psi^i_\mu~,~\chi^i~.
\end{equation}
As we mentioned before, in order to have a closed algebra it is
necessary to include compensators, {\it i.e.} auxiliary fields. For the
Weyl multiplet, the non-propagating fields are an antisymmetric two-form tensor
$v_{ab}$, a scalar field $D$ and an $SU(2)$ Majorana spinor
$\chi^i$. The ${\bf Q}$-${\bf S}$ supersymmetry and ${\bf K}$
transformation rules for the bosonic fields in the multiplet are
\begin{eqnarray}\label{bc}
\delta e^a_\mu&=&-2i{\bar \epsilon}\gamma^a\psi_\mu~,\nonumber \\
\delta
V_{\mu}^{ij}&=&-3i{\bar\epsilon}^{i}\phi^{j}_\mu+2i{\bar\epsilon}^{i}\gamma\cdot
v\psi^{j}_\mu-{i\over8}{\bar\epsilon}^{i}\gamma_\mu\chi^{j}+3i{\bar\eta}^{i}\psi^{j}_\mu
+(i\leftrightarrow j)~,\nonumber\\
\delta b_\mu&=&-2i{\bar \epsilon}\phi_\mu-2i{\bar\eta}\psi_\mu-2\xi_{K\mu}~,\nonumber\\
\delta
v_{ab}&=&-{i\over8}{\bar\epsilon}\gamma_{ab}\chi-{3\over2}i{\bar\epsilon}\hat{R}_{ab}(Q)~,\nonumber\\
\delta D&=&-i{\bar\epsilon}\gamma^a\hat{\cal
D}_a\chi-8i{\bar\epsilon}\hat{R}_{ab}(Q)v^{ab}+i{\bar\eta}\chi~,
\end{eqnarray}
and for the fermionic fields we have
\begin{eqnarray}\label{bca}
\delta \psi^i_\mu&=&{\cal D}_\mu\epsilon^i+{1\over
2}v^{ab}\gamma_{\mu a b}\epsilon^i-\gamma_\mu\eta^i~,\nonumber
\\
\delta\chi^i&=&D\epsilon^i-2\gamma^c\gamma^{ab}\hat{\cal
D}_av_{bc}\epsilon^i+\gamma\cdot\hat{R}^i_{~j}(U)\epsilon^j-2\gamma^av^{bc}v^{de}\epsilon_{abcde}\epsilon^i
+4\gamma\cdot v\eta^i~,
\end{eqnarray}
where $\delta\equiv\bar{\epsilon}^i {\bf Q}_i+\bar{\eta}^i{\bf
S}_i+\xi^a_K{\bf K}_a$. The superconformal covariant derivative $\hat{\cal D}_a$ appearing in
(\ref{bc}) and (\ref{bca}) is defined in (\ref{ba}). The un-hatted derivative ${\cal D}_a$ is defined similarly but the sum is only over ${\bf X}_A = \big\{{\bf M}_{ab}, {\bf D}, {\bf U}_{ij}\big\}$.

\subsubsection{Matter multiplets}

In this section we will describe the properties and transformation
rules for matter multiplets coupled to five-dimensional
conformal supergravity. The three matter multiplets relevant for our
purposes are the vector multiplet, hypermultiplet and linear
multiplet.\\
\begin{itemize}
\item{\bf Vector multiplet:} The off-shell components of the vector multiplet ${\bf V}$
are,
\begin{equation}\nonumber
M^I~,~A^I_\mu~,~Y^{I}_{ij}~,~\Omega^{I}_i~.
\end{equation}
$M^I$ are scalar fields and $A^I_\mu$ gauge fields. The multiplet
also contains a $SU(2)$ triplet auxiliary field $Y^I_{ij}$ and the
$SU(2)$ Majorana spinor $\Omega^I_i$. The index $I$ labels the
generators of the gauge group $G$. For brevity, we consider $G$ as
$n_V+1$ copies of $U(1)$; the generalization to non-Abelian gauge groups is discussed in Refs.~\refcite{Bergshoeff:2001hc} and \refcite{Fujita:2001kv}. The
${\bf Q}$ and ${\bf S}$ transformation rules for the fermion in the
vector multiplet is,
\begin{equation}\nonumber
\delta\Omega^{Ii}=-{1\over4}\gamma\cdot\hat{F}^I\epsilon^i-{1\over2}\gamma^a\hat{\cal
D}_aM^I\epsilon^i +Y^{Ii}_{~j}\epsilon^j-M^I\eta^i~,
\end{equation}

with the field strength given by
\begin{equation}\nonumber
\hat{F}^I_{\mu\nu}=2\partial_{[\mu}A^I_{\nu]}+4i{\bar\psi}_{[\mu}\gamma_{\nu]}\Omega^I
-2i{\bar\psi_\mu}{\psi_\nu}M^I~.
\end{equation}

\item{\bf Hypermultiplet:} The components of ${\bf H}$, the hypermultiplet, are
\begin{equation}\nonumber
{\cal A}^\alpha_i~,~\zeta^\alpha~,~{\cal F}^i_\alpha~,
\end{equation}
where the index $\alpha=1\cdots 2r$ represents $USp(2r)$. The
scalars ${\cal A}^\alpha_i$ are anti-hermitian, $\zeta^\alpha$ is a
Majorana spinor and ${\cal F}^i_\alpha$ are auxiliary fields. For
our discussion, the relevant supersymmetry transformation is given
by
\begin{equation}\label{ce}
\delta \zeta^\alpha=\gamma^a\hat{\cal D}_a{\cal
A}_j^\alpha\epsilon^j-\gamma\cdot v{\cal
A}^\alpha_j\epsilon^j+3{\cal A}^\alpha_j\eta^j~.
\end{equation}

As we will discuss shortly, (\ref{ce}) will allow us to consistently
preserve the Poincar\'e gauge by performing a compensating  ${\bf
S}^i$ transformation, {\it i.e.} fix $\eta^i$ in terms of $\epsilon^i$.
The covariant derivative appearing in (\ref{ce}) is
\begin{equation}\nonumber
\hat{\cal D}_\mu{\cal
A}^\alpha_i=\left(\partial_\mu-{3\over2}b_\mu\right){\cal
A}^\alpha_i-V_{\mu i}^{~j}{\cal A}^{\alpha}_j-2i\bar{\psi}_{\mu
i}\zeta^{\alpha}~.
\end{equation}

\item{\bf Linear multiplet:} The components of the linear multiplet ${\bf L}$
are
\begin{equation}\nonumber
L^{ij}~,~E^{a}~,~N~,~\varphi^i~.
\end{equation}
The scalar $L^{ij}$ is symmetric in $SU(2)$ indices, $E_a$ is a
vector, $N$ is a scalar and $\varphi^i$ is a $SU(2)$ Majorana
spinor. In addition, the algebra will close if the vector satisfies
$\hat{\cal D}^aE_a=0$. The transformation rule for the scalar
$L^{ij}$ reads
\begin{equation}\label{cf}
\delta L^{ij}=i{\bar\epsilon}^i\varphi^j+(i\leftrightarrow j)~.
\end{equation}
An interesting property of the linear multiplet is that it can be
used as a ``kinetic multiplet'', {\it i.e.} the components of a
matter (or Weyl) multiplet can be embedded into the linear
multiplet. These embedding formulae are constructed by noticing that
any symmetric, real bosonic combination of fields which is invariant
under ${\bf S}$-supersymmetry leads to the transformation rule
in (\ref{cf}) with the appropriate identification of $\varphi^i$.
The construction of the remaining components is done by repeated
supersymmetry transformations. We denote such an embedding of a multiplet ${\bf X}$ into a linear multiplet as ${\bf L}[{\bf X}]$. This is the key ingredient that
allows the construction of invariants, which is discussed in the
next section.
\end{itemize}

\subsection{Constructing invariant actions}

The first step towards constructing Lagrangians is done by
identifying a quantity invariant under supersymmetry
transformations. As discussed in Ref.~\refcite{Fujita:2001kv}, the
contraction of some given linear and vector multiplets
\begin{eqnarray}\label{dc}\nonumber
{\cal L}({\bf L\cdot V})\equiv & &Y^{ij}\cdot
L_{ij}+2i\bar{\Omega}\cdot\varphi+2i\bar{\psi}_i^a\gamma_a\Omega_j\cdot
L^{ij}\\&-&{1\over2}A_a\cdot\left(E^a-2i\bar{\psi}_b\gamma^{ba}\varphi+
2i\bar{\psi}_b^{(i}\gamma^{abc}\psi^{j)}_cL_{ij}\right)\nonumber\\
&+&{1\over2}M\cdot\left(N-2i\bar{\psi}_b\gamma^b\varphi-
2i\bar{\psi}^{(i}_a\gamma^{ab}\psi^{j)}_bL_{ij}\right)~,
\end{eqnarray}
transforms as a total derivative under all gauge transformations in
(\ref{ab}). By exploiting the kinetic property of the linear
multiplet, {\it i.e.} embedding of the Weyl or matter multiplets
into ${\bf L}$, the invariant density (\ref{dc}) is the building
block for constructing supersymmetric actions. Here we will only present the
construction of the invariants relevant for $N=2$ ungauged
supergravity.

First, let us consider the dynamics of the vector multiplets. The
action we are pursuing should describe a Yang-Mills system coupled
to gravity. In five dimensions the gauge field interactions will
have a Chern-Simons term of the form $A\wedge F\wedge F$. By
inspecting (\ref{dc}), the Chern-Simons term can be included in
$A_a\cdot E^a$ by appropriately embedding ${\bf L}[{\bf V}]$, where
the field $E_a$ will take the form
\ben
E_a\sim\epsilon_{abcde}F^{bc}F^{de}+\ldots~.
\een
After carefully
performing this embedding, the bosonic terms in (\ref{dc}) gives the
following Lagrangian for the vector multiplet,
\begin{eqnarray}\label{da}
{\cal L}_{B,V}({\bf L\cdot
V})&=&-Y^{ij}\cdot L_{ij}[{\bf V}]+{1\over2}A_a\cdot E^a[{\bf V}]-{1\over2}M\cdot N[{\bf V}]\nonumber\\
&=&{\cal N}\left({1\over2}D-{1\over4}R+3v^2\right)+2{\cal
N}_Iv^{ab}F^I_{ab}+{1\over4}{\cal N}_{IJ}F^I_{ab}F^{Jab}\nonumber\\&
-&{\cal N}_{IJ}\left({1\over2}{\cal D}^aM^I{\cal
D}_aM^J+Y^I_{ij}Y^{Jij}\right)+{1\over24e}c_{IJK}A^I_aF^J_{bc}F^K_{de}\epsilon^{abcde}~.
\end{eqnarray}
At this level, the function ${\cal N}$ is defined as an arbitrary
cubic function of the scalars $M^I$, a condition that arises such
that the embedding of the linear multiplet preserves the symmetry
transformations. Using the same notation as in the two-derivative
on-shell theory, we can write
\begin{equation}\label{Ndef}
{\cal N}={1\over6}c_{IJK}M^IM^JM^K~,
\end{equation}
where $c_{IJK}$ are constants, and ${\cal N}_I$ and ${\cal N}_{IJ}$
are derivatives of ${\cal N}$ with respect to $M^I$
\ben\label{Ndeftwo}
{\cal N}_I \equiv \frac{\p{\cal N}}{\p M^I}  = \frac12 c_{IJK}M^JM^K~, ~~~~~ {\cal N}_{IJ} \equiv \frac{\p^2{\cal N}}{\p M^I \p M^J}  =  c_{IJK}M^K~.
\een
Note that in
the off-shell theory ${\cal N}$ is not fixed, in contrast to (\ref{constraint}) in the on-shell theory.

The hypermultiplet can also be embedded in the linear multiplet,
${\bf L}[{\bf H}]$. The resulting bosonic sector of the Lagrangian
contains the kinetic terms for the hyperscalar ${\cal A}^i_\alpha$ and its
coupling to the Weyl multiplet,
\begin{eqnarray}\label{dab}
{\cal L}_{B,H}({\bf L\cdot V})
&=&-Y^{ij}\cdot L_{ij}[{\bf H}]+{1\over2}A_a\cdot E^a[{\bf H}]-{1\over2}M\cdot N[{\bf H}]\nonumber\\
&=&2{\cal D}^a{\cal A}^\alpha_i{\cal D}_a{\cal A}^i_\alpha+{\cal
A}^2\left({1\over4}D+{3\over8}R-{1\over2}v^2\right)~.
\end{eqnarray}
The hypermultiplet can be decoupled from the remaining fields in the
theory. In this framework, the decoupling is understood as the gauge
fixing that will reduce the superconformal symmetries to the
super-Poincar\'e group. In a suitable gauge, we will find that (\ref{da}) and
(\ref{dab}) lead to a canonical normalization for the Ricci scalar, {\it i.e.} the conventional Einstein-Hilbert term.

As we mentioned at the beginning of this section, we are interested
in studying the supersymmetric Lagrangian containing the mixed
gauge-gravitational Chern-Simons term (\ref{lcs}). Similar to the
construction of the vector Lagrangian, the term $A_a\cdot E^a$
guides the form of the embedding, where now we need
\ben
E_a\sim\epsilon_{abcde}\hbox{Tr}(\hat{R}^{bc}\hat{R}^{de})+\ldots~.
\een
This requires an embedding of the Weyl multiplet into the linear multiplet, ${\bf L}[{\bf W}^2]$, and the invariant (\ref{dc}) turns
into
\begin{eqnarray}\label{dac}
{\cal L}_{B,W}({\bf L\cdot V}) &=&{c_{2I}\over24}\left(-Y^{Iij}\cdot
L_{ij}[{\bf W^2}]+{1\over2}A^I_a\cdot E^a[{\bf W^2}]
-{1\over2}M^I\cdot N[{\bf W^2}]\right)\nonumber\\
&=& {c_{2I}\over24}\bigg[  {1\over 16} \epsilon_{abcde} A^{Ia}
C^{bcfg}C^{de}_{~~fg}
-{1\over12}\epsilon_{abcde}A^{Ia}\hat{R}^{bcij}(U)\hat{R}^{de}_{~~ij}(U)
\nonumber\\
&&~~~~~~+ {1\over8}M^I C^{abcd}C_{abcd}-{1 \over 3}M^I
\hat{R}^{abij}(U)\hat{R}_{abij}(U)
-{4\over3}Y^I_{ij}v_{ab}\hat{R}^{abij}(U) \nonumber\\
&&~~~~~~+{1 \over 3}M^I C_{abcd} v^{ab}v^{cd}+{1\over 2}F^{Iab} C_{abcd}
v^{cd} +{1 \over 12}M^I D^2+{1 \over 6}F^{Iab}v_{ab}D \nonumber\\
&&~~~~~~+{8\over 3}M^I v_{ab} \hat{\cal D}^b \hat{\cal D}_c v^{ac}
+{4\over 3} M^I {\hat{\cal D}}^a v^{bc} {\hat{\cal D}}_a v_{bc}+
{4\over 3} M^I {\hat{\cal D}}^a v^{bc} {\hat{\cal D}}_b v_{ca} \nonumber\\
&&~~~~~~ -{2\over 3} M^I \epsilon_{abcde}v^{ab}v^{cd}{\hat{\cal D}}_f
v^{ef}+
{2\over 3} F^{Iab}\epsilon_{abcde}v^{cf} {\hat{\cal D}}_f v^{de}\nonumber\\
&&~~~~~~
 +F^{Iab}\epsilon_{abcde}v^c_{~f}{\hat{\cal D}}^d v^{ef} -{4 \over 3}F^{Iab}v_{ac}v^{cd}v_{db}
 -{1 \over 3} F^{Iab} v_{ab}v^2\nonumber\\
&&~~~~~~ +4 M^I v_{ab} v^{bc}v_{cd}v^{da}-M^I (v^{ab}v_{ab})^2 \bigg]~,
\end{eqnarray}
where the overall coefficient in ${\cal L}_{B,W}$ is fixed by the
anomaly cancellation condition. The double covariant derivative of
$v_{ab}$ reads
\begin{equation}
v_{ab}\hat{\cal D}^b\hat{\cal D}_cv^{ac}=v_{ab}{\cal D}^b{\cal
D}_cv^{ac}-{2\over3}v^{ac}v_{cb}R^b_{~a}-{1\over12}v^2R~,
\end{equation}
where ${\cal D}_a$ is the covariant derivative with respect to ${\bf
M}_{ab}~,~{\bf D}~,~{\bf U}^{ij}$. In (\ref{dac}), $C_{abcd}$ is the
Weyl tensor
\begin{equation}
C_{abcd}=R_{abcd}-{2\over3}\left(g_{a[c}R_{d]b}-g_{b[c}R_{d]a}\right)+{1\over6}g_{a[c}g_{d]b}R~,
\end{equation}
with $R_{abcd}$ the Riemann tensor, $R_{ab}$ and $R$ the Ricci
tensor and scalar, respectively. The bosonic components of the
curvature tensor associated to the $SU(2)$ symmetry is given by
\begin{equation}
\hat{R}_{\mu\nu}^{~~ij}(U)=\partial_{\mu}V^{ij}_{\nu}-V_{\mu
k}^{~i}V_{\nu}^{kj}-(\mu\leftrightarrow \nu)~.
\end{equation}
Finally, the bosonic terms of the five dimensional Lagrangian
for the superconformal theory are given by
\begin{equation}\label{db}
{\cal L}_B={\cal L}_{B,V}+{\cal L}_{B,H}+{\cal L}_{B,W}~.
\end{equation}

\subsection{Poincar\'e supergravity}

Our main interest is five dimensional Poincar\'e supergravity.
Starting from the superconformal theory, it is possible to gauge fix
the additional conformal symmetries and consistently obtain an
off-shell representation of $N=2$ supergravity. This requires
choosing the {\it vevs} of certain fields associated with the
conformal group and the $R$-symmetry, which spontaneously breaks the
superconformal symmetry. The procedure does not make use of the
equations of motion, and the number of symmetries and degrees of
freedom eliminated is balanced. This makes the process reversible
and therefore, the conformal theory is gauge equivalent to
Poincar\'e supergravity.\cite{de Wit:1983tq}

We will start by considering the Weyl multiplet coupled to $n_V+1$
vector fields and one hypermultiplet.\footnote{One could include
additional hyper multiplets (or other matter fields not discussed
here), which would require the inclusion of non-dynamical multiplets
in order to consistently eliminate the extra gauge symmetries,
obscuring the procedure.} The Lagrangian describing the bosonic
sector of the conformal theory is given by (\ref{db}). The first
step towards gauge fixing the theory is to notice that the
dilatational field $b_\mu$ only appears in the Lagrangians (\ref{da}), (\ref{dab}),
(\ref{dac}) through the covariant derivatives of the matter fields.
This allows us to fix special conformal transformations by choosing
the gauge $b_\mu=0$.

In order to have the canonical normalization for the Ricci scalar in
(\ref{db}), our gauge choice for the dilatational group is ${\cal
A}^2=-2$. Notice that in the two-derivative theory this gauge
choice, combined with the equations of motion of the auxiliary field
$D$, gives the very special geometry constraint ${\cal N}=1$.

The $SU(2)$ symmetry is fixed by identifying the indices in the
hypermultiplet scalar, {\it i.e.} ${\cal
A}^i_\alpha=\delta^i_\alpha$. Finally, since we restricted the
discussion to an Abelian gauge group for the vector multiplet, the
auxiliary fields $V^{ij}_\mu$ and $Y^{I}_{ij}$ will only appear
quadratically in (\ref{da}) and (\ref{dab}). Therefore, it is
appropriate for the ungauged theory to set both $Y^{I}_{ij}$
and $V^{ij}_\mu$ to zero.

Summarizing, our gauge choice is given by:
\begin{eqnarray}\label{ea}
{\cal A}^i_{\alpha}=\delta^i_\alpha~,~{\cal A}^2=-2~,\cr
b_\mu=0~,~V^{ij}_\mu=0~,~Y^{I}_{ij}=0~.
\end{eqnarray}
Substituting (\ref{ea}) into (\ref{da}) and (\ref{dab}) gives rise to the
two-derivative Lagrangian
\begin{eqnarray}\label{offtwod}
{\cal L}_0&=& -{1\over2}D-{3\over4}R+v^2+{\cal
N}\left({1\over2}D-{1\over4}R+3v^2\right)+2{\cal
N}_Iv^{ab}F^I_{ab}\nonumber\\
& &+{\cal
N}_{IJ}\left({1\over4}F^I_{ab}F^{Jab}+{1\over2}\partial_aM^I\partial^aM^J\right)+{1\over24}
c_{IJK}A^I_{a}F^{J}_{bc}F^{K}_{de}\epsilon^{abcde}~.
\end{eqnarray}
Similarly, the higher-derivative Lagrangian (\ref{dac}) becomes
\bea\label{offfourd}
{\cal L}_1& = {c_{2I} \over 24}
\Big(& {1\over 16} \epsilon_{abcde} A^{Ia} R^{bcfg}R^{de}_{~~fg} +
{1 \over 8}M^I C^{abcd}C_{abcd} +{1 \over 12}M^I D^2 +{1 \over
6}F^{Iab}v_{ab}D \nonumber\\& &+{1 \over 3}M^I C_{abcd}
v^{ab}v^{cd}+{1\over 2}F^{Iab} C_{abcd} v^{cd}
+{8\over 3}M^I v_{ab} {\cal D}^b {\cal D}_c v^{ac} \nonumber\\
&&-\frac{16}{9} M^I v^{ac}v_{cb}R_a^{~b} -\frac29 M^I v^2 R +{4\over 3} M^I {{\cal D}}^a v^{bc} {{\cal D}}_a v_{bc} + {4\over
3} M^I {{\cal D}}^a v^{bc} {{\cal D}}_b v_{ca}\nn\\
& & -{2\over 3} M^I
\epsilon_{abcde}v^{ab}v^{cd}{{\cal D}}_f v^{ef}+
{2\over 3} F^{Iab}\epsilon_{abcde}v^{cf} {{\cal D}}_f v^{de}
 +F^{Iab}\epsilon_{abcde}v^c_{~f}{{\cal D}}^d v^{ef}\nonumber\\
& & -{4 \over 3}F^{Iab}v_{ac}v^{cd}v_{db}-{1 \over 3} F^{Iab}
v_{ab}v^2 +4 M^I v_{ab} v^{bc}v_{cd}v^{da}-M^I (v^2)^2\Big)~,
\eea
where ${\cal N}$, ${\cal N}_I$ and ${\cal N}_{IJ}$ are defined in (\ref{Ndef}) and (\ref{Ndeftwo}). The symbol ${\cal D}_a$ now refers to the usual covariant derivative of general relativity and should not be confused with the conformal covariant derivatives of the previous sections. Indeed, the presence of the auxiliary fields $D$ and $v_{ab}$ are the only remnants of the superconformal formalism.

As we mentioned before, the supersymmetry transformations are also
affected by the gauge fixing. In particular the parameter $\eta^i$
associated to ${\bf S}$-supersymmetry is fixed. The BPS condition
for the hypermultiplet fermion follows from (\ref{ce})
\begin{equation}\label{eaaa}
\gamma^a\hat{\cal D}_a{\cal A}_j^\alpha\epsilon^j-\gamma\cdot v{\cal
A}^\alpha_j\epsilon^j+3{\cal A}^\alpha_j\eta^j=0~.
\end{equation}
For the field configuration (\ref{ea}), we can solve (\ref{eaaa}) for $\eta^i$,
\begin{equation}\label{eb}
\eta^i={1\over3}\gamma\cdot v\epsilon^i~.
\end{equation}
Replacing (\ref{eb}) in the transformation rules for the remaining
fermionic fields, we obtain the following the residual supersymmetry transformations\footnote{We now leave the $i$ indices implicit since they play very little role in what follows.  See Ref. \refcite{Lapan:2007jx} for a
discussion of this point.}
\begin{eqnarray}\label{susy}
\delta\psi_\mu&=&\left({\cal D}_\mu+{1\over2}v^{ab}\gamma_{\mu
ab}-{1\over3}\gamma_\mu\gamma\cdot v\right)\epsilon~, \nonumber\\
\delta\Omega^{I}&=&\left(-{1\over4}\gamma\cdot
F^I-{1\over2}\gamma^a\partial_aM^I-{1\over3}M^I\gamma\cdot v
\right)\epsilon~,\nonumber\\ \delta \chi
&=&\left(D-2\gamma^c\gamma^{ab}{\cal
D}_av_{bc}-2\gamma^a\epsilon_{abcde}v^{bc}v^{de}+
{4\over3}(\gamma\cdot v)^2\right)\epsilon~.
\end{eqnarray}
It is the vanishing of these transformations which constitute the BPS conditions in the off-shell Poincar\'e supergravity.

\section{Off-shell Poincar\'e Supergravity}

In the last section we provided the off-shell Lagrangian and supersymmetry transformations for five-dimensional supergravity with $R^2$ terms. This forms the starting point for our detailed analysis of corrections to black holes and similar objects, so let us briefly summarize the theory and make some general comments before investigating any specific solutions.

The degrees of freedom from the Weyl (gravity) multiplet are \ben
e_\mu^{~a}~, \psi_\mu ~, v_{ab}~, D~, \chi~, \een where $e_\mu^{~a}$
is the vielbein, $\psi_\mu$ is the gravitino, $v_{ab}$ is an
anti-symmetric two-form, $D$ is a scalar, and $\chi$ a Majorana
fermion. The last three fields are auxiliary fields, representing
non-physical degrees of freedom. Coupled to the above fields are a
number of $U(1)$ vector multiplets containing \ben A^I_{\mu}~, M^I~,
\O^I~, \een which are the gauge fields, K\"ahler moduli, and
gauginos, respectively. The index $I=1\dots n_V+1$ runs over all of
the gauge fields in the theory, although only $n_V$ of them are dynamically independent.

The theory is described by the action
\ben
S= {1\over4\pi^2}\int ~d^5 x \sqrt{g}\lt({\cal L}_0 +{\cal L}_1\rt)~,
\een
where the bosonic part of the leading (two-derivative) Lagrangian is (\ref{offtwod}) and the bosonic higher derivative corrections are described by (\ref{offfourd}). The supersymmetry variations of the fermionic fields around bosonic backgrounds are given by (\ref{susy}).

Note that the four-derivative Lagrangian (\ref{offfourd}) is proportional to the constants $c_{2I}$, which can be thought of as the effective expansion parameters of the theory. Furthermore, the expansion coefficients $c_{2I}$ make no appearance in the supersymmetry transformations (\ref{susy}) for the supersymmetry algebra is completely off-shell, {\it i.e.} independent of the action of the theory.

\subsection{Integrating out the auxiliary fields}

We have termed the fields $v_{ab}$, $D$, and $\chi$ as auxiliary fields. This nomenclature is clear from the viewpoint of the superconformal symmetry of Section \ref{confsect}, where these fields were added to compensate for the mismatch between the number of bosonic and fermionic degrees of freedom. However, focusing on the bosonic fields, from the point of view of the leading-order action (\ref{offtwod}) the fields $v_{ab}$ and $D$ are also auxiliary variables in the sense of possessing algebraic equations of motion. It can be easily seen that substituting the equations of motion for $v_{ab}$ and $D$ into (\ref{offtwod}) leads to the on-shell two-derivative supergravity Lagrangian (\ref{onshelllag}), complete with the very special geometry constraint (\ref{constraint}).

When the higher-derivative corrections encapsulated in (\ref{offfourd}) are taken into account, the two-form $v_{ab}$ no longer has an algebraic equation of motion. It seems fair to now ask in what sense it is still an auxiliary field. To sensibly interpret this, we must recall that the Lagrangian including stringy corrections should be understood as an effective Lagrangian, {\it i.e.} part of a derivative expansion suppressed by powers of the five-dimensional Planck scale. Thus, it is only sensible to integrate out the auxiliary fields iteratively, in an expansion in inverse powers of the Planck mass or, equivalently, in powers of the constants $c_{2I}$.

\subsection{Comments on field redefinitions}

In higher-derivative theories of gravity, the precise form of the Lagrangian is ambiguous due to possible field redefinitions. For example, one may consider
\ben\label{redef}
g_{\mu\nu} \rightarrow g_{\mu\nu} + a R g_{\mu\nu} + bR_{\mu\nu} +\dots~,
\een
for some dimensionful constants $a$ and $b$, or generalizations involving the matter fields.
Field redefinitions leave the leading order Einstein-Hilbert action invariant, but can change the coefficients and form of the $R^2$ terms. Since they mix terms of different orders in
derivatives it is generally ambiguous to label certain terms as ``two-derivative'' or
``higher-derivative''.

One of the advantages of the off-shell formalism we employ is that it addresses these ambiguities.
The reason is that the off-shell supersymmetry transformations are independent of
the action, yet they do not mix different orders in derivatives (if we assign the auxiliary fields $v_{ab}$ and $D$ derivative orders of one and two, respectively). General field redefinitions of the form (\ref{redef}) would modify the supersymmetry algebra and mix orders of derivatives. Thus if we restrict
to variables where the supersymmetry transformations take their off-shell forms, {\it e.g.} in (\ref{susy}), then most of the field redefinition ambiguity is fixed. In our formalism it is therefore meaningful
to label terms by their order in derivatives.\cite{Hanaki:2006pj}

\subsection{Modified very special geometry}

In the on-shell theory, there is a constraint imposed by hand
\ben\label{vsg} {\cal N} \equiv \frac16 c_{IJK}M^IM^JM^K =1~. \een
This is known as the {\it very special geometry} constraint and
indicates that all of the K\"ahler moduli are not independent
fields. Interestingly, the off-shell formalism does not require this
to be imposed externally. Rather, the equation of motion for $D$
following from the two-derivative Lagrangian (\ref{offtwod}) is
precisely this condition. This immediately implies that (\ref{vsg})
does not hold in the presence of higher-derivative corrections since
$D$ also appears in the four-derivative Lagrangian (\ref{offfourd}).
Indeed, the $D$ equation of motion following from the full
Lagrangian ${\cal L}_0 + {\cal L}_1$ is \ben\label{mvsg} {\cal N} =
1- \frac{c_{2I}}{72}\lt[ F^I_{ab}v^{ab} +M^I D \rt] ~. \een

Very special geometry is an interesting mathematical structure in its own right and the modified very special geometry is also likely to be an interesting structure, but we will have little to say about that here. Indeed, it would be of much interest to explore this topic further. For the present purposes, we use (\ref{mvsg}) as just another equation in specifying our solutions.

\subsection{Isometries and projections on Killing spinors}

In the following we will be investigating supersymmetric solutions
to the theory described above. While we will consider maximally
supersymmetric solutions, for which the supersymmetry parameter
$\epsilon$ in the BPS conditions is understood to be unconstrained,
we will also discuss asymptotically flat solutions such as black
holes and black strings. These asymptotically flat solutions break
some fraction of supersymmetry and so $\epsilon$ is expected to
satisfy some sort of projective constraint(s). We can derive this
projection in the following way (analogous to that of
Ref.~\refcite{Gauntlett:2002nw} in the on-shell formalism) which is
generally applicable. Assume the existence of some spinor $\epsilon$
satisfying the BPS condition from the gravitino variation
\ben\label{kill} \lt\{{\cal D}_\mu +\frac16 v^{ab} e_\mu^{~c}
\lt(\g_{abc} -4 \eta_{ac}\g_b\rt)\rt\}\epsilon=0~. \een Now define
the vector, $V_\mu = -\bar{\epsilon} \g_\mu \epsilon$ and use
(\ref{kill}) to compute its covariant derivative \bea
{\cal D}_\mu V_\nu &=& -\frac16 v^{ab} e_\mu^{~c} e_\nu^{~d} \bar{\epsilon}\lt([\g_{abc} ,\g_d] + 4 \eta_{ac} \{ \g_b , \g_d \}\rt)\epsilon~, \nn \\
&=&-\frac16 v^{ab} e_\mu^{~c} e_\nu^{~d} \bar{\epsilon}\lt(\g_{abcd} + 8 \eta_{ac} \eta_{bd}\rt)\epsilon~.
\eea
The right-hand side in the second line is anti-symmetric under exchange of $\mu$ and $\nu$, thus $V_\mu$ is a Killing vector. One can now use various
Fierz identities\cite{Gauntlett:2002nw} to derive a projection obeyed by the Killing spinor
\ben\label{projection}
V^\mu \g_\mu \epsilon = -f \epsilon~,
\een
where $f = \sqrt{V^\mu V_\mu}$. Since there is only one condition on $\epsilon$, this argument leads to solutions which preserve half of the supersymmetries.

The details of the supersymmetry analysis are qualitatively
different for solutions with a null isometry ($f^2=0$) and those
with a timelike isometry ($f^2>0$). We will study these two cases in
turn in subsequent sections. The analysis proceeds as follows. One
introduces a metric {\it ansatz} with an isometry that we identify
with $V_\mu = -\bar{\epsilon} \g_\mu \epsilon$. This determines a
projection obeyed by the Killing spinor via (\ref{projection}). One
then uses the BPS conditions to obtain as much information as
possible about the undetermined functions of the {\it ansatz}. In
the off-shell formalism, the results of this analysis are completely
independent of the action. Equations of motion, which do of course
depend on the precise form of the action, are then imposed as needed
to completely specify the solution.

\section{Attractor Solutions}

An important property of extremal black holes
is  attractor behavior. The literature on the attractor mechanism is extensive but the original works which first explored the phenomenon include Refs.~\refcite{Ferrara:1996dd}, \refcite{Chamseddine:1996pi} and \refcite{Ferrara:1995ih}--\refcite{Ferrara:1996um}. Furthermore, useful reviews which approach the subject from different viewpoints include Refs.~\refcite{Larsen:2006xm}, \refcite{Sen:2007qy}, \refcite{Moore:2004fg}, and \refcite{Bellucci:2007ds}.

In general, there exist BPS and
non-BPS extremal black holes, and both  display  attractor
behavior. The non-BPS branch is quite interesting, but it will not
be discussed here (see Refs.~\refcite{Kraus:2005vz}, \refcite{Dabholkar:2006tb}, and \refcite{Goldstein:2005hq}--\refcite{Andrianopoli:2006ub} for discussion).

We would like to reconsider BPS attractors within the higher derivative
setting developed in this review. First, recall that attractor
behavior involves two related aspects:
\begin{itemize}
\item
{\bf Attractor mechanism:} Within a fixed basin of attraction, the scalar fields flow to constants at the black hole
horizon which depend on the black hole charges alone. In particular the
endpoint of the attractor flow is independent of the initial conditions, {\it i.e.}
the values of the asymptotic moduli.

\item
{\bf The attractor solution:} The limiting value of the geometry
(and the associated matter fields) near the black hole horizon
constitutes a solution in its own right, independently of the flow.
One remarkable feature is that the attractor solution has enhanced, in fact maximal,
supersymmetry. This property is highly constraining,  and so the solution
can be analyzed in much detail.

\end{itemize}
We will ultimately derive complete, asymptotically flat solutions, from which attractor solutions are extracted
by taking appropriate near horizon limits. But since this method  obscures the intrinsic simplicity
of the attractor solutions,   it is
instructive  to construct the attractor solutions directly. This is what we do
in this section.

\subsection{Maximal supersymmetry in the off-shell formalism}
As we have emphasized, the attractor solution has maximal supersymmetry. Thus
we consider the vanishing of the supersymmetry variations (\ref{susy}), which
we repeat for ease of reference
\bea\label{gsusy}
0&=&\left({\cal D}_\mu+{1\over2}v^{ab}\gamma_{\mu ab}-{1\over3}\gamma_\mu\gamma\cdot
v\right)\epsilon~, \cr
0&=&\left(-{1\over4}\gamma\cdot
F^I-{1\over2}\gamma^a\partial_aM^I-{1\over3}M^I\gamma\cdot v
\right)\epsilon~,\cr
0&=&\left(D-2\gamma^c\gamma^{ab}{\cal
D}_av_{bc}-2\gamma^a\epsilon_{abcde}v^{bc}v^{de}+
{4\over3}(\gamma\cdot v)^2\right)\epsilon~.
\eea
The supersymmetry parameter $\epsilon$ should be subject to  no projection conditions if the solution is to preserve maximal supersymmetry. Therefore, terms with different structures
of $\gamma$-matrices cannot cancel each other on the attractor solution.
The gaugino variation (the middle equation in (\ref{gsusy})) therefore demands
\ben
M^I = {\rm constant}~,
\een
and also
\ben\label{gFI}
F^I = - {4\over 3} M^I v~.
\een
Constancy of the scalar fields is a familiar feature of attractors in
two-derivative gravity. The values of the constants will be determined below.
The second result (\ref{gFI}) is special to the off-shell formalism in that it
identifies the auxiliary two-form $v$ with the graviphoton field strength.

We next extract the information from the third equation in (\ref{gsusy}). We can
write it as
\ben \left[ ( D - {8\over 3} v^2) - 2\gamma^{abc} {\cal D}_a
v_{bc} + 2\gamma^a ( {\cal D}^b v_{ba} - {1\over 3}\epsilon_{abcde}
v^{bc} v^{de})\right]\epsilon =0~, \een
by using the algebraic identities
\bea\label{galgid} \gamma^{ab}\gamma^{cd} & = & - (\eta^{ac}
\eta^{bd}-\eta^{ad}\eta^{bc}) -(\gamma^{ac}
\eta^{bd}-\gamma^{bc}\eta^{ad} + \gamma^{bd}
\eta^{ac}-\gamma^{ad}\eta^{bc}) +\gamma^{abcd} ~,\cr \gamma^a
\gamma^{bc} & = & \eta^{ab}\gamma^c - \eta^{ac}\gamma^b +
\gamma^{abc}~,\cr \gamma_{abcde}&=&\epsilon_{abcde}~. \eea
Again, maximal supersymmetry precludes any cancellation between different
tensor structures of the $\gamma$-matrices.
We therefore determine the value of the $D$-field as
\ben\label{gD}
D = {8\over 3} v^2~,
\een
and we find that the auxiliary two-form $v$ must satisfy
\bea\label{gveom}
\epsilon^{abcde} {\cal D}_a v_{bc} &= & 0~,\cr
{\cal D}^b v_{ba}  - {1\over 3}\epsilon_{abcde} v^{bc} v^{de}&=&0~.
\eea
Both equations support the identification of the auxiliary field $v$ with the graviphoton field
strength in minimal supergravity. The first equation is analogous to the Bianchi identity and
the second is analogous to the two-derivative equation of motion for a gauge field with Chern-Simons coupling.  Note that $v$ satisfies the
equations of motion of two-derivative minimal supergravity even though, here, we have
not assumed an action yet.

The final piece of information from maximal supersymmetry is the
vanishing of the gravitino variation, corresponding to the first
equation in (\ref{gsusy}). This equation is identical to the
gravitino variation of minimal supergravity, with the auxiliary
two-form $v$ taking the role of the graviphoton field strength. The
solutions to minimal supergravity have been classified
completely.\cite{Gauntlett:2002nw} Adapted to our notation, the
solutions with maximal supersymmetry are:
\begin{itemize}
\item Flat space.
\item A certain class of pp-waves.
\item Generalized G\"{o}del space-times.
\item AdS$_3\times S^2$ with geometry
\ben\label{glals}
ds^2=\ell_A^2ds^2_{AdS}-\ell_S^2d\Omega^2_2~, \quad \hbox{with}\quad
\ell_A=2\ell_S~. \een
Note that supersymmetry relates the two radii.
Additionally,  $v$ is proportional to the volume form on $S^2$
\ben\label{gvstwo}
v={3\over 4}\ell_S\epsilon_{S^2}~. \een
\item AdS$_2\times S^3$ with geometry
\ben ds^2=\ell_A^2ds^2_{AdS}-\ell_S^2d\Omega^2_3~, \quad
\hbox{with}\quad\ell_A=\half\ell_S~, \een
Again, supersymmetry relates the two radii. In this case $v$ is proportional to the volume form on  AdS$_2$
\ben\label{gvstw}
v={3\over 4}\ell_A\epsilon_{AdS_2}~. \een
\item The near horizon BMPV solution or the {\it rotating attractor}.
\end{itemize}
The computations leading to the above classification of solutions to
minimal supergravity use the equations of motion and the Bianchi
identity for the field strength, as well as the on-shell
supersymmetry transformations. Presently, we analyze the
consequences of maximal supersymmetry in the off-shell formalism,
but do not wish to apply the equations of motion yet, because they
depend on the action. Fortunately, we found in (\ref{gveom}) that
supersymmetry imposes the standard two-derivative equation of motion
and Bianchi identity for the auxiliary two-tensor $v$, which in turn
can be identified with the graviphoton in minimal supergravity. In
our context, the classification therefore gives precisely the
conditions for maximal supersymmetry, with no actual equations of
motion imposed.

We will not repeat the general classification of
Ref.~\refcite{Gauntlett:2002nw} but just explain why the
possibilities are so limited and derive the quantitative results
given above. First recall that there exists an integrability
condition obtained from the commutator of covariant derivatives
acting on a spinor \ben\label{intspinor} \lt[ {\cal D}_\mu , {\cal
D}_\nu \rt] \epsilon = \frac14 R_{\mu\nu ab} \g^{ab} \epsilon~. \een
We can evaluate the left-hand side of (\ref{intspinor}) by
differentiating and then antisymmetrizing the BPS condition
resulting from the gravitino variation, {\it i.e.} the first
equation in (\ref{gsusy}). The resulting equations are rather
unwieldy, but they can be simplified to purely algebraic conditions
by using reorderings akin to (\ref{galgid}) along with the
supersymmetry conditions (\ref{gveom}). The terms proportional to
the tensor structure $\gamma^{ab}$ give the Riemann tensor
\bea\label{gcurv}
R_{\mu\nu\rho\sigma}
 = &
  -{16\over 9} v_{\mu\nu} v_{\rho\sigma}
- {4\over 3} ( v_{\mu\rho}  v_{\nu\sigma} -v_{\mu\sigma}
v_{\nu\rho}) + {2\over 9} (g_{\mu\rho}g_{\nu\sigma} -
g_{\nu\rho}g_{\mu\sigma}) v^2\cr & -{4\over 9} \left(
g_{\mu\sigma} v_{\rho\tau} v^\tau_{~\nu} - g_{\mu\rho}v_{\sigma\tau}
v^\tau_{~\nu} - g_{\nu\sigma} v_{\rho\tau} v^\tau_{~\mu} +
g_{\nu\rho}v_{\sigma\tau} v^\tau_{~\mu} \right)~. \eea
Conceptually, we might want to start with a $v$ that solves equations (\ref{gveom}), since then the geometry is completely
determined by (\ref{gcurv}).  But the two sets of
equations are of course entangled. Also, one must further check that
the gravitino variation does in fact vanish, and not just its commutator.

The most basic solutions for the study of black holes and strings are the AdS$_3\times S^2$ and
AdS$_2\times S^3$ geometries. For these, the  solutions to
the supersymmetry conditions (\ref{gveom}) are given by magnetic and electric fluxes,
as in (\ref{gvstwo}) and (\ref{gvstw}).
In each case we can insert in (\ref{gcurv}) and verify that the geometry is
in fact maximally symmetric and that the scales $\ell_A, \ell_S$ are
related to those of $v$ in the manner indicated.

\subsection{The magnetic attractor solution}
\label{sec:magattr}
So far we have just analyzed the consequences of supersymmetry.
In order to determine the solutions completely we also need
information from the equations of motion. We next show how this
works in the case of the simplest nontrivial attractor solution, the
AdS$_3\times S^2$ that is interpreted as the near horizon geometry of a magnetic
string.

The key ingredient beyond maximal supersymmetry is the modified very special geometry
constraint
\bea\label{gspecgeom}
{1\over 6} c_{IJK} M^I M^J M^K &=& 1 - {c_{2I}\over 72}\left( F^I\cdot v + M^I D\right)~,\cr
&=& 1 - {c_{2I} \over 54} M^I v^2~,\cr
&=& 1 - {c_{2I} \over 12} {M^I\over\ell^2_A}~.
\eea
We first used the $D$ equation of motion (\ref{mvsg}) and then simplified using (\ref{gFI}) and
(\ref{gD}). In the last line we used
\ben
v^2 = {9\over 8\ell^2_S}= {9\over 2\ell^2_A}~,
\een
from (\ref{gvstwo}) and (\ref{glals}).

In AdS$_3\times S^2$ the field strengths (\ref{gFI}) become
\ben F^I = - {4\over 3} M^I v =  -{1\over 2} M^I \ell_A
\epsilon_{S^2}~. \een
In our normalization the magnetic
fluxes are fixed as
\ben F^I = -{p^I\over 2} \epsilon_{S^2}~, \een
so we determine the scalar fields as
\ben M^I = {p^I\over \ell_A}~. \een
Inserting this into the modified very special geometry constraint (\ref{gspecgeom}) we finally
determine the precise scale of the geometry
\ben\label{gscale}
\ell^3_A = {1\over 6}c_{IJK} p^I p^J p^K +
{1\over 12}c_{2I}p^I \equiv p^3 + {1\over 12} c_2\cdot p~. \een
The preceding three equations specify the attractor solution
completely in terms of magnetic charges $p^I$.

\subsection{The electric attractor solution}
\label{sec:elecattr}

The electric attractor solution is the AdS$_2\times S^3$ near horizon geometry of a
non-rotating 5D black hole. The scales of the geometry are
\ben
\ell \equiv \ell_A = {1\over 2}\ell_S~,
\een
and the auxiliary two-form $v$ in (\ref{gvstw}) gives
\ben
v^2 = - {9\over 8\ell^2}~,
\een
so that the modified very special geometry constraint becomes
\ben
{1\over 6}c_{IJK} M^I M^J M^K  = 1 - {c_{2I}M^I\over 54} v^2
= 1 + {c_{2I}M^I\over 48\ell^2}~.
\een
We can write this in a more convenient way by introducing the rescaled
moduli
\ben
{\hat M}^I = 2\ell M^I~,
\een
so that
\ben\label{elecscale}
\ell^3 = {1\over 8}\left[ {1\over 6} c_{IJK} {\hat M}^I {\hat M}^J {\hat M}^K
- {c_{2I}\over 12}{\hat M}^I\right]~.
\een
This equation gives the scale of the geometry in terms of the rescaled moduli.

We would often like to specify the solution in terms its electric
charges, rather than the rescaled moduli. Electric charges may be
defined as integration constants in Gauss' law, a step that depends
on the detailed action of the theory. We will carry out this
computation in Section \ref{sec:rotattr} and find that
\ben
q_I = {1\over 2} c_{IJK} {\hat M}^J {\hat M}^K - {c_{2I}\over 8}~.
\een
If the $q_I$ are given, this relation determines the rescaled moduli ${\hat M}^I$ and so,
through (\ref{elecscale}), the scale $\ell$.

\subsubsection{Rotating attractors}
\label{sec:rotatt}

Just as the non-rotating 5D black hole can be generalized to include rotation, the electric attractor just studied is a special case of a more general
rotating attractor.  Here we just briefly state the result, deferring more discussion
until after we have derived the full, corrected, rotating black hole solution.
The attractor solution is
\begin{eqnarray}\label{ggeom}
ds^2 &=& w^2 \left[ (1 + (e^0)^2)(\rho^2 d\tau^2 -
{d\rho^2\over\rho^2} - d\theta^2 -\sin^2\theta d\phi^2) -
(dy+\cos\theta d\phi)^2\right]~,\cr v &=& - {3\over 4} w (d\tau
\wedge d\rho - e^0 \sin\theta d\theta \wedge d\phi)~.
\end{eqnarray}
The geometry describes a spatial circle nontrivially fibered over AdS$_2 \times S^2$.
The two-form $v$ is a  solution to the supersymmetry conditions (\ref{gveom}) written in the convenient form
\ben d \star v + {4\over 3} v\wedge v = 0~. \een
The modified special geometry constraint follows by inserting $v$ in
the second line of (\ref{gspecgeom}).
As we discuss in more detail in Section \ref{sec:rotattr}, the parameters $w$ and $e^0$ specify the scale
sizes and angular momentum of the solution.

\section{Black Strings and Null Supersymmetry}

We now begin investigating asymptotically flat solutions which preserve only a fraction of the supersymmetry of the theory. We begin with black string solutions, which were first discussed in Ref.~\refcite{Gibbons:1994vm}. In particular, we will study corrections to the Calabi-Yau black strings studied in Ref.~\refcite{Chamseddine:1999qs}. These solutions each have at least one null isometry so we will determine the off-shell supersymmetry conditions for any such spacetime. The conditions from supersymmetry do not completely specify the solution and we will require more
conditions on the functions in our {\it ansatz}, including equations of motion from the full higher-derivative Lagrangian. We will comment some on the general case, then specialize to purely magnetically charged strings which carry no momentum along their length; this is precisely the case
studied in Ref.~\refcite{Castro:2007hc}. Under these assumptions we will only need to use the equation of motion for $D$ and the Bianchi identity for
$F^I$ to completely specify the solution.

\subsection{Metric ansatz}

As argued in Ref.~\refcite{Gauntlett:2002nw} the most general metric, up to diffeomorphisms, with lightlike killing vector $V= \p_{y^+}$,  is
\ben\label{stringansatz}
ds^2 = e^{2U_1} \lt( {\cal F} (dy^-)^2 + 2dy^+ dy^-\rt)-e^{-4U_2} \delta_{ij}\lt(dx^i + a^i dy^-\rt) \lt(dx^j + a^j dy^-\rt)~,
\een
where $i,j = 1,2,3$ and the undetermined functions $U_{1,2}$, ${\cal F}$, and $a^i$ are independent of $y^+$. We choose for our vielbeins
\ben
e^{\plh} = e^{U_1}(dy^+ + \half {\cal F} dy^-) ~, ~~~~ e^{\mih} = e^{U_1} dy^- ~, ~~~~ e^\ih = e^{-2U_2}(dx^i +a^i dy^-)~.
\een
The spin connections which follow from this choice of vielbeins are
\bea
\o^\plh_{~\plh} &=& e^{-U_1}\lt(a^i\p_i U_1 -\p_- U_1\rt)e^\mih ~, \cr
\o^\plh_{~\ih} &=& \half e^{2U_2} \p_i \F e^\mih + \p_i U_1 e^{2U_2} e^\plh - e^{-U_1}\lt(S_{ij} +2 \lt(\p_- U_2- a^k \p_k U_2 \rt)\d_{ij} \rt) e^\jh ~, \cr
\o^\mih_{~\ih} &=& \p_i U_1 e^{2U_2} e^\mih~, \cr
\o^\ih_{~\jh} &=& -e^{-U_1} A_{ij}e^\mih +2e^{2U_2}\lt(\p_i U_2 \d_{jk} - \p_j U_2 \d_{ik}\rt)e^\kh ~,
\eea
where we have defined
\bea
S_{ij} &=&\half \lt(\p_i a^j + \p_j a^i\rt)  ~,\nn\\
A_{ij} &=& \half \lt(\p_i a^j - \p_j a^i\rt) ~.
\eea

\subsection{Supersymmetry conditions}
\label{sec:magsusy}
We now substitute our {\it ansatz} into the supersymmetry conditions (\ref{susy}) to determine the bosonic backgrounds which preserve supersymmetry. Following (\ref{projection}), we look for Killing spinors which satisfy the projection
\ben\label{lproj}
\g_\plh \epsilon = \g^\mih \epsilon = 0~.
\een
Equivalent forms of the projection are
\bea
\g_{\plh\mih} \epsilon &=&  \epsilon~, \nn\\
\g_{\ih\jh\kh} \epsilon &=& -\veps_{ijk} \epsilon ~,\nn\\
\g_{\ih\jh} \epsilon &=& \veps_{ijk} \g_\kh \epsilon~,
\eea
where we have used the gamma matrix and orientation conventions
\bea
\g_{abcde}&=& \veps_{abcde}~, \nn\\
\veps_{\plh\mih\ih\jh\kh} &=&\veps_{ijk}~.
\eea
with $\veps_{123}=1$. Furthermore, the isometry $V= \p_{y^+}$ ensures that $\epsilon$ has no $y^+$ dependence.

Around bosonic backgrounds, the condition for supersymmetry from the variation of the gravitino is
\ben
\d\psi_\mu = \lt[\p_\mu +{1\over 4} \o_\mu^{ab}\g_{ab}+{1\over6} v^{ab} e_\mu^{~c} \lt(\g_{abc} -4 \eta_{ac}\g_b\rt)\rt]\epsilon=0 ~.
\een
The $y^+$ component is rather simple. After applying the projection (\ref{lproj}) we find
\ben
-\frac13 e_+^{~\plh}v^{\mih\ih}  \g_\ih\epsilon =0~,
\een
from which it follows $v^{\mih\ih}=-v_{\plh\ih}=0$.

The $y^-$ component gives
\bea
0&=&\lt[\lt(\p_- +{1\over 2} \o_-^{\plh\mih}-{1\over6}\veps_{ijk}e_-^{~\ih}v^{\jh\kh}\rt)  +\lt({1\over2} \o_-^{\mih\kh} +{1\over6} v^{\ih\jh}
e_-^{~\mih} \veps_{ijk}  \rt)\g_{\mih\kh} \rt.\\
&&~~~\lt. +\lt(-{2\over3} v^{\plh\mih} e_-^{~\mih} \rt)
\g_\mih+\lt({1\over4} \o_-^{\ih\jh}\veps_{ijk} - v^{\plh\kh}
e_-^{~\mih}+{1\over3} v^{\plh\mih} e_-^{~\kh} -{2\over3}v^{\ih\kh}
e_-^{~\ih} \rt)\g_\kh  \rt]\epsilon~. \nn \eea Vanishing of
$\g_\mih$ terms imply $v^{\plh\mih}=0$. Vanishing of $\g_{\mih\ih}$
terms requires \ben v^{\ih\jh} = {3\over2}e^{2U_2}\veps_{ijk}\p_k
U_1 ~. \een
Cancellation of $\g_\ih$ terms yields \ben v^{\plh\ih} =
-v_{\mih\ih} = {1\over4} e^{-U_1} \veps_{ijk}\p_j a^k +e^{-U_1}
\veps_{ijk}a^j \p_k \lt(U_1-U_2\rt)~. \een
Finally, cancellation of terms without $\g$'s yields
\ben\label{dminuseps} \lt(\p_- -\half \p_- U_1\rt) \epsilon=0~. \een

The $x^i$ components of the gravitino variation yield the additional constraints that $U_1=U_2\equiv U$ and
\ben
\lt( \p_i -\half \p_i U\rt)\epsilon=0~.
\een
Combining the above equation with (\ref{dminuseps}), we can solve for the Killing spinor
\ben
\epsilon = e^{U/2} \epsilon_0~,
\een
where $\epsilon_0$ is a constant spinor satisfying the projection (\ref{lproj}).

We have now exhausted the conditions following from stationarity of
the gravitino under supersymmetry transformations. Requiring the
gaugino to be stationary about a bosonic background yields the
condition \ben \d\O^I =  \lt(-{1\over4} \gamma\cdot F^I -{1\over2}
\gamma^a \partial_a M^I - {1\over 3} M^I \gamma\cdot v\rt) \epsilon
=0~. \een For the present \textit{ansatz}, and utilizing the
previously found values for $v_{ab}$, we can write the above as \ben
\lt[ -\frac14 \lt( 2 F^I_{\mih\plh}-2F^I_{\plh\ih}\g_{\mih\ih}
+F^I_{\ih\jh}\veps_{ijk}\g_{\kh}\rt) +\half e^{2U} \p_i M^I \g_\ih -
e^{2U} \p_k U \g_\kh\rt]\epsilon = 0~. \een Requiring terms with
like $\g$-matrix structure to cancel we find that the only
non-trivial field strength components are \ben\label{nullgauge}
F^I_{\ih\jh} = \veps_{ijk}e^{4U} \p_k \lt( e^{-2U} M^I\rt) ~, \een
and $F^I_{\mih\ih}$, which is unconstrained.

The final supersymmetry condition is obtained by imposing stationarity of the auxiliary fermion
\ben
\d\chi = \left[D-2\gamma^c\gamma^{ab}{\cal
D}_av_{bc}-2\gamma^a\varepsilon_{abcde}v^{bc}v^{de}+{4\over3}(\gamma\cdot
v)^2\right]\epsilon=0~.
\een
Using the projection (\ref{lproj}) and the known values for $v_{ab}$ yields the single condition
\ben
D= 6 e^{4U} \nabla^2 U ~,
\een
where $\nabla^2 = \d_{ij}\p_i \p_j$.

\subsubsection{Summary of supersymmetry conditions}

As a quick summary, we now restate the results of the above
supersymmetry analysis. The metric {\it ansatz} with lightlike
isometry $V=\p_{y^+}$ which is consistent with supersymmetry is
\ben\label{nullsusymetric} ds^2 = e^{2U} \lt(\F \lt(dy^-\rt)^2 +
2dy^+ dy^- \rt) -e^{-4U} \d_{ij}\lt(dx^i +a^i dy^-\rt)\lt(dx^i +a^i
dy^-\rt)~, \een
or in a null orthonormal frame
\ben
e^{\plh} = e^{U}(dy^+ + \half {\cal F} dy^-) ~, ~~~~ e^{\mih} = e^{U} dy^- ~, ~~~~ e^\ih = e^{-2U}(dx^i +a^i dy^-)~.
\een
The non-trivial auxiliary fields are
\bea\label{nullauxiliary} v^{\ih\jh} &=& \frac32 e^{2U}\veps_{ijk}
\p_k U~,\cr v^{\plh\ih} &=& -v_{\mih\ih} = \frac14 e^{-U}
\veps_{ijk}a^j \p_k U ~,\cr D &=& 6 e^{4U} \nabla^2 U~. \eea The
gauge field strengths are given by \ben\label{nullgaugeform} F^I =
F^I_{\mih\ih} e^\mih \wedge e^\ih + \half \veps_{ijk}e^{4U}\p_k
\lt(e^{-2U}M^I\rt)e^\ih \wedge e^\jh ~. \een All of the undetermined
functions in the above (that is $U$, $\F$, $a^i$, $M^I$ and
$F^I_{\mih\ih}$) are independent of the isometry coordinate $y^+$,
but otherwise unconstrained.

\subsection{Bianchi identity}

We have shown in the previous section how supersymmetry partially determines the gauge field strengths. However, these field strengths (\ref{nullgaugeform}) do not manifestly follow from exterior differentiation of some one-form potentials. Therefore we must impose the Bianchi identity
\ben
dF^I =0~,
\een
which should result in further non-trivial conditions. Physically, this is because supersymmetry is
consistent with any extended distribution of magnetic charges,
while here we are considering solutions away from their isolated sources.

The two non-trivial conditions from the Bianchi identity are
\bea\label{bianchicoord}
0 &=& \p_{[k} F^I_{ij]}~,\cr
0 &=& \p_- F^I_{ij} + \p_j F^I_{-i}-\p_i F^I_{-j}~.
\eea
In coordinate frame the field strength
components can be written as
\bea\label{nullgaugecoord}
F^I_{ij} &=& \veps_{ijk} \p_k \lt( e^{-2U} M^I\rt)~,\cr
F^I_{-i} &=& e^{-U} F_{\mih\ih} - \veps_{ijk} a^j \p_k \lt(e^{-2U} M^I\rt)~.
\eea
The relations
(\ref{bianchicoord}) are more usefully expressed after contraction
with $\veps_{ijk}$ and substitution of (\ref{nullgaugecoord}). The
first equation becomes \ben \nabla^2 \lt( e^{-2U} M^I\rt)=0~, \een
allowing us to write the moduli in terms of harmonic functions $H^I$
on $\mathbb{R}^3$
\ben\label{nullmoduli} M^I = e^{2U} H^I~. \een It is important to
note that the $H^I$ have arbitrary dependence on the null coordinate
$y^-$. The second equation in (\ref{bianchicoord}) then becomes
\ben\label{nullbianchitwo} \p_- \p_k H^I =\veps_{kij} \p_i
\lt(e^{-U}F^I_{\mih\jh}\rt) -\p_i a^k \p_i H^I + \p_i\lt(a^i \p_k
H^I\rt)~. \een Thus $F^I_{\mih\jh}$ is determined up to integration
constants by the Bianchi identities in terms of metric functions and
the harmonic functions which describe the moduli.

\subsection{Modified very special geometry}

As determined in (\ref{mvsg}), the equation of motion for $D$ yields the modified very special geometry constraint
\ben
{\cal N} = 1-{c_{2I}\over72}\lt(F^I_{ab}v^{ab} +M^I D\rt)~.
\een
Substituting in the results from supersymmetry and the Bianchi identity, specifically equations (\ref{nullauxiliary}), (\ref{nullgaugeform}), and (\ref{nullmoduli}), we find
\ben\label{nullmvsg}
e^{-6U} = {1\over6}c_{IJK}H^IH^JH^K+{c_{2I}\over24}\lt(\nabla U \cdot \nabla H^I + 2 H^I \nabla^2 U\rt)~,
\een
where $\nabla_i = \p_i$ are derivatives on $\mathbb{R}^3$. This equation thus specifies the metric function $U(y^-,x^i)$ in terms of the harmonic functions $H^I$.

\subsection{Other equations of motion and the general solution}

So far, we have used only one equation of motion in addition to the
BPS conditions and Bianchi identity. Supersymmetry puts constraints
on the metric and determines the auxiliary fields in terms of the
moduli $M^I$ and metric functions $U$ and $a^i$ as summarized in
equations (\ref{nullsusymetric})--(\ref{nullauxiliary}).
Supersymmetry also constrains the gauge field strengths $F^I$ to be
of the form (\ref{nullgaugeform}). The Bianchi identity for $F^I$
provides two relations. First, it specifies the moduli in terms of
$U$ and some arbitrary harmonic functions $H^I$ through
(\ref{nullmoduli}). The second relation (\ref{nullbianchitwo})
completes the specification of the gauge field strengths (up to
integration constants) in terms of $U$, $a^i$ and the harmonic
functions $H^I$. Now examining the equation of motion for $D$
specifies $U$ in terms of the harmonic functions $H^I$ through
(\ref{nullmvsg}).

There remain the functions $\F$ and $a^i$ which are not yet specified. No solutions with higher-derivative corrections have yet been found which have non-zero values for these fields, but we can make a few comments. Having already examined the Bianchi identity, we have fixed completely the magnetic part of the gauge field strength. The electric part, $F^I_{\mih\ih}$, should be further constrained by the Maxwell equation, {\it i.e.} the equation of motion for the gauge field. Combining this with (\ref{nullbianchitwo}) should be enough to completely specify $F^I_{\mih\ih}$ and $a^i$.

To specify the final undetermined function $\F$, we turn to the equation of motion from the metric. Since the Ricci tensor of the metric is
\ben
R_{\mih\mih} = \half e^{4U}\nabla^2 \F + \ldots~,
\een
we expect that $\F$ is determined by the $(\mih\mih)$ component of the Einstein equation.

As stated already, the most general solutions with null supersymmetry are not yet known. In the next section, we will examine the full solution for a certain special case.

\subsection{The magnetic string solution}

We now simplify our analysis by specializing to string solutions which carry no momentum or electric charge ($M2$-brane charge) and have an additional null isometry $\tilde{V}=\p_{y^-}$. In terms of our \textit{ansatz} this corresponds to setting
\bea
\F &=& 0~, \nn \\
a^i &=&0~, \cr
F^I_{\mih\ih} &=& 0~,
\eea
and assuming that the remaining undetermined functions of the solution are also independent of the other null coordinate $y^-$. Furthermore, the dependence of undetermined functions on the $x^i$ is assumed to be spherically symmetric, \textit{i.e.} dependent only on the radial variable $r^2 = \d_{ij}x^i x^j$. These are the solutions discussed in Ref.~\refcite{Castro:2007hc}.

It is convenient to now use spherical coordinates so that the metric takes the form
\ben\label{magneticmetric}
ds^2 = 2e^{2U}dy^+ dy^- -e^{-4U}\lt(dr^2 +r^2 d\O_2^2\rt)~.
\een
In these coordinates, the gauge fields and auxiliary field $v_{ab}$ are given by
\bea\label{bda}
F^I_{\theta\phi} &=& \p_r \lt(e^{-2U} M^I\rt) r^2 \sin\theta ~, \nn\\
v_{\theta\phi} &=& \frac32 e^{-2U} r^2 \sin\theta \p_r U~,
\eea
and the auxiliary scalar is
\ben\label{magneticd}
D= 6 e^{4U} \nabla^2 U ~,
\een
where $\nabla^2 = r^{-2} \p_r \lt(r^2 \p_r \rt)$ is the Laplacian on $\mathbb{R}^3$.

\subsubsection{Maxwell equation and Bianchi identity}

Since we have narrowed our search to solutions with no electric charge, we do not expect to have any constraints from the Maxwell equation. Indeed, it can be straightforwardly verified that the equations of motion for the $A^I_\mu$ are identically satisfied for the {\it ansatz} described by equations (\ref{magneticmetric})--(\ref{magneticd}). Thus we get no new information from these equations of motion.

For magnetic solutions the nontrivial condition arises from the Bianchi identity $dF^I=0$. As found in (\ref{nullmoduli}), this determines the moduli to be
\ben
M^I = e^{2U}H^I~,
\een
where $H^I$ is some $y^-$ independent function which is harmonic on the three-dimensional base $\mathbb{R}^3$. Here we look for single-center solutions on $\mathbb{R}^3$ so
\ben\label{bie}
M^Ie^{-2U}=H^I=M^I_\infty+{p^{I}\over 2r}~,
\een
with $M^I_\infty$ the value of $M^I$ in the asymptotically flat region
where $U=0$ and $p^{I}$ is some constant. By using (\ref{bda}) we see that
the field strengths are given by
\ben
F^I=-{p^I\over2}\epsilon_{S^2}~.
\een
In our units, this identifies $p^I$ as the integer quantized magnetic flux of $F^I$
\ben
p^I =-{1\over2\pi}\int_{S^2} F^I~.
\een
It is worth noting that the magnetic charge does not get modified after including higher derivatives since it
is topological, \textit{i.e.} the Bianchi identity is not corrected by higher-order effects. We will find in Section \ref{sec:timelikesusy} that this does not hold for electric charges which are instead governed by the equations of motion for the gauge fields.

\subsubsection{$D$ equation of motion}

So far, by imposing the conditions for supersymmetry and integrating
the Bianchi identity, we have been able to write our solution in terms
of one unknown function $U(r)$. To determine this remaining
function we use the equation of motion for the auxiliary field $D$.
As stated earlier in (\ref{nullmvsg}) this is given by
\ben\label{bka}
e^{-6U}={1\over
6}c_{IJK}H^IH^JH^K+{c_{2I}\over24}\left(\nabla H^I\cdot\nabla
U+2H^I\nabla^2U\right)~.
\een
Here $H^I$ are the harmonic functions defined in (\ref{bie})  and we used
\ben
\label{bkab}
{\cal N}={1\over 6}c_{IJK}H^IH^JH^Ke^{6U}~.
\een
The $D$ constraint (\ref{bka}) is now an ordinary differential equation
that determines $U(r)$. Its solution specifies the entire geometry
and  all the matter fields.

\subsubsection{Magnetic attractors}

We can solve (\ref{bka}) exactly in the near horizon region. This case
corresponds to vanishing integration constants in (\ref{bie}) so that
\ben\label{bmd}
H^I = {p^I\over 2r}~.
\een
Then (\ref{bka}) gives
\ben\label{bmdb}
e^{-6U}={1\over8r^3}\left(p^3+{1\over12}c_{2}\cdot
p\right)={\ell_S^3\over r^3}~,
\een
where $p^3 = {1\over 6} c_{IJK} p^I p^J p^K$. The geometry in this
case is AdS$_3\times S^2$ with the scale $\ell_S$ in agreement with the magnetic attractor solution developed in
Section \ref{sec:magattr}.

\subsubsection{Corrected geometry for large black strings}
One way to
find solutions to (\ref{bka}) is by perturbation theory. This strategy
captures the correct  physics when the solution is regular already
in the leading order theory, \textit{i.e.} for large black strings.
Accordingly, the starting point is the familiar solution
\ben\label{bm}
e^{-6U_0}={1\over6}c_{IJK}H^IH^JH^K~,
\een
to the two-derivative theory. This solves (\ref{bka}) with $c_{2I}=0$.

Although $c_{2I}$ is not small it will be multiplied by terms that
are of higher order in the derivative expansion. It is therefore
meaningful to expand the full solution to (\ref{bka}) in the form
\ben\label{bme}
e^{-6U}=e^{-6U_0}+c_{2I}\varepsilon^I+
{1\over2}c_{2I}c_{2J}\varepsilon^{IJ}+\ldots~,
\een
where $\varepsilon^I(r), \varepsilon^{IJ}(r),\ldots$ determine the
corrected geometry with increasing precision.

Inserting (\ref{bme}) in (\ref{bka}) and keeping only the terms linear in
$c_{2I}$ we find the first order correction\footnote{It is understood
that the correction $\varepsilon^I$ is only defined in the
combination $c_{2I}\varepsilon^I$.}
\ben\label{bmb}
\varepsilon^I={1\over24}(\nabla H^I\cdot\nabla
U_0+2H^I\nabla^2U_0)~.
\een
Iterating, we find the second order correction
\ben\label{bmg}
\varepsilon^{IJ}=-{1\over72}\left(\nabla
H^I\cdot\nabla(e^{6U_0}\varepsilon^J)+2H^I\nabla^2(e^{6U_0}\varepsilon^J)
\right)~,
\een
where the first order correction $\varepsilon^I$ is given by (\ref{bmb}).
Higher orders can be computed similarly. In summary, we find that
starting from a smooth solution to the two-derivative theory we
can systematically and explicitly compute the higher order
corrections. The series is expected to be uniformly convergent.

In the near horizon limit (\ref{bmd}), the full solution
(\ref{bmdb}) is recovered exactly when taking the leading correction
(\ref{bmb}) into account. As indicated in (\ref{bmdb}) the effect of
the higher derivative corrections is to expand the sphere by a
specific amount (which is small for large charges). The perturbative
solution gives approximate expressions for the corrections also in
the bulk of the solution. Numerical analysis indicates that the
corrections remain positive so at any value of the isotropic
coordinate $r$ the corresponding sphere is expanded by a specific
amount.

In this section we have focused on large black strings, that is, those which are non-singular in the leading supergravity description. We will later turn to {\it small strings}, particularly the important case of {\it fundamental strings}, in Section \ref{sec:small}.

\section{Timelike Supersymmetry -- Black Holes and Rings}\label{sec:timelikesusy}

We now turn to the case in which the Killing vector $V^\mu$ is timelike over some region of the
solution.  This class of solutions includes 5D black holes and black rings.     The analysis that follows
is mainly taken from Refs.~\refcite{Castro:2007hc} and \refcite{Castro:2007ci}, with some further generalizations included.

\subsection{Metric ansatz}

We start with a general metric {\it ansatz} with Killing vector ${\p \over \p t}$,
\ben
ds^2 = e^{4U_1(x)}(dt+\omega)^2 -e^{-2U_2(x)} ds_B^2~.
\een
Here $\omega$ is a 1-form on the 4D base $B$ with coordinates $x^i$
with $i=1,\dots,4$. We choose vielbeins
\ben
e^\th= e^{2U_1}(dt+\omega)~,\quad e^\ih = e^{-U_2} \et^\ih~,
\een
where $\et^\ih$ are vielbeins for $ds_B^2$.   The corresponding spin connection is
\bea\label{spincon}
\omega^\th_{~\ih} &=& 2 e^{U_2} \nt_\ih U_1 e^\th+{1\over 2} e^{2U_1+U_2} d\omega_{\ih\jh}\et^\jh~,   \nn\\
\omega^\ih_{~\jh} &=&  \omt^\ih_{~\jh}+ {1\over 2} e^{2U_1 +2U_2} d\omega_{\ih\jh} e^\th + e^{U_2} \nt_\ih U_2 e^\jh -e^{U_2} \nt_\jh U_2 e^\ih~.
\eea

We will adopt the following convention for hatted indices.  Hatted indices of five dimensional tensors
are orthonormal with respect to the full 5D metric, whereas those of tensors  defined on the base
space are orthonormal with respect to $ds_B^2$.   For example, $d\omega$ is defined to live on the
base, and so obeys  $d\omega_{ij}= \et^\kh_i \et^\lh_j d\omega_{\kh\lh}$.   Furthermore,  the tilde
on $\nt_\ih$ indicates that the $\ih$ index is orthonormal with respect to the base metric.    To avoid
confusion, we comment below when two different types of hatted indices are used in a single equation.

The Hodge dual on the base space is defined as
\ben\label{wb}\star_4\alpha_{\ih\jh}={1\over 2}\epsilon_{\ih\jh\kh\lh}\alpha^{\kh\lh}~,
\een
with $\epsilon_{\hat{1}\hat{2}\hat{3}\hat{4}}=1$. A 2-form on the
base space can be decomposed into self-dual and anti-self-dual
forms,
\ben\label{wbb}\alpha=\alpha^++\alpha^-~,
\een
where $\star_4\alpha^{\pm}=\pm\alpha^{\pm}$.

Equation (\ref{projection}) tells us to look for supersymmetric solutions with a Killing spinor obeying the projection
\ben\label{tproj}
 \gamma^\th \epsilon = - \epsilon~,
\een
with a useful alternative form being
\ben
\alpha^{-\ih\jh}\g_{\ih\jh} \epsilon =0~,
\een
where $\alpha^{-\ih\jh}$ is any two-form that is anti-self-dual on the 4D base space. The strategy we employ is the same as for the null projection discussed in the previous section:
we first exhaust the conditions implied by unbroken supersymmetry, and then impose some of the
equations of motion or other constraints.

\subsection{Supersymmetry conditions}
\label{sec:electsusy}

There are three supersymmetry conditions we need to solve.
Following the same procedure as in the previous section we first impose a
vanishing gravitino variation,
\ben\label{wca}
\delta \psi_\mu=\left[{\cal
D}_\mu+{1\over 2}v^{ab}\gamma_{\mu
ab}-{1\over3}\gamma_{\mu}\gamma\cdot v\right]\epsilon=0~.
\een
Evaluated in our background, the time component of equation (\ref{wca})
reads
\ben\label{wcb}
\left[\partial_t-e^{2U_1+U_2}\partial_iU_1\gamma_{\ih}-{2\over3}e^{2U_1}v^{\th\ih}\gamma_{\ih}
-{1\over4}e^{4U_1+2U_2}d\omega_{ij}\gamma^{\ih\jh}-
{1\over6}e^{2U_1}v_{\ih\jh}\gamma^{\ih\jh}\right]\epsilon=0~,
\een
where we used the projection (\ref{tproj}). The terms proportional to
$\gamma_{\ih}$ and $\gamma_{\ih\jh}$ give the conditions
\bea\label{wcc}v_{\th\ih}&=&{3\over 2}e^{U_2}\nt_\ih U_1~,\cr
v^+_{\ih\jh}&=&-{3\over 4}e^{2U_1}d\omega^+_{\ih\jh}~.
\eea
The spatial component of the gravitino variation (\ref{wca}) simplifies to
\ben\label{wcd}
\left[\nt_i+{1\over2}\partial_jU_2\gamma_{\ih\jh}
+v^{\th\kh}e^{~\jh}_i\left(\gamma_{\jh\kh}-{2\over3}\gamma_{\jh}\gamma_{\kh}\right)
-e^{~\kh}_{i}\left(v^{-}_{\kh\jh}+{1\over4}e^{2U_1}d\omega^-_{\kh\jh}\right)\gamma^{\jh}\right]\epsilon=0~,
\een
where we used the results from (\ref{wcc}). The last term in (\ref{wcd}) relates
the anti-self-dual pieces of $v$ and $d\omega$,
\ben\label{wce}
v^-_{\ih\jh}=-{1\over 4}e^{2U_1}d\omega^-_{\ih\jh}~.
\een
To forestall confusion, we note that in equations (\ref{wcc}) and (\ref{wce}) the indices on $v$
are orthonormal with respect to the full 5D metric, while those on $d\omega$ are orthonormal with
respect to the base metric.

The remaining components of (\ref{wcd}) impose equality of the two
metric functions $U_1=U_2\equiv U$ and determine the Killing spinor as
\ben\label{wcf}\epsilon=e^{U(x)}\epsilon_0~,
\een
with $\epsilon_0$ a covariantly  constant spinor on the base, $\nt_i \epsilon_0 =0$.  This implies
that the base space is
hyperK\"ahler.\footnote{Recall that there is an implicit $SU(2)$ index on the spinor $\epsilon$. One can then construct three distinct two-forms, $\Phi^{ij}_{ab} = \bar{\epsilon}^i \g_{ab} \epsilon^j$, which enjoy an $SU(2)$ algebra. This algebra defines the hyperK\"ahler structure of the base space $B$; see Ref.~\refcite{Gauntlett:2002nw} for details.}

The gaugino variation is given by
\ben\label{wda}\delta\Omega^I=\left[-{1\over 4}\gamma\cdot
F^I-{1\over 2}\gamma^a\partial_aM^I-{1\over 3}M^I\gamma\cdot
v\right]\epsilon=0~.
\een
This condition  determines the electric and self-dual pieces of
$F^I_{ab}$,
\bea\label{wdb}F^{I\th\ih}&=&e^{-U}\nt_\ih(e^{2U}M^I)~,\cr
F^{I+}&=&-{4\over3}M^Iv^+~.
\eea
Defining the anti-self-dual form
\ben\label{wdc}
\Theta^I=-e^{2U}M^Id\omega^-+F^{I-}~,
\een
the field strength can be written as
\ben\label{wdd}F^I=d(M^Ie^{\hat{t}})+\Theta^I~.
\een
The Bianchi identity implies that $\Theta^I$ is closed.
We emphasize that $\Theta^I$, or more precisely $F^{I-}$, is undetermined
by supersymmetry. These anti-self-dual components are important for
black ring geometries but vanish for rotating black holes. %

Finally, the variation of the auxiliary fermion is
\ben\label{we}\delta\chi=\left[D-2\gamma^c\gamma^{ab}{\cal
D}_av_{bc}-2\gamma^a\epsilon_{abcde}v^{bc}v^{de}+{4\over 3}(\gamma\cdot
v)^2\right]\epsilon=0~.
\een
Using equations (\ref{wcc})
and (\ref{wce}), the terms proportional to one or two gamma matrices cancel identically. The terms independent of $\gamma_{\ih}$ give an
equation for $D$, which reads
\ben\label{wea}D=3e^{2U}(\nt^2U-6(\nt
U)^2)+{1\over 2}e^{8U}(3d\omega^+_{\ih\jh}d\omega^{+\ih\jh}
+d\omega^-_{\ih\jh}d\omega^{-\ih\jh})~.
\een

\subsection{Maxwell equations}

The part of the action containing the gauge fields is
\ben\label{xaa}S^{(A)}={1\over 4\pi^2}\int d^5x\sqrt{g}\left({\cal L}_0^{(A)}+{\cal L}_1^{(A)}\right)~,
\een
where the two-derivative terms are
\ben\label{xa}{\cal L}_0^{(A)}=2{\cal N}_Iv^{ab}F^I_{ab}+{1\over 4}{\cal
N}_{IJ}F^I_{ab}F^{Jab}+{1\over24}c_{IJK}A^I_aF^J_{bc}F^K_{de}\epsilon^{abcde}~,
\een
and the four-derivative contributions are
\bea\label{xb}{\cal
L}_1^{(A)}&=&{c_{2I}\over 24}\bigg({1\over16}\epsilon^{abcde}
A^I_aR_{bc}^{\phantom{bc}fg}R_{defg}
+{2\over 3}\epsilon_{abcde}F^{Iab}v^{cf}{\cal
D}_fv^{de}+\epsilon_{abcde}F^{Iab}v^{c}_{~f}{\cal
D}^dv^{ef} \cr &&~~~~~~~~+{1\over 6}F^{Iab}v_{ab}D
+{1\over 2}F^{Iab}C_{abcd}v^{cd}-{4\over 3}F^{Iab}v_{ac}v^{cd}v_{db}
-{1\over 3}F^{Iab}v_{ab}v^2\bigg)~.
\eea
Variation of (\ref{xaa}) with respect to $A^I_{\mu}$ gives,
\bea\label{xc}\nabla_\mu\left(4{\cal N}_Iv^{\mu\nu}+{\cal
N}_{IJ}F^{J\mu\nu}+2{\delta {\cal L}_1\over \delta
F^I_{\mu\nu}}\right)={1\over 8}c_{IJK}
F^J_{\alpha\beta}F^K_{\sigma\rho}\epsilon^{\nu\alpha\beta\sigma\rho}
+{c_{2I}\over 24\cdot16}\epsilon^{\nu\alpha\beta\sigma\rho}
R_{\alpha\beta\mu\gamma}R_{\sigma\rho}^{\phantom{\sigma\rho}\mu\gamma}~,
\eea
with
\bea\label{xca}2{\delta {\cal L}_1\over \delta
F^{Iab}}={c_{2I}\over 24}\bigg(&&{1\over 3
}v_{ab}D-{8\over 3}v_{ac}v^{cd}v_{db}-{2\over 3}v_{ab}v^2
+C_{abcd}v^{cd}\cr&&~~~~~~~~~~~+{4\over 3}\epsilon_{abcde}v^{cf}{\cal
D}_fv^{de}+2\epsilon_{abcde}v^c_{~f}{\cal D}^dv^{ef}\bigg)~,
\eea
and
\ben\label{xcb}{\delta {\cal L}_1\over \delta
F^I_{\mu\nu}}=e_{a}^{~\mu}e_{b}^{~\nu}{\delta {\cal L}_1\over \delta
F^I_{ab}}~.
\een

A lengthy computation is now required in order to expand and simplify (\ref{xc}).  After making heavy
use of the conditions derived from supersymmetry,  we eventually
find that the spatial components of (\ref{xc}) are satisfied identically, while the time component
reduces to
\bea\label{max}\nt^2 \Bigg[ M_I e^{-2U}&-&{c_{2I} \over 24} \left(
3(\nt U)^2 -{1 \over 4} e^{6U}d\omega^{+\ih\jh}d\omega^+_{\ih\jh} -
{1 \over 12} e^{6U}d\omega^{-\ih\jh}d\omega^-_{\ih\jh}\right)\Bigg]
\cr &&={1 \over 2}c_{IJK} \Theta^J \cdot \Theta^K +{c_{2I} \over 24}{1\over8}
\Rt^{\ih\jh\kh\lh}\Rt_{\ih\jh\kh\lh}~, \eea
where $ \Theta^I \cdot \Theta^J=\Theta^{I\ih\jh}\Theta^J_{\ih\jh}$
and $\Rt_{\ih\jh\kh\lh}$ is the Riemann tensor of the metric on the
base.   Note also that the indices on $\Theta^I_{\ih\jh}$ are
defined to be orthonormal with respect to the metric on the base.

\subsection{$D$ equation}

The equation of motion for the auxiliary field $D$ was given in (\ref{mvsg}).   In the present case it
becomes
\bea\label{cons}
{\cal N}&=& 1- {c_{2I} \over 24} e^{2U} \Bigg[  M^I \left( \nt^2 U -4(\nt U)^2\right) + \nt_\ih M^I \nt_\ih U
\cr  &&~~~~~~~~~~~~\quad +{1\over 4} e^{6U} M^I \left( d\omega^{+\ih\jh} d\omega^+_{\ih\jh} +{1 \over 3} d\omega^{-\ih\jh} d\omega^-_{\ih\jh}
\right)-{1\over 12} e^{4U} \Theta^I_{\ih\jh} d\omega^{-\ih\jh} \Bigg]~.
\eea

\subsection{$v$ equation}

The final ingredient needed to completely determine the general solution is the $v$ equation of motion.
In fact, for the explicit solutions considered in this review, namely the spinning black holes,  this information is not
needed.   It is however needed to determine the black ring solution, and so we display the result.
The full $v$ equation of motion is rather forbidding, and so we simplify by considering just a flat base space.
Furthermore,  simplifications result upon contracting the $v$ equation with $d\omega$.  It turns out
that the $v$ equation contracted with $d\omega^+$ is automatically satisfied given our prior results,
and so, after a lengthy calculation,  we are left with
\bea\label{veq} &&
 {1\over 4} d\omega^{-\ih\jh} d\omega^-_{\ih\jh} +{1\over 8} e^{-2U} M_I \Theta^{I\ih\jh}
d\omega^-_{\ih\jh} = -{c_{2I} \over 16\cdot 24} d\omega^{-\ih\jh} \Big[  -{1\over 6} e^{-6U} \nt^2 (e^{6U} \Theta^I_{\ih\jh})\cr
&&\quad\quad\quad\quad\quad\quad\quad\quad\quad\quad+4 \nt_\jh \nt_\kh
(e^{2U} M^I  d\omega^+_{\ih\kh})  +{1\over 3} \nt^2 (e^{2U} M^I d\omega^-_{\ih\jh})
\cr && \quad\quad\quad\quad\quad\quad\quad\quad\quad\quad+{1\over 6}e^{6U} \Theta^I_{\ih\jh} \left((3 d\omega^{+\kh\lh}d\omega^+_{\kh\lh}
+d\omega^{-\kh\lh}d\omega^-_{\kh\lh} \right) \Big]~.
\eea

\subsection{Spinning black holes on Gibbons-Hawking space}
\label{ghbh}

We now focus on 5D electrically charged spinning black hole
solutions.   The main simplification here is that we
take\footnote{The two-derivative BMPV solution enjoys such a
property. We include (\ref{bmpvansatz}) as part of our {\it ansatz}
to investigate higher-derivative corrections to this solution.}
\ben\label{bmpvansatz} d\omega^- = \Theta^I =0~. \een Then, to
determine the full solution the relevant equations (\ref{max}) and
(\ref{cons}) become \bea\label{BMPVeqs1} &&\!\!\!\!\nt^2 \left[ M_I
e^{-2U}-{c_{2I} \over 24} \left( 3(\nt U)^2 -{1 \over 4}
e^{6U}d\omega^{+\ih\jh}d\omega^+_{\ih\jh} \right)\right]
= {c_{2I} \over 24\cdot 8} \Rt^{\ih\jh\kh\lh}\Rt_{\ih\jh\kh\lh}~, \\
\label{BMPVeqs2} \!\!\!\!\!\!\!\!\!\!\!\! {\cal N} &&=1- {c_{2I}
\over 24} e^{2U} \Big[  M^I \left( \nt^2 U -4(\nt U)^2\right) +
\nt_\ih M^I \nt_\ih U
 +{1\over 4} e^{6U} M^I  d\omega^{+\ih\jh} d\omega^+_{\ih\jh} \Big]~.
\eea

The base space is now taken to be a Gibbons-Hawking space with metric
\ben\label{ghmetric}
ds_B^2 = (H^0)^{-1}(dx^5 +\vec{\chi}\cdot d\vec{x})^2 +H^0 d\vec{x}^2~,
\een
with $H^0$ and $\vec{\chi}$ satisfying\footnote{Take care not confuse $\vec{\nabla}$, the gradient on $\mathbb{R}^3$, with $\nt$, the covariant derivative on the four-dimensional Gibbons-Hawking space.}
\ben
\vec{\nabla} H^0 = \vec{\nabla}\times\vec{\chi}~,
\een
which in turn implies that $H^0$ is harmonic on $\mathbb{R}^3$, up to isolated singularities. The $x^5$ direction is taken to be compact, $x^5 \cong x^5 +4\pi$, and an isometry direction for the entire solution.
We note a few special cases.
Setting $H^0 =1/|\vec{x}|$ yields the flat metric on $\mathbb{R}^4$ in Gibbons-Hawking
coordinates.   Taking $H^0=1$ yields a flat metric on $\mathbb{R}^3 \times S^1$. A more interesting choice is the charge $p^0$ Taub-NUT space with
\ben\label{jhzero}
H^0 = H^0_{\infty} + \frac{p^0}{|\vec{x}|} ~.
\een
For general $p^0$ the geometry has a conical singularity but the $p^0=1$ case is non-singular.

With this choice of base space we find
\ben
\Rt^{\ih\jh\kh\lh}\Rt_{\ih\jh\kh\lh} = 2 \nt^2 \left( {(\nt H^0)^2 \over (H^0)^2}\right) + \dots~,
\een
where the dots represent $\delta$-functions due to possible isolated singularities in $H^0$, such as in (\ref{jhzero}). Thus we can write $\Rt^{\ih\jh\kh\lh}\Rt_{\ih\jh\kh\lh}$ as a total Laplacian
\ben\label{rsquared}
\Rt^{\ih\jh\kh\lh}\Rt_{\ih\jh\kh\lh} = \nt^2 \Phi \equiv  \nt^2 \left( 2{(\nt H^0)^2 \over (H^0)^2}+\sum_i {a_i \over |\vec{x}-\vec{x}_i|}\rt)~,
\een
for some coefficients $a_i$. We can now solve (\ref{BMPVeqs1}) as
\ben\label{Meq}
M_I e^{-2U}-{c_{2I} \over 24} \left[ 3(\nt U)^2 -{1 \over 4} e^{6U}d\omega^{+\ih\jh}d\omega^+_{\ih\jh} \right]
-{c_{2I} \over 24\cdot 8} \Phi   = H_I~,
\een
where $\nt^2 H_I =0$. The choice
\ben\label{jhi}
H_I = 1 +{q_I \over 4\rho}~,\quad  \rho = |\vec{x}|~,
\een
identifies the $q_I$ as the conserved 5D electric charges in the case that the base is $\mathbb{R}^4$. In Section \ref{q4D5D} we will closely study the case of a Taub-NUT base space and see that there are modifications to the asymptotics controlled by the coefficients $a_i$ in (\ref{rsquared}).

The rotation, as encoded in $d\omega^+$, is determined uniquely from closure and self-duality to be
\ben
d\omega^+ = -{J\over 8 \rho^2}x^m \left( \et^{\hat{5}} \wedge \et^{\hat{m}} + {1 \over 2} \epsilon_{\hat{m}\hat{n}\hat{p}}
 \et^{\hat{n}}\wedge \et^{\hat{p}}\right)~,
\een
where $\et^a$ are the obvious vielbeins for the Gibbons-Hawking metric (\ref{ghmetric}) and the orientation is $ \epsilon_{\hat{5}\hat{m}\hat{n}\hat{p}}=1$.
With this normalization, $J$ is the angular momentum of the 5D spinning black hole (note that
for the supersymmetric black hole the two independent angular momenta in 5D must be equal.)

Now that $d\omega^+$ has been specified the full solution can found as follows.   After using
(\ref{Meq}) to find $M_I$ we determine $M^I$ by solving $M_I ={1\over 2}c_{IJK} M^J M^K$ (this can  be done explicitly only for special choices of
$c_{IJK}$).   We then insert $M^I$ into (\ref{BMPVeqs2}) to obtain a nonlinear, second order, differential equation for
$U= U(\rho)$. This last equation typically can be solved only by numerical integration.
However, the near horizon limit of the solution can be computed analytically as we do next.

\subsection{The rotating attractor revisited}
\label{sec:rotattr}
We can employ these formulae to make the rotating attractor discussed in Section
\ref{sec:rotatt} more explicit. For this we take the base to be flat $\mathbb{R}^4$,
deferring the Taub-NUT case to the next section.

The attractor solution corresponds to dropping the constant in the harmonic
functions (\ref{jhzero},\ref{jhi}) and considering a metric factor of  the form
\ben
e^{2U} = {\rho\over\ell^2}~.
\een
The resulting geometry takes the form of a circle fibered over an AdS$_2 \times S^2$,
as detailed in (\ref{ggeom}). Inserting the various functions into the modified very special
geometry constraint (\ref{BMPVeqs2}) and the relation expressing flux conservation (\ref{Meq})
we verify that this {\it ansatz} gives an exact solution. Importantly, we also find the relation
between the parameters of the attractor solution and the charges measured at infinity.

In more detail, to display the near horizon solution it is useful to define the rescaled quantities
\ben
\Mh^I = 2\ell M^I~,\quad \Jh ={1\over 8 \ell^3} J~,
\een
where $\ell$ can be identified with the radii of the AdS$_2$ and $S^2$ factors in string frame.

We then have the following procedure: given asymptotic charges $(J,q_I)$
we find the rescaled variables $({\hat J}, {\hat M}^I)$ by solving the
equations (\ref{BMPVeqs2},\ref{Meq}) written in the form
\bea\label{bk}
J & = &\left( {1\over 3!} c_{IJK} {\hat M}^I {\hat M}^J {\hat M}^K
- {c_{2I}{\hat M}^I\over 12}( 1 - 2{\hat J}^2)\right){\hat J}~,\cr
q_I & =& {1\over 2} c_{IJK} {\hat M}^J {\hat M}^K - {c_{2I}\over 8}\left( 1 - {4\over 3} {\hat J}^2\right)~.
\eea
With the solution in hand we compute
\bea\label{bl}
\ell^3 &=& {1\over 8}\left( {1\over 3!} c_{IJK} {\hat M}^I {\hat M}^J {\hat M}^K
- {c_{2I}{\hat M}^I\over 12}( 1 - 2{\hat J}^2)\right)~,\cr
M^I & =& {1\over 2\ell}{\hat M}^I~,
\eea
to find the values for the physical scale of the solution $\ell$
and the physical moduli $M^I$, written as functions of $(J,q_I)$.  A novel feature of the
higher derivative attractor mechanism is that the fixed values of the moduli depend on
the angular momentum as well as the electric charges.  From (\ref{bk}) it is clear that the
$J$ dependence only appears through the higher derivative terms.

In general it is of course rather difficult to invert (\ref{bk})
explicitly. This is the situation also before higher derivative
corrections have been taken into account and/or if angular
momentum is neglected.  However, in the large charge regime we can make the dependence on the higher derivative
corrections manifest in an inverse charge expansion.   Let us
define the dual charges $q^I$ through
 \ben\label{bla} q_I = {1\over 2}
c_{IJK} q^J q^K~.
\een
We also define\footnote{The ${3\over 2}$ power  is introduced so that $Q$ has the same
dimension as the physical charges $q_I$.}
\ben\label{blb} Q^{3/2} = {1\over
3!} c_{IJK} q^I q^J q^K~,
\een
\ben\label{blc}  C_{IJ} = c_{IJK} q^K~.
\een
Each of these quantities depend on charges and Calabi-Yau data
but not on moduli.

With the definitions (\ref{bla})-(\ref{blc}) we can invert (\ref{bk}) for large
charges ({\it i.e.} expand to first order in $c_{2I}$) and find
\bea\label{bld}{\hat M}^I &=& q^I + {1\over 8} \left( 1 -
{4\over 3} {J^2\over Q^3}\right)C^{IJ}c_{2J} + \ldots~,\cr {\hat
J} & =& {J\over Q^{3/2}} \left( 1 + {c_{2}\cdot q\over 48
Q^{3/2}}\left[1-4{J^2 \over Q^3}\right] \right)+\ldots~,
\eea
where $C^{IJ}$ is the inverse of the matrix $C_{IJ}$ defined in (\ref{blc}). Then (\ref{bl}) gives the physical
scale of the geometry and the physical moduli as
\bea\label{ble}\ell & =& {1\over 2}Q^{1/2}\left( 1  -
{c_{2}\cdot q\over 144 Q^{3/2}}\left[1-4{J^2 \over Q^3}\right]
\right)+\ldots ~,\cr M^I & =& {q^I\over Q^{1/2}}\left( 1  +
{c_2\cdot q\over 144 Q^{3/2}}\left[1-4{J^2 \over Q^3}\right]
\right) +{ 1 \over 8Q^{1/2}}\left( 1 - {4\over 3}{J^2\over
Q^3}\right)C^{IJ}c_{2J} +\ldots ~.
\eea

\subsection{Example: $K3\times T^2$ compactifications}

We can find more explicit results
in the special case of $K3\times T^2$ compactifications. In this case $c_{1ij} =
c_{ij}$, $i,j=2,\ldots 23$ are the only nontrivial intersection
numbers and $c_{2,i}=0$, $c_{2,1}=24$  are the 2nd Chern-class coefficients.   We define
$c^{ij}$ to be the inverse of the $K3$ intersection matrix $c_{ij}$.

We first derive the attractor solution. Our procedure instructs us
to first find the hatted variables in terms of conserved charges by
inverting (\ref{bk}). In the present case we find
\ben\label{ga}
{\hat M}^1 = \sqrt{\half c^{ij} q_i q_j + {4J^2\over (q_1+1)^2}\over q_1 + 3}~, \quad\quad
{\hat M}^i = \sqrt{ q_1 + 3\over \half c^{ij} q_i q_j + {4J^2\over (q_1+1)^2}} c^{ij} q_j~,
\een
and
\ben\label{gb}
{\hat J}  = \sqrt{ q_1 + 3\over \half c^{ij} q_i q_j + {4J^2\over (q_1+1)^2}}{J\over q_1+1}~.
\een
All quantities of interest are given in terms of these  variables.
%
%
For completeness, we display the entropy of this solution here, although
it will be derived later in Section \ref{sec:sext} for an arbitrary
Calabi-Yau compactification. For a spinning black hole the entropy
is given by (\ref{elent}) which, after substitution of (\ref{ga}) and (\ref{gb}) and the intersection numbers and Chern class coefficients for
$K3\times T^2$, becomes
\bea\label{gc} S &=& 2\pi \sqrt{ \half c^{ij} q_i q_j (q_1 + 3)
-  {(q_1-1) (q_1+3)\over (q_1+1)^2}J^2}~.
\eea
In the case of $K3\times T^2$ the charge $q_1$ corresponding to
M2-branes wrapping $T^2$ is apparently special; the higher order corrections to the entropy are encoded entirely in the modified functional dependence on $q_1$.

We now turn to the full asymptotically flat solution in the static case $J=0$.    The full solution can be expressed
explicitly in terms of the function $U$,  which obeys a nonlinear ODE requiring a numerical
treatment.   We first invert $M_I = {1 \over 2} c_{IJK} M^J M^K$ as
\ben\label{Msol} M^1 = \sqrt{{ c^{ij} M_i M_j \over 2M_1}}~,\quad M^i =c^{ij} M_j \sqrt{{2 M_1 \over
c^{kl} M_k M_l}}~.
\een
Substituting into (\ref{Meq}) gives
\ben\label{K3eq1} M^1 = \left({e^{2U} c^{ij} H_i H_j \over 2H_1 +6 U'^2}\right)^{1/2}~,\quad M^i  =\left({e^{2U} c^{ij} H_i H_j \over 2H_1 +6 U'^2}\right)^{-1/2}e ^{2U} c^{ij} H_j~,
\een
where $' $ denotes differentiation with respect to $ r= 2 \sqrt{\rho}$ (in terms of $r$ the base space metric
becomes $dr^2 +r^2 d\Omega_3^2$).   The special geometry
constraint (\ref{BMPVeqs2}) is
\ben\label{K3eq2} \half c_{ij}M^i M^j M^1 -1 +e^{2U}\left[ (U'' +{3\over r}U'-4 U'^2)M^1 +U' {M^1}' \right] =0~.
\een
The problem is now to insert (\ref{K3eq1}) into (\ref{K3eq2}) and solve for $U(r)$.

\begin{figure}[pb]
\centerline{\psfig{file=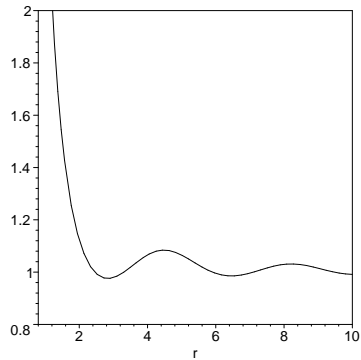,width=6cm}} \vspace*{8pt}
\caption{\label{smbh}Numerical solution of equation (\ref{K3eq3}); the curve represents $e^{-2U(r)}$ for small values of $r$. The oscillatory behavior is characteristic of higher derivative theories and will be discussed further in Section \ref{farstring}.}
\end{figure}

This is straightforward to solve numerically given specific choices of charges.  Consider a small
black hole, $q_1=0$ with $q_2=q_3=1,~ c^{23}=1$.   We also assume $H=H_2=H_3 =1+{1 \over r^2}$
are the only harmonic functions not equal to unity.   Then (\ref{K3eq2}) becomes
\ben\label{K3eq3} HU'' +(1+3 U'^2)\left[ \left({3\over r}+{1\over
r^3}\right) U' +H\right] -e^{-3U}(1+3 U'^2)^{3/2}=0~. \een
The boundary conditions are fixed by matching to the small $r$
behavior
\ben e^{-2U} \sim {\ell_S^2 \over r^2}~, \een with $\ell_S =
3^{-1/6}$.
The result of the numerical solution for $U(r)$ is shown in
Fig.~\ref{smbh}.

\subsection{Comments on black rings}

Black rings\cite{Elvang:2004rt,Bena:2004de} incorporate nonzero
$\Theta^I$ and $d\omega^-$.  After choosing the base space, the
two-form $\Theta^I$ can be determined by the requirements of closure
and anti-self duality.   In the two-derivative limit $d\omega^-$ can
be determined from the $v$ equation of motion according to \ben
d\omega^- = {1 \over 2} e^{-2U} M_I \Theta^I~. \een In the higher
derivative case there is instead equation (\ref{veq}), which has not
yet been solved.    The full black ring solution is therefore not
available at present.   We can, however, find the near horizon
geometry of the black ring and an expression for its entropy. This
question will be revisited in the Section \ref{sec:bring}.

\section{Black Hole Entropy and Extremization Principles}\label{sec:entropy}
An important application of the solutions we construct is to the study of gravitational
thermodynamics. The higher derivative corrections to the supergravity
solutions are interesting for this purpose because they
are sensitive to details of the microscopic statistical description.

The black hole entropy is famously given by the Bekenstein-Hawking area law
\ben\label{ha}
S= {1\over 4G_D} A_{D-2}~.
\een
This expression applies only when the gravitational action is just
the standard Einstein-Hilbert term. In general, one must use instead
the Wald entropy formula\footnote{Theories with gravitational
Chern-Simons terms may violate diffeomorphism invariance. Then
Wald's formula does not apply and one must use a further
generalization due to Tachikawa.\cite{Tachikawa:2006sz}}
\ben\label{hb}
S = -{1\over 8G_D}
\int_{\rm hor} d^{D-2}x \sqrt{h} {\delta{\cal L}_D\over \delta R_{\mu\nu\rho\sigma}}\epsilon^{\mu\nu} \epsilon^{\rho\sigma}~.
\een
This reduces to (\ref{ha}) for the two-derivative action, but
generally the density one must integrate over the event horizon is
more complicated than the canonical volume form. In practice, it is
in fact rather cumbersome to evaluate (\ref{hb}) and evaluate the
requisite integral but there is a short-cut that applies to black
holes with near horizon geometry presented as a fibration over
AdS$_2\times S^2$. Then the Wald entropy (\ref{hb}) is the Legendre
transform of the on-shell action\footnote{This refers to the usual notion of ``on-shell action'', {\it i.e.} the action evaluated on a solution to all of the equations of motion. This is not to be confused with the sense of ``on-shell'' that we have been using throughout this review, {\it i.e.} with only the auxiliary field equations of motion imposed.} up to an overall numerical factor.
This general procedure is known as the entropy function formalism.\cite{Sen:2005wa} In Section \ref{sec:sext} we apply
the entropy function formalism to our five dimensional black hole solutions
with AdS$_2\times S^3$ near horizon geometry.

Although we analyze a theory in five dimensions, we can discuss
four dimensional black holes by adding excitations to black strings
with AdS$_3 \times S^2$ near string geometry. For large
excitation energy the black hole entropy is given by Cardy's
formula
\ben\label{hc} S = 2\pi \left[ \sqrt{{c_L\over 6}\left(h_L-{c_L\over
24}\right)} + \sqrt{{c_R\over 6}\left(h_R-{c_R\over
24}\right)}~\right]~, \een
where $h_L, h_R$ are eigenvalues of the AdS$_3$ energy generators
$L_0, {\bar L}_0$. Since Cardy's formula can be justified in both
the gravitational description\footnote{In fact Cardy's formula
(\ref{hc}) agrees with the Wald entropy whenever diffeomorphism
invariance applies ($c_L=c_R$),\cite{Kraus:2005vz,Saida:1999ec} or with Tachikawa's generalization\cite{Tachikawa:2006sz} when $c_L \neq c_R$.} and also in the dual CFT, the
central charge becomes a proxy for the entropy in the AdS$_3\times
S^2$ setting. It is therefore the central charge that we want to
compute for our solutions. The central charge is convenient to
compute because it is just the on-shell action, up to an overall
numerical factor. This methodology is known as
$c$-extremization.\cite{Kraus:2005vz} In Section \ref{sec:cext} we
apply $c$-extremization to our five dimensional black string
solutions with AdS$_3\times S^2$ near horizon geometry.

The entropy function formalism and the $c$-extremization procedure can be
carried out while keeping arbitrary the scales of the AdS and sphere
geometries, as well as matter fields consistent with the symmetries.
These parameters are then determined by the extremization procedure
in a manner independent of supersymmetry. The computations therefore
constitute an important consistency check on the explicit Lagrangian
and other parts of the framework.

\subsection{Black strings and $c$-extremization}
\label{sec:cext} The general $c$-extremization procedure considers a
AdS$_3\times S^{D-3}$ solution to a theory with action of the form
\ben
S= {1 \over 16\pi G_D} \int\! d^D x \sqrt{g}{\cal L} +S_{CS}+
S_{\rm bndy} ~.
\een
The Chern-Simons terms (if any) are collected in the term $S_{CS}$,
and $S_{\rm bndy}$ are the terms regulating the infrared divergences
at the boundary of AdS$_3$. The total central charge
\ben\label{hctot} c = \half(c_L + c_R)~, \een
is essentially the trace anomaly of the CFT, which in turn is encoded in the
on-shell action of the theory. The precise relation is
\ben
c=  -{3\Omega_{D-3}\over 8G_D}\ell_A^3\ell_S^{D-3} {\cal L}_{\rm ext}~,
\een
with the understanding that the action must be extremized over all parameters, with
magnetic charges through $S^{D-3}$ kept fixed.

We want to apply this formalism to the black string attractor
solution found in Section \ref{sec:magattr}. The isometries of the
near horizon region determines the form of the solution as
\bea\label{hd} ds^2 & =& \ell_A^2  ds^2_{AdS} - \ell_S^2
d\Omega_2^2~, \cr F^I &=& -{p^I \over 2}\epsilon_2~, \cr v&=& V
\epsilon_2~,\cr M^I & = & mp^I~. \eea
In Section \ref{sec:magattr} we used maximal supersymmetry and the modified very special geometry constraint to determine the parameters $\ell_A, \ell_S, V, m$ and the
auxiliary scalar $D$ in terms of the magnetic charges $p^I$ as
\bea\label{he} V & =&  {3\over 8}\ell_A~,~ D = {12\over\ell^2_A}~,~
m = {1\over\ell_A}~,\cr \ell_S & =& {1\over 2} \ell_A~,~~  \ell_A^3
= p^3+{1\over12}c_2\cdot p~, \eea
where
\ben p^3 \equiv {1\over 6}c_{IJK}p^I p^J p^K~. \een
However, it is instructive to use just the {\it ansatz} (\ref{hd})
for now. Inserting this {\it ansatz} into the leading order action
(\ref{offtwod}) we find
\bea\label{hlzero} {\cal L}_0 &=&
2\bigg({1\over4}(p^3m^3-1)D-{1\over4}(p^3m^3+3)\left({3\over
\ell_A^2}-{1\over \ell_S^2}\right)\cr&&~~~~~~+{1\over
\ell_S^4}\left((3p^3m^3+1)V^2+3p^3m^2V\right)+{3p^3\over
\ell_S^4}{m\over8}\bigg)~, \eea
and the four derivative action (\ref{offfourd}) yields
\bea\label{hlone}
 {\cal L}_1={c_2\cdot p\over
24}\bigg[&\frac{m}{4}\left(\frac{1}{ \ell_A^2}-\frac{1}{
\ell_S^2}\right)^2+{2\over3}{V^3\over \ell_S^8}+4m{V^4\over
\ell_S^8}+m{D^2\over12}+{D\over6}{V\over \ell_S^4}\cr&
-{2\over3}m{V^2\over \ell_S^4}\left({3\over \ell_A^2}+{5\over
\ell_S^2}\right)+{1\over2}{V\over \ell_S^4}\left({1\over
\ell_A^2}-{1\over \ell_S^2}\right)\bigg]~. \eea
According to $c$-extremization we now need to extremize the $c$-function
\ben\label{hcfct}
c(\ell_A,\ell_S,V,D,m) = -6\ell_A^3\ell_S^2 ({\cal L}_0 + {\cal L}_1 )~,
\een
with respect to all variables. The resulting extremization conditions are quite involved.
For example, the variation of (\ref{hcfct}) with respect to $m$ gives
\bea\label{hmext}
&{3p^3m^2\over4}\left(D-{3\over \ell_A^2}+{1\over
\ell_S^2}\right)+{3p^3\over \ell_S^4}\left(3m^2V^2+2mV+{1\over
8}\right)+\cr &+{c_2\cdot p\over 48}\left[{1\over4}\left({1\over
\ell_A^2}-{1\over \ell_S^2}\right)^2+4{V^4\over
\ell_S^8}+{D^2\over12}-{2\over3}{V^2\over \ell_S^4}\left({3\over
\ell_A^2}+{5\over \ell_S^2}\right)\right]=0~.
\eea
It would be very difficult to solve equations with such complexity
without any guidance. Fortunately we already determined the
attractor solution (\ref{he}) and it is straightforward to verify
that it does indeed satisfy (\ref{hmext}). We can similarly vary the
$c$-function (\ref{hcfct}) with respect to $\ell_A$, $\ell_S$, $V$,
$D$ and show that the resulting equations are satisfied by the
attractor solution (\ref{he}). Thus the attractor solution
extremizes the $c$-function (\ref{hcfct}) as it should.

Since we have proceeded indirectly we have not excluded the
possibility that $c$-extremization could have other solutions with
the same charge configuration. Such solutions would not be
supersymmetric. This possibility further imposes the point that
$c$-extremization is logically independent from the considerations
using maximal supersymmetry that determined the attractor solution
in the first place. The success of $c$-extremization therefore
constitutes a valuable consistency check on the entire framework.

At this point we have verified that the $c$-function is extremized
on the attractor solution (\ref{he}). The central charge is now
simply the value of the (\ref{hcfct}) on that solution. The
computation gives
\ben\label{hstringc} c=6p^3+{3\over4}c_2\cdot p~. \een

In order to put this result in perspective, let us recall the
microscopic interpretation of these black
strings.\cite{Maldacena:1997de} We can interpret $N=2$
supergravity in five dimensions as the low energy limit of M-theory
compactified on some Calabi-Yau threefold $CY_3$.  The black string in five dimensions
corresponds to a M5-brane wrapping a 4-cycle in $CY_3$ that has
component $p^I$ along the basis four-cycle $\omega_I$. The central
charges of the effective string CFT are known to
be\cite{Maldacena:1997de,Harvey:1998bx}
\ben\label{hcanom}
c_L= 6p^3 + \half c_2 \cdot p~,\quad c_R= 6p^3 +  c_2 \cdot
p~,
\een
where $c_{IJK}$ are the triple intersection numbers of the $CY_3$,
and $c_{2I}$ are the expansion coefficients of the second Chern
class. Computing the total central charge (\ref{hctot}) from
(\ref{hcanom}) we find precise agreement with our result
(\ref{hstringc}) found by $c$-extremization.

It is worth noting that the simple form of the central charge comes
about in a rather nontrivial way in the $c$-extremization procedure.
The radius of curvature $\ell_A$ from the last line of (\ref{he})
introduces powers of $(p^3 + {1\over 12} c_2\cdot p)^{1/3}$ in the
denominator of the Lagrangian (\ref{hlzero}-\ref{hlone}). It is only
due to intricate cancellations that the final result
(\ref{hstringc}) becomes a polynomial in the charges $p^I$.

\subsection{Black hole entropy}
\label{sec:sext} We want to compute the entropy of black hole
solutions with AdS$_2\times S^3$ near horizon geometry. As mentioned
in the introduction to this section the most efficient method to
find the entropy is by use of the entropy function,\cite{Sen:2005wa}
which amounts to computing the Legendre transform of the Lagrangian
density evaluated on the near horizon solution.  Some care is needed
because the 5D action contains non-gauge invariant Chern-Simons
terms while the entropy function method applies to gauge invariant
actions.

We first review the general procedure for determining the entropy
from the near horizon solution, mainly following
Ref.~\refcite{Goldstein:2007km}. The general setup is valid for
spinning black holes as well as black rings.

The near horizon geometries of interest take the form of a circle
fibered over an AdS$_2 \times S^2$ base:
\bea
ds^2& =& w^{-1}\Big[v_1 \Big(\rho^2 d\tau^2
-{d\rho^2 \over \rho^2}\Big) -v_2 (d\theta^2 +\sin^2 \theta
d\phi^2) \Big] -w^2 \Big(dx^5 + e^0 \rho d\tau + p^0 \cos \theta
d\phi\Big)^2~, \cr A^I & =& e^I \rho d\tau+p^I\cos \theta
+a^I\Big(dx^5 + e^0 \rho d\tau + p^0 \cos\theta d\phi\Big)~, \cr
v&=& -{1 \over 4 \Nc} M_I F^I~.
\eea
The parameters $w$, $v_{1,2}$, $a^I$ and all scalar fields are
assumed to be constant.     Kaluza-Klein reduction along $x^5$ yields a 4D
theory on AdS$_2 \times S^2$.  The solution carries the magnetic
charges $p^I$, while $e^I$ denote electric potentials.\footnote{An
important point, discussed at length in Section \ref{sec:fourdfived},
is that $e^I$ are
conjugate to 4D electric charges, which differ from the 5D
charges.}

Omitting  the Chern-Simons terms for the moment, let the action be
\ben\label{cb} I = {1 \over 4\pi^2}\int\! d^5x \sqrt{g}{\cal L}~.
\een
Define
\ben\label{cc} f= {1 \over 4\pi^2}\int\! d\theta d\phi dx^5
\sqrt{g}{\cal L}~.
\een
Then the black hole entropy is
\ben\label{cd} S= 2\pi \Big(e^0 {\p f \over \p e^0} + e^I {\p f \over \p
e^I} -f \Big)~.
\een
Here $w$, $v_{1,2}$ etc. take their on-shell values.  One way to
find these values is to extremize $f$ while holding fixed the
magnetic charges and electric potentials.  The general
extremization problem would be quite complicated given the
complexity of our four-derivative action.   Fortunately, in
the cases of interest we already know the values of all fields
from the explicit solutions.

The Chern-Simons term is handled by first reducing the action along
$x^5$ and then adding a total derivative to ${\cal L}$ to restore
gauge invariance in the resulting 4D action (it is of course not
possible to restore gauge invariance in 5D).\cite{Castro:2007ci}

 The result of this computation is the entropy formula
\ben\label{elent}
S = 2\pi \sqrt{1-\Jh^2} \left({1\over 6}c_{IJK} \Mh^I \Mh^J \Mh^K +{1 \over 6} \Jh^2 c_{2I} \Mh^I \right)~,
\een
where the rescaled moduli are evaluated at their attractor values (\ref{bk}).

We can also express the entropy in terms of the conserved charges. We first
use (\ref{bl}) to find an expression in terms of geometrical variables
\ben\label{cya}S = 2\pi \sqrt{(2\ell)^6-J^2} \Big( 1   +{c_{2I}
M^I \over 48\ell^2}\Big)~,
\een
and then expand to first order in $c_{2I}$ using (\ref{ble}) to find
\ben\label{cyb} S=
2\pi \sqrt{Q^3-J^2} \Big( 1   +{c_{2}\cdot q\over 16}{Q^{3/2}\over
(Q^3-J^2)}  + \cdots\Big)~.
\een

From the standpoint of our 5D supergravity action (\ref{elent}) is an exact expression for the entropy.
But as a statement about black hole entropy in string theory it is only valid to first order in $c_{2I}$,
since we have only kept terms in the effective action up to four derivatives.    The situation here
is to be contrasted with that for 5D black strings, where anomaly arguments imply that the
entropy is uncorrected by terms beyond four derivatives.  The anomaly argument relies on the
presence of an AdS$_3$ factor, which is absent for the 5D black holes considered in this section.

\subsection{Black ring entropy}
\label{sec:bring}
Although we have not yet determined the complete black ring solution we can
compute its entropy by applying the entropy function formalism to the black ring attractor.

For the black ring the near horizon solution is
\bea ds^2& = &w^{-1}v_3\Big[ \Big(\rho^2 d\tau^2 -{d\rho^2 \over
\rho^2}\Big) - d\Omega^2 \Big] -w^2 \Big(dx^5 + e^0 \rho d\tau
\Big)^2~, \cr A^I & =&-{1 \over 2} p^I \cos \theta d\phi-{e^I \over
e^0} dx^5 ~. \eea
Further details of the solution follow from the fact that the near
horizon geometry is a magnetic attractor.
The near horizon geometry is a product of a BTZ black hole and an
$S^2$,  and there is enhanced supersymmetry.  These
conditions imply
\bea\label{ec} M^I & =& {p^I \over 2w e^0}~, \cr v_3&=&
w^3(e^0)^2~,\cr D& =& {3 \over   w^2 (e^0)^2}~, \cr v&=&-{3 \over
4}w e^0\sin\theta d\theta \wedge d\phi~. \eea

The computation of the  entropy in terms of the entropy function proceeds as in the
case of the spinning black hole. The result is
\ben\label{ee}
S ={2\pi\over e^0} \left({1\over 6} c_{IJK} p^I p^J p^K +{1 \over 6} c_{2I}p^I\right)~.
\een
The entropy is expressed above in terms of magnetic charges $p^I$
and the potential $e^0$, but the preferred form of the entropy would
be a function of the conserved asymptotic charges. To get a formula
purely in terms of the charges $(p^I,q_I)$ and the angular momenta
we need to trade away $e^0$.  But for this one needs knowledge of
more than just the near horizon geometry, which, as we noted above,
is not available at present.

Let us finally note that the entropy can be expressed in geometric variables as
\ben\label{ef} S = (2-\Nc){A \over \pi}=(2-\Nc){A \over 4 G_5}~,
\een
where $A$ is the area of the event horizon. In the two-derivative
limit we have $\Nc=1$ and we recover the Bekenstein-Hawking
entropy.

\section{Small Black Holes and Strings}\label{sec:small}

One of the main motivations for studying  higher derivative
corrections is their potential to regularize geometries that are
singular in the lowest order supergravity approximation.\cite{Kraus:2005vz,Dabholkar:2004dq}\cdash\cite{Hubeny:2004ji,Dabholkar:2004yr} One version of this phenomenon occurs for black holes possessing a nonzero entropy, where
the effect of the higher derivative terms is not to remove the black hole singularity,
but rather to shield it with an event horizon.   The resulting spacetime is then
qualitatively similar to that of an ordinary ``large" black hole.  Examples of this
occur for both four and five dimensional black holes in string theory.
A second, and in many ways more striking, example pertains to the case in which
the solution has a vanishing entropy.   In this case the singularity, instead of
being shielded by a finite size event horizon, is smoothed out entirely.  Our five
dimensional string solutions provide an explicit realization of this.

To realize  the latter type of solutions, we  consider magnetic string solutions
whose charge configurations satisfy
$p^3 = {1\over 6}c_{IJK}p^I p^J p^K=0$. We refer to these as {\it small strings}.
Recall from Section \ref{sec:magattr} that our string solutions had a near horizon AdS$_3 \times S^2$ geometry
with AdS scale size given by
\ben
\label{scalel}
\ell^3_A = p^3 + {1\over 12} c_2\cdot p~.
\een
For small strings the geometry is singular in the two derivative approximation, since
$\ell^3_A=0$.  Conversely,  $\ell^3_A\neq 0$ when the correction proportional
to $c_{2I}$ is taken into account. Thus it appears that a spacetime singularity has
been resolved. To understand the
causal structure we can note that our metric is a particular example
of the general class of
geometries studied in Ref.~\refcite{Gibbons:1994vm}.    The resulting
Penrose diagram is
like that of the M5-brane in eleven dimensions.  In particular, the
geometry is completely smooth,
and there is no finite entropy event horizon.

We should, however, close one potential loophole.
In principle, it could be that the actual near string geometry realized in the full
asymptotically flat solution is not the regular solution
that is consistent with the charges, but instead a deformed but still singular
geometry. In order to exclude this possibility we must construct the complete
solution that smoothly interpolates between the regular near horizon geometry
and asymptotically flat space. In this section we present such an interpolating solution,
thereby confirming that the singularity is indeed smoothed out.

Since the near string geometry after corrections are taken into account has an
AdS$_3$ factor, it is natural to ask whether the AdS/CFT correspondence
applies, and to determine what special features the holography
might exhibit. This question has attracted significant attention recently and
remains an active area of
inquiry.\cite{Dabholkar:2007gp}\cdash\cite{Kraus:2007vu}

A particularly important example of a small string is obtained when the Calabi-Yau is
$K3\times T^2$, and the only magnetic charge that is turned on is that
corresponding to an $M5$-brane wrapping the $K3$. The resulting 5D string
is then dual, via IIA-heterotic duality,  to the fundamental heterotic
string.\cite{Sen:1995cj,Harvey:1995rn}    We will focus on this particular
example in this section.

\subsection{The small string: explicit solution}

Let $M^1$ be the single modulus on the torus and $M^i$ be the
moduli of $K3$ where $i=2,\ldots,23$. The charge configuration of
interest specifies the harmonic functions as
\bea\label{bn}
H^1&=&M^1_{\infty}+{p^1\over2r}~,\cr
H^{i}&=&M^i_\infty~,~~~~~~i=2,\ldots,23~.
\eea
The only nonvanishing intersection numbers are $c_{1ij}=c_{ij}$
where $c_{ij}$ is the intersection matrix for $K3$.   To simplify, we choose
$M^i_\infty$ consistent with ${1\over 2} c_{ij} M^i_\infty M^j_\infty =1$, so that (\ref{bkab}) becomes
\ben\label{bna}
{\cal N}e^{-6U}={1\over6}c_{IJK}H^IH^JH^K=H^1~.
\een
The master equation (\ref{bka}) now becomes
\ben\label{bnb}
H^1 = e^{-6U}- \left[ \partial_r H^1\partial_r
U+2H^1~{1\over r^2}\partial_r (r^2 \partial_r U)\right] ~,
\een
where we used $c_2(K3)=24$ and $c_{2i}=0$. We can write this more
explicitly as
\ben\label{bnc}
1 + {p^1\over 2r} = e^{-6U}  - 2(1 + {p^1\over 2r})
U^{\prime\prime} -{4\over r}\left( 1 + {3p^1\over 8r} \right)
U^\prime~,
\een
where primes denote derivatives with respect to $r$.  Note that we set $M^1_\infty=1$;
a  general value can be restored by a rescaling  of $p^1$ and a shift of $U$.

In our units distance $r$ is measured in units of the 5D Planck length.
The parameter $p^1$ is a pure number counting the fundamental strings.
For a given $p^1$, it is straightforward to integrate (\ref{bnc}) numerically.
Instead, to gain some analytical insight we
will take $p^1\gg 1$ so as to have an expansion parameter.
We will analyze the problem one region at a time.

\subsubsection{The AdS$_3\times S^2$-region}

This is the leading order behavior close to the string. According
to our magnetic attractor solution in the form (\ref{bmdb}) we expect the precise
asymptotics
\ben\label{bnd}
e^{-6U} \to {\ell^3_S\over r^3}~~~,~r\to 0~,
\een
where the $S^2$-radius is given by
\ben\label{bne}
 \ell_S = \left(
{p^1\over 4}\right)^{1/3}~.
\een
For $p^1\gg 1$ this is much smaller than the scale size of a large string,  which from (\ref{scalel}) has scale $\sim p$.
However, it is nevertheless
much larger than the 5D Planck scale. The modulus describing the
volume of the internal $T^2$ is
\ben\label{bnea}
 M^1 = {p^1\over
2\ell_S} = 2^{-1/3}(p^1)^{2/3}~,
\een
which also corresponds to the length scale $(p^1)^{1/3}$.

\subsubsection{The near-string region}

We next seek a solution in the entire range $r\ll p^1$ which
includes the scale (\ref{bne}) but reaches further out. In fact, it may
be taken to be all of space in a scaling limit where
$p^1\to\infty$.

In the near string region (\ref{bnc}) reduces to
\ben\label{bnf}
{p^1\over 2r}
= e^{-6U}  - {p^1\over r} U^{\prime\prime} -  {3p^1\over 2r^2}
U^\prime~.
\een
We can scale out the string number $p^1$ by
substituting
\ben\label{bng}
 e^{-6U(r)} = {p^1\over 4r^3}
e^{-6\Delta(r)}~,
\een
which amounts to
\ben\label{bnh}
 U(r) = {1\over 2}
\ln {r\over\ell_S} + \Delta(r)~.
\een
This gives
\ben\label{bni}
\Delta^{\prime\prime} + {3\over 2r} \Delta^\prime   + {1\over
4r^2} ( 1 - e^{-6\Delta}) + {1\over 2}=0~,
\een
which describes the
geometry in the entire region $r\ll p^1$. The asymptotic behavior
at small $r$ is
\ben\label{bnj}
 \Delta(r) = -{1\over 13}r^2 + {3\over (13)^3}r^4 +{20\over 9(13)^4}r^6+\cdots~.
\een
Since $\Delta(r)\to 0$ smoothly as $r\to 0$
we have an analytical description of the approach to the
AdS$_3\times S^2$ region.

\begin{figure}[pb]
\centerline{\psfig{file=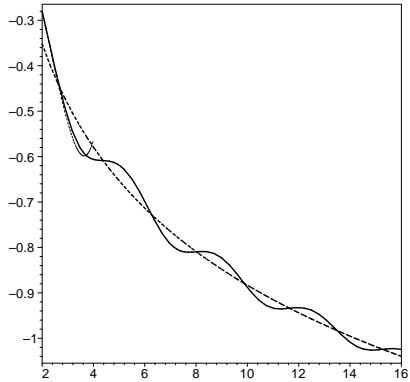,width=6cm}} \vspace*{8pt}
\caption{Analytical and numerical results for $\Delta(r)$ in the
near string region. The solid curve describes the numerical solution
of (\ref{bni}). The dotted curve represent the analytical solution
for small values of r given by (\ref{bnj}), and the dashed curved is
the approximate solution for large values of r (\ref{bnk}).
\label{FS}}
\end{figure}

The asymptotic behavior for large $r$ is also smooth. Expanding in
$u={1\over r}$ we find
\ben\label{bnk} \Delta(r) = -{1\over 6} \ln (2r^2) -  {1\over
36}{1\over r^2}+{13\over12\cdot36}{1\over r^4}+\cdots~. \een
It is straightforward to solve (\ref{bni}) numerically.
Fig.~\ref{FS} shows the curve that interpolates between the
asymptotic forms (\ref{bnj}) and (\ref{bnk}). The oscillatory
behavior in the intermediate region is characteristic of higher
derivative theories. We comment in more detail below.

In the original variable $U(r)$ the approximation (\ref{bnk}) gives
\ben\label{bnl} e^{-6U} = {p^1\over 2r} \left( 1 + {1\over
6r^2}-{1\over6r^4}+\ldots\right)~,
\een
for large $r$. The leading behavior, $e^{-6U} = H^1\sim {p^1\over 2r}$, agrees
with the near string behavior   in
two-derivative supergravity. In the full theory this singular
region is replaced by a smooth geometry.

\subsubsection{The approach to asymptotically flat
space}\label{farstring}

We still need to analyze the region where $r$ is large, meaning
$r\sim p^1$ or larger.   Here we encounter some subtleties in matching the
solution on to the asymptotically flat region.

In the asymptotic region the full equation (\ref{bnb}) simplifies to
\ben\label{bnm}
 1 + {p^1\over 2r} = e^{-6U} - 2( 1+ {p^1\over 2r})
U^{\prime\prime}~.
\een
Terms with explicit factors of $1/r$ were
neglected, but we kept derivatives with respect to $r$ so as to allow
for  Planck scale structure, even though $r\sim p^1\gg 1$.
Changing variables as
\ben\label{bnn}
 e^{-6U} = (1 + {p^1\over 2r})
e^{-6W}~,
\een
we find
\ben\label{bno}
 W^{\prime\prime}= {1\over 2}(
e^{-6W} -1) \simeq -3W~.
\een
The expansion for small $W$ is justified because
(\ref{bnl}) imposes the boundary condition $W\to 0$ for $r\ll p^1$.

The solution $W=0$ expected from two-derivative supergravity is in fact a
solution to (\ref{bno}), but there are also more general solutions of the
form
\ben\label{bnp}
 W = A \sin ( \sqrt{3} r +\delta)~.
\een
The amplitude of this solution is undamped, so it is not really an
intrinsic feature of the localized string solution we consider.
Instead it is a property of fluctuations about flat space, albeit an
unphysical one. The existence of such spurious solutions is a
well-known feature of theories with higher derivatives, and is
related to the possibility  of field redefinitions.\cite{Dabholkar:2004dq}\cdash\cite{Hubeny:2004ji,Zwiebach:1985uq} In the present
context the issue is that the oscillatory solutions can be mapped to zero by
a new choice of variables, such as  $\tilde{W} =
(\nabla^2 - 3)W$.


To summarize, modulo the one subtlety associated with field redefinitions,
we have found a smooth solution interpolating between the near horizon AdS$_3 \times S^2$ attractor and asymptotically flat space.  The solution is completely regular,
the causal structure being the same as that of an M5-brane in eleven dimensions.  While
our result is highly suggestive of the existence of a smooth solution of the full theory with all higher derivative corrections included, we cannot establish with certainty that this is the case.  The reason is that for small strings there is no small parameter
suppressing even higher derivative terms.  Indeed, it is easy to check that in the
near horizon region terms in the action with more than four derivatives contribute
at the same order as those included in the present analysis.  As a result, the  precise
numerical results for the attractor moduli and scale size are expected to receive
corrections of order unity.      On the other hand, it seems highly plausible
that the solution will remain smooth even after these additional corrections have been
taken into account.

\subsection{Holography for the fundamental string}
The small  string solutions are not merely regular, they exhibit an AdS$_3\times S^2$
near string geometry. This raises the possibility that the AdS/CFT correspondence
applies to small strings.\cite{Dabholkar:2007gp}\cdash\cite{Kraus:2007vu}
As noted already, in a particular duality frame these solutions correspond
to $n=p^1$ fundamental heterotic strings.  The underlying theory in this case is
Heterotic string theory compactified on $T^5$, with the fundamental strings
extended along one of the noncompact spatial directions.  The near horizon
solution in this frame has string coupling $g_s \sim {1 / \sqrt{n}}$ and
curvature of order the string scale.  Therefore, for a large number of strings,
quantum effects are suppressed while $\alpha'$ corrections are of order unity.  This implies that the proper description of these solutions is really in terms of a
worldsheet CFT, since this will capture all of the $\alpha'$ corrections rather than
just the leading ones used in our supergravity approach.

In trying to establish an AdS/CFT correspondence in this case, the following
features are expected:
\begin{itemize}
\item
The attractor geometry has an $SL(2,\mathbb{R})\times OSp(4^*|4) $ isometry group\cite{Lapan:2007jx} (for a recent discussion on the super-isometry group of 5D small black hole attractors see Ref.~\refcite{Alishahiha:2007ap}).
  It is not completely clear whether the full theory also is   $SL(2,\mathbb{R})\times OSp(4^*|4) $
invariant.  If it is, then since the general argument of Brown and Henneaux\cite{Brown:1986nw} implies that the $SL(2,\mathbb{R})$ symmetries are enhanced to Virasoro algebras,
the symmetry algebra of the theory will be some superconformal extension of
$OSp(4^*|4)$, with affine $SU(2)\times Sp(4)$ R-currents.\cite{Lapan:2007jx,Kraus:2007vu}
\item It is expected that the superconformal algebra has $(0,8)$ supersymmetry.  In our supergravity
solution based on M-theory on $K3\times T^2$ (or equivalently Heterotic on $T^5$),
only half of the supersymmetry is manifest since we work in an $N=2$ formalism.
\item
Based on the worldsheet structure of the heterotic string,
the central charges of the theory are expected to be $c_R=12n$ and $c_L=24n$, possibly with subleading $1/n$ corrections.
For our supergravity solution, we can infer that $c_L-c_R=12n$, since this combination is determined by a diffeomorphism anomaly.\cite{Kraus:2005zm}
To determine $c_L + c_R$ via $c$-extremization one needs to use the full set of
higher derivative corrections, which are not known.  As we saw in Section \ref{sec:cext}, the four-derivative
action yields the expected $c_L + c_R = 36n$, but this result  is not reliable for small strings.  Alternatively, once the precise superconformal algebra has been
established, we can use the R-symmetry anomaly to determine
$c_L$.   If this algebra contains a $(0,4)$ subalgebra, then the desired $c_L=24n$
will follow.
\end{itemize}
The most immediate difficulty with all these expectations is the absence of a
conventional superconformal algebra  with $(0,8)$ supersymmetry.  In particular, one can
start with currents corresponding to Virasoro, R-symmetry, and local supersymmetry, and
then look for a consistent operator product expansion in which only these currents (plus central terms) appear in the singular parts.  One finds that it is impossible to
satisfy the Jacobi identities. Faced with this problem, one is  led to consider the {\it nonlinear
superconformal algebras}.\cite{Henneaux:1999ib} These are algebras with
bilinears of the R-currents appearing in the OPEs of
supercurrents.\cite{Knizhnik:1986wc,Bershadsky:1986ms} One of the appealing features
of these algebras is that their central charges are completely determined
algebraically in terms of the level of the R-current algebra. In particular, this would give exact expressions for nontrivial
quantum corrections to the spacetime central charges.\cite{Kraus:2007vu}

Unfortunately there are serious difficulties with this optimistic scenario.
There are at least two major problems:
\begin{itemize}
\item
The nonlinear algebras of interest do not permit unitary highest weight representations. The Jacobi identities
either imply that the central charge is negative, or else one of the R-currents has
negative level. Either way, this would seem unacceptable for the spacetime theory.
Firstly, unitarity is of course sacred in quantum mechanical descriptions. Secondly, it would be extremely surprising to find
instabilities in a BPS system with so much supersymmetry.
\item
The nonlinear $(0,8)$ supersymmetry does
not have any obvious $(0,4)$ subalgebra. The actual central charge
determined from the nonlinear algebra does not agree with the expectations.
\end{itemize}
There is no logical inconsistency in this state of affairs, since we were careful to emphasize
that some our assumptions are optimistic and not backed by explicit computations.
Obviously, we must be cautious in extending our usual AdS/CFT expectations to
the unfamiliar terrain of small strings.

The simplest consistent modification of the expectations is that
the superconformal symmetry of this theory is something other than the
nonlinear algebra based on $OSp(4^*|4)$.  There might instead be some more
exotic W-algebra, which perhaps contains an $(0,4)$ subalgebra.   This remains
a fascinating direction for future research.  To motivate further study, we note
that understanding holography for fundamental strings could lead to an example
of AdS/CFT in which both   the bulk and boundary sides of the duality are analytically
tractable.

\section{Comparing 4D and 5D Solutions}
\label{sec:fourdfived}

There is a rich web of interconnections between supergravity theories in diverse dimensions, and it is illuminating to consider the relations between solutions
to these different theories.  A solution with a spacelike isometry can be converted to a lower
dimensional one by Kaluza-Klein reduction along the isometry direction. Conversely,
a solution can be uplifted to one higher dimension by interpreting a gauge field
as the off-diagonal components of a higher dimensional metric.

Here we will be concerned with the relation between 4D and 5D solutions.  The BPS equations governing general 4D supersymmetric solutions are well established, including the contributions from a class of four-derivative corrections.\cite{LopesCardoso:2000qm}  On the other hand, in this review we have obtained the corresponding 5D BPS equations.   What is the relation between the two?

We first address this question at the two-derivative level, and show (following Ref.~\refcite{BLM}) that the 4D BPS equations can be mapped to a special case of the
5D BPS equations.  That is to say, the general 4D BPS solution can be interpreted
as the rewriting of a 5D solution.  Note, though, that the space of solutions is larger   in 5D in the sense that the general 5D solution has no spacelike isometry and hence
can't be reduced to 4D.

We then turn to the generalization of this correspondence with four-derivative corrections included, and find that there is  apparently no simple relation between
the two sets of solutions.  We discuss the likely reason for this mismatch.

An interesting application of this circle of ideas is to the so-called 4D/5D connection,
which gives a relation between the entropies of black holes in four and five dimensions.\cite{Gaiotto:2005gf}  This connection involves 5D solutions whose
base metric is a Taub-NUT.   The Taub-NUT geometry interpolates between $\mathbb{R}^4$ at the origin and   $\mathbb{R}^3\times S^1$ at infinity, and the size of the $S^1$ is  freely adjustable.   By placing a black hole at the origin and dialing the $S^1$ radius we can thereby interpolate between black holes with 4D and 5D asymptotics.  Since the
attractor mechanism implies that the BPS entropy is independent of moduli, we conclude
that the 4D and 5D black hole entropy formulas are closely related.   Higher derivative
corrections turn out to introduce an interesting twist to this story.\cite{Castro:2007ci}  The relation between the 4D and 5D black hole charges is not the naive one expected from the lowest
order solutions, but is rather modified due to the fact that higher derivative terms
induce delocalized charge densities on the Taub-NUT space.   We work out the corrected
charge dictionary explicitly.

\subsection{Relation between 4D and 5D BPS equations}

We now show how to relate the BPS equations governing 4D and 5D solutions.
At the two-derivative level, we find that the two sets of equations are equivalent.
On the other hand, we shall see that this is apparently no longer the case once higher derivative terms are included.

\subsubsection{Two-derivative BPS equations}

The field content of 4D  $N=2$ supergravity consists of the metric, gauge fields $a^A$,
and complex moduli $Y^A$ (we neglect the hypermultiplet fields, which decouple in the
context of BPS black holes).   The $A$ label runs over the values $(0,I)$, where $I$ denotes
the corresponding 5D label.   The extra $a^0$ vector corresponds to the Kaluza-Klein gauge field  that arises
upon reduction from 5D to 4D.    The action is fixed by the prepotential $F$,  which we take to be
of the form arising from compactification on  $CY_3$,
\ben  F = -{1 \over 6}{C_{IJK} Y^I Y^J Y^K \over Y^0}~.
\een
See, {\it e.g.}, Ref.~\refcite{Mohaupt:2000mj}.

Supersymmetry fixes the metric and field strengths to be of the form
\begin{eqnarray} \label{4dansatz}
ds_{4d}^2 &=& e^{2g}(dt+\sigma)^2-e^{-2g}dx^m dx^m~, \cr
f^A&=& d[e^{2g}(Y^A+\Yb^A)(dt+\sigma)] +
{i \over 2}\epsilon_{mn}^{~~~p}\nabla_p (Y^A-\Yb^A)dx^m dx^n~,
\end{eqnarray}
with $\sigma=\sigma_mdx^m$, and $\epsilon_{mnp}$ is the volume form
of $dx^m dx^m$.

The moduli are determined in terms of harmonic functions $(h_A,h^A)$
on $\mathbb{R}^3$ by
\begin{eqnarray}
\label{ad} Y^A -\Yb^A &=& ih^A~, \\
F_A-\Fb_A&=&ih_A~,
\label{ada} \end{eqnarray}
with $F_A = {\p F \over \p Y^A}$.
We also have
\begin{eqnarray}\label{ae}
i[\Yb^A F_A-\Fb_A Y^A]- e^{-2g}&=&0~, \\ \label{aea}
 h^A\vec{\nabla} h_A- h_A \vec{\nabla} h^A - \vec{\nabla}\times\vec{\sigma}&=& 0~.
 \end{eqnarray}

 The above system of equations fixes the form of the general BPS solution.   We now review
 the solution of these equations. Equation (\ref{ad}) is trivially solved  as
 \ben Y^A = {\rm Re} Y^A +{i \over 2}h^A~.
 \een
 We next solve (\ref{ada}) as
\ben\label{afb}Y^I= -{|Y^0| \over \sqrt{h^0}} x^I +{Y^0\over h^0}h^I~,
\een
where
\ben\label{afc} {1\over 2} C_{IJK}x^J x^K = h_I +{1\over 2}{C_{IJK}
h^J h^K\over h^0}~. \een Similarly, $Y^0$ can be found from the
equation $F_0-\Fb_0=ih_0$.

The metric function $e^{-2g}$ is now determined from (\ref{ae}) as
\ben\label{afe} e^{-2g} = i[\Yb^A F_A - \Fb_A Y^A] =
 {(h^0)^2 h_0 +h^0 h_I h^I +{1\over 3}C_{IJK}h^I h^J h^K \over 2 {\rm Re} Y^0}~,
\een
and $\sigma$ is determined from (\ref{aea}).

\subsubsection{4D-5D dictionary}

The dictionary between 4D and 5D solutions has been studied in Ref.~\refcite{BLM}.
5D solutions are related to those in 4D by performing Kaluza-Klein reduction along the $x^5$ circle, which
amounts to writing the 5D solution as
\begin{eqnarray} \label{5Dred} ds_{5d}^2& =& e^{2\phi}\Big(e^{2g}(dt+\omh)^2 - e^{-2g}dx^m dx^m\Big)  -e^{-4\phi} (dx^5+ a^0)^2~,  \\ \label{5Dreda} A^I&=& \Phi^I (dx^5+a^0)-2^{-{1\over 3}} a^I~.
\end{eqnarray}
Comparing with the 5D solutions as discussed in Section \ref{sec:twodghsol}, we can read off
\begin{eqnarray}
 e^{-4\phi} &=&{e^{-2U} \over H^0} -e^{4U} \omega_5^2~, \\
 e^{2g}&=& {e^{2U+2\phi} \over H^0}~, \\
  a^0 &=& \chi -e^{4U+4\phi} \omega_5 (dt +\omh)~.
 \end{eqnarray}

To relate the 4D and 5D BPS equations we make the further  identifications
\begin{eqnarray}\label{afcc}  \sigma &=& \omh~, \cr
Y^0 &=& -{1 \over 2}\left( e^{2U +2\phi}\omega_5-i \right) H^0~, \cr
Y^I &=& 2^{-{4\over 3}} \left[ -e^{-2U+2\phi} M^I +{1\over 2} (e^{2U+2\phi}\omega_5-i)H^I \right]~, \cr
h^I &=& -2^{-{4\over 3}} H^I~, \cr
h_I &=& 2^{-{2\over 3}} H_I~.
\end{eqnarray}

It is then straightforward to check that the 4D and 5D equations are mapped to each under this
identification.  For example, (\ref{ada}) with $A=I$ and $A=0$ yield, respectively, (\ref{Msoltwo}) and
(\ref{apf}).    Also,  (\ref{ae}) maps to the special geometry constraint (\ref{spec}).   This
establishes the equivalence of 4D and 5D BPS solutions at the two-derivative level.

\subsection{Higher derivative case}

Above, we demonstrated  the equivalence of the  two-derivative BPS equations in 4D and 5D.
Does this equivalence extend to the higher derivative BPS equations?

Four-derivative corrections are included into the 4D BPS equations via the generalized
prepotential
\ben\label{af} F = -{1\over 6} {c_{IJK} Y^I Y^J Y^K \over Y^0} -{c_{2I} \over 24\cdot 64} {Y^I \over Y^0}\Upsilon~.
\een
Equations (\ref{4dansatz})-(\ref{ada}) remain valid, while (\ref{ae})-(\ref{aea}) are modified to\cite{LopesCardoso:2000qm}
\begin{eqnarray}\label{aez}i[\Yb^I F_I-\Fb_I Y^I]-e^{-2g}=&& 128i e^g \vec{\nabla}\cdot[(\vec{\nabla} e^{-g})(F_\Upsilon -\Fb_\Upsilon)] \cr && -32i e^{4g}(\vec{\nabla}\times \vec {\sigma})^2(F_\Upsilon-\Fb_\Upsilon) \cr && -64 e^{2g} (\vec{\nabla}\times \vec {\sigma})\cdot \vec{\nabla} (F_\Upsilon+\Fb_\Upsilon)~, \\
H^I\nabla_p H_I- H_I \nabla_p H^I - (\vec{\nabla}\times \vec {\sigma})_p=&&-128i \nabla^q[\nabla_{[p}(e^{2g}\vec({\nabla}\times \vec {\sigma})_{q]} (F_\Upsilon -\Fb_\Upsilon) )] \cr &&-128\nabla^q[2\nabla_{[p}g\nabla_{q]} (F_\Upsilon +\Fb_\Upsilon) ]~,
\end{eqnarray}
where $F_\Upsilon = {\p F \over \p \Upsilon}$, and after taking the derivative we are instructed
to set
\ben  \Upsilon =-64 (\vec{\nabla} g-{1 \over 2}ie^{2g}\vec{\nabla}\times \vec {\sigma} )^2~.
\een

We can now try to compare with the four-derivative BPS equations in 5D, as presented in Section \ref{sec:timelikesusy}. Without going into the details, it turns out that if we continue to use the same
dictionary as in two-derivative case, then we find that the 4D and 5D BPS equations do not agree.
In particular, while the 5D equations are expressed covariantly in terms of the 4D base space,
this property is not realized upon writing the 4D BPS equations in 5D language.

Before concluding that the equations are indeed physically different we should consider the possibility
of including corrections to the dictionary.    A little thought shows that this is unlikely to work.
Any such correction would have to involve $c_{2I}$, since we have already demonstrated
agreement when $c_{2I}=0$.   But both the 4D and 5D BPS equations are linear in $c_{2I}$,
and this property will be upset upon introducing a $c_{2I}$ dependent change of variables.
In the same vein, we note that both the 4D and 5D field strengths are determined just by
supersymmetry, without using the explicit form of the action, and furthermore precisely map into each
other under the uncorrected dictionary.      Again, this feature will be disturbed by including corrections
to the dictionary.

We therefore conclude that the two sets of BPS equations are in fact different.     This is actually
not so surprising for the reason that it is known that the 4D supergravity action usually used does not
include the full set of terms relevant for black hole solutions.   This follows from the observation
in Ref.~\refcite{Sahoo:2006pm} that this action gives the wrong result for the entropy of extremal non-BPS
equations.  On the other hand, the 5D action used here does give the right answer,\cite{Sahoo:2006pm} in accord
with the general arguments based on anomalies.\cite{Kraus:2005vz,Kraus:2005zm}     There is therefore no reason why the BPS
equations derived from these actions should agree.    Indeed, what is surprising is that the
4D action does manage to give the right answer for the entropy of 4D BPS black holes, even if not
for the full geometry.       An interesting open question is to determine what terms in the 4D action
are missing, and to then verify agreement with the 5D BPS equations.

\subsection{Quantum/string corrections to the 4D/5D connection}

We now turn to the relation between the entropies of four and five dimensional black holes.  To illustrate the salient issues we consider the simplest case of electrically charged, non-rotating, 5D black holes, and their 4D analogues.  At the two-derivative
level the following relation holds\cite{Gaiotto:2005gf}
\ben \label{S4d5d} S_{5D}(q_I) = S_{4D}(q_I, p^0=1)~.
\een
This formula is motivated by placing the 5D black hole at the tip of Taub-NUT.
Since Taub-NUT is a unit charge Kaluza-Klein monopole, this yields a 4D black
hole carrying magnetic charge $p^0=1$.   On the other hand, suppose that we sit at
a fixed distance from the black hole and then expand the size of the Taub-NUT circle to
infinity.   Since Taub-NUT looks like $\mathbb{R}^4$ near the origin it is clear
that this limiting process gives back the original 5D black hole.  Finally, the
moduli independence of the entropy yields (\ref{S4d5d}).

The preceding argument contains a hidden assumption, namely that the act of
placing the black hole in Taub-NUT does not change its electric charge.  But why should this be so?  In fact it is not, as was first noticed in Ref.~\refcite{Castro:2007hc} and further studied in Ref.~\refcite{Castro:2007ci}.  The reason is that higher derivative
terms induce a delocalized charge density on the Taub-NUT, so that the charge
carried by the 4D black hole is actually that of the 5D black hole plus that of the
Taub-NUT.

To expand on this point, let us return to the general solutions of Section \ref{ghbh}, \textit{i.e.} spinning black holes on a Gibbons-Hawking base. The Maxwell equations led to (\ref{BMPVeqs1}) which demonstrates that the curvature on the base space provides a delocalized source for the gauge field. This effect should be expected simply from the fact that we deal with an action with a $\int A^I \wedge R\wedge R$ Chern-Simons term.

We now proceed to make explicit the relation between the charges.

\subsubsection{Relation between 4D and 5D charges}
\label{q4D5D}
Consider a general action of gauge fields in the language of forms
\ben
S= {1\over4\pi^2}\int_{{\cal M}_5} \star_5 {\cal L}\lt(A^I, F^I\rt)~.
\een
The Euler-Lagrange equations of motion are
\ben
d\star_5 \frac{\p {\cal L}}{\p F^I} = \star_5 \frac{\p {\cal L}}{\p A^I}~.
\een
Since the left side is exact, we see that this identifies a divergenceless current
\ben
j_I= \frac{\p {\cal L}}{\p A^I}~.
\een
The conserved charge is obtained by integrating $\star_5 j_I$ over a spacelike slice $\Sigma$, suitably normalized. Through the equations of motion and Stoke's theorem this can be expressed as an integral over the asymptotic boundary of $\Sigma$
\ben\label{ya}
Q_I = -{1\over4\pi^2}\int_{\p\Sigma} \star_5 \frac{\p {\cal L}}{\p F^I}~,
\een
which clearly reproduces the conventional $Q \sim \int \star F$ for the Maxwell action.

For the present case, we consider  solutions where the gauge fields fall off sufficiently fast that only the two-derivative terms in the Lagrangian lead to non-zero contributions to the surface integral in (\ref{ya}). Our charge formula is then
\ben
Q_I = -{1 \over 2\pi^2} \int_{\p\Sigma} \left( {1 \over 2}
{\cal N}_{IJ}  \star_5F^J +2 M_I  \star_5v \right)~.
\een
For our solutions with timelike supersymmetry, we can identify $\Sigma$ with the hyperK\"ahler base space.

Now let us compare the charge computations for two distinct
solutions, one with a flat $\mathbb{R}^4$ base space and another on
Taub-NUT, considering just the non-rotating black hole  for
simplicity. As mentioned previously, both $\mathbb{R}^4$ and
Taub-NUT can be written as
\ben ds^2 = (H^0)^{-1}(dx^5
+\vec{\chi}\cdot d\vec{x})^2 +H^0 \lt(d\rho^2 +\rho^2 d\Omega_2
\rt)~, \een
where
\ben \label{taub}
H^0= \eta + \frac{1}{\rho}~,
\een
with
$\eta=0$ for $\mathbb{R}^4$ and $\eta=1$ for Taub-NUT. The
coordinate $x^5$ is compact with period $4\pi$, and we choose the
orientation $\epsilon_{\hat{\rho}\hat{\theta}\hat{\phi}\hat{5}}=1$.
Using the formulas from Section \ref{sec:timelikesusy} for the gauge
fields and auxiliary field, we see that in Gibbons-Hawking
coordinates \ben\label{yb} Q_I = \lim_{\rho \to \infty} \lt[
-4\rho^2 \p_\rho \lt( e^{-2U} M_I\rt)\rt]~, \een which is
independent of the Gibbons-Hawking function $H^0$.

Now recall the
result from the higher-derivative Maxwell equation (\ref{Meq})
\ben \label{Meqtwo}
 M_I e^{-2U}-{c_{2I} \over 8} (\nt U)^2
-{c_{2I} \over 24\cdot 8} \Phi = 1+ \frac{q_I}{4\rho}~,
\een
where $\Phi$ is defined in (\ref{rsquared}) as
\ben\label{rsquaredtwo}
\Rt^2_{\ih\jh\kh\lh} = \nt^2 \Phi \equiv \nt^2 \lt(2{(\nt H^0)^2 \over (H^0)^2}+\sum_i {a_i \over |\vec{x}-\vec{x}_i|}\rt)~.
\een
Both $\mathbb{R}^4$ and $p^0=1$ Taub-NUT are completely smooth geometries and so there are no singularities in the corresponding $\Rt^2$. On the other hand, (\ref{rsquaredtwo}) has manifestly singular terms unless the $a_i$ are chosen to cancel singularities in the $H^0$ dependent term. Comparing with (\ref{taub}), we see that smoothness is assured when the $a_i$ are chosen such that
\ben\label{rsquaredtaub}
\Rt^2_{\ih\jh\kh\lh} = \nt^2 \lt(2{(\nt H^0)^2 \over (H^0)^2} - {2 \over \rho}\rt)~.
\een
The solution (\ref{Meqtwo}) is now fully specified as
\ben
 M_I e^{-2U}-{c_{2I} \over 8} (\nt U)^2
-{c_{2I} \over 24\cdot 4} \lt({(\nt H^0)^2 \over (H^0)^2} - {1 \over \rho}\rt) = 1+ \frac{q_I}{4\rho}~.
\een

In the absence of stringy corrections, \textit{i.e.} for $c_{2I}=0$, we have $M_I e^{-2U}= 1+ \frac{q_I}{4\rho}$ which gives $Q_I =q_I$ independent of the base space geometry. However, including these corrections we see that aymptotically\footnote{We are ignoring the ${c_{2I} \over 8} (\nt U)^2$ term since one can check that it falls off too rapidly as $\rho\to\infty$ to contribute to $Q_I$.}
\ben
M_I e^{-2U} = 1+ {1\over4}\left(q_I -\eta \frac{c_{2I}}{24}\rt)\rho^{-1} + O(\rho^{-2})~,
\een
yielding the asymptotic charge
\ben\label{yc}
Q_I = q_I -\eta \frac{c_{2I}}{24}~.
\een

The preceding computation tells us that formula (\ref{S4d5d}) gets modified to
\ben \label{snew} S_{5D}(q_I) = S_{4D}(q_I-{c_{2I}\over 24}, p^0=1)~.
\een
An analogous, and more complicated, relation holds in the case of rotating black holes; see Ref.~\refcite{Castro:2007ci} for the details.

An interesting open question is whether (\ref{snew}) is further corrected by terms
with even more than  four derivatives.


\section{Discussion}
We conclude this review with a discussion of several open problems and future directions.

\subsection{Black rings}

In the two-derivative gravity theory the most general known supersymmetric
black solution with a connected horizon is a black ring, from which black holes and black
strings can be obtained as special cases. We therefore would  like
to find a black ring solution taking into account  the higher derivative
corrections considered in this review.

While the full black ring solution is not yet known, most
ingredients have been identified. First, the supersymmetry analysis
in Section \ref{sec:electsusy} was carried out for the general black ring.
Subsequently, we determined Gauss' law for the ring solution and
also found the modified very special geometry constraint. The only
missing ingredient is the equation of motion for the auxiliary two
form $v$, needed to relate the magnetic dipole charges in $\Theta^I$
with the angular momentum described by $d\omega$. Once all the
pieces are in place we must of course integrate the equations
of motion, repeating the steps for obtaining the  black ring solution in
two-derivative gravity. It is not obvious that the higher derivative equations of motion
will be integrable, but it is encouraging that we were able to write Gauss'
law for the black ring (\ref{max}) in a form that is manifestly
integrable.

\subsection{Some other approaches}

In this review we have collected our results for black hole entropy obtained from
an off-shell supersymmetric action in five dimensions.  To put things in
perspective, it is useful to compare and contrast our results with those
obtained using other actions.  The comparison is easiest to make in the case of
5D black strings, or equivalently, 4D black holes with $p^0=0$, since this is
where the most results are available.  To keep matters simple, we also set $q_{I\neq 0 }=0$.  These black holes have near horizon geometry AdS$_3\times S^2$ in 5D, and
AdS$_2 \times S^2$ in 4D, the latter obtained by Kaluza-Klein reduction along
the 5D string.

We first recall our results for this case.  The entropy is given by
\ben \label{Scardy} S = 2\pi \sqrt{ {c_L \over 6}(h_L - {c_L \over 24}) } +
2\pi \sqrt{ {c_R \over 6}(h_R - {c_R \over 24}) }~,
\een
where the central charges are
\ben \label{cencharge}  c_L =c_{IJK}p^I p^J p^K + c_{2I}p^I    ~,\quad c_R = c_{IJK}p^I p^J p^K +{1 \over 2} c_{2I} p^I~,
\een
and
\ben \label{q0}   q_0 =  h_R - h_L - \frac{c_R-c_L}{24}~.
\een
In general, the result (\ref{Scardy}) corresponds to a near extremal black hole.
An extremal BPS black hole corresponds to $( h_L> {c_L \over 24},~h_R={c_R \over 24})$, while $( h_L={c_L \over 24},~h_R> {c_R \over 24} )$ yields an  extremal non-BPS black hole.  In all cases, the entropy formula
is in precise agreement with the microscopic entropy counting obtained in Ref. \refcite{Maldacena:1997de}.  This agreement, and in particular the fact that
terms with more than four derivatives do not contribute, is explained by
anomaly based reasoning,\cite{Kraus:2005vz,Kraus:2005zm}  which, we emphasize,
only applies to black holes with a near horizon AdS$_3$ factor.  The validity
of (\ref{Scardy}) also requires that one of $h_{L,R}$ be much larger than the
central charge, a parameter region which may or may not be realizable after using the
freedom to perform duality rotations.\cite{Nampuri:2007gw}  Another way
to understand why terms with more than four derivatives do not contribute is
by noting that all such terms either vanish on the solutions or can be removed by a
field redefinition.\cite{David:2007ak}

Now we turn to the results using other actions.  One well known approach consists of using a 4D action that is the off-shell supersymmetric completion of the square of the
Riemann tensor.\cite{Behrndt:1996jn}\cdash\cite{LopesCardoso:2000qm}   This action gives agreement with (\ref{Scardy}),
and hence with the microscopic entropy, for extremal BPS black holes, but disagreement
otherwise.  As discussed in Ref. \refcite{Sahoo:2006rp}, the likely reason for the
failure of this action in the non-BPS case is that it is just the minimally supersymmetric
action for the gravity multiplet, and does not take into account the extra terms
allowed by the presence of vector multiplets.    Given that this action is apparently incomplete, it is somewhat surprising that it nevertheless gives the correct results
in the BPS case.

Another  action that has attracted significant attention is the six-dimensional
action\cite{Exirifard:2006qv,Sahoo:2006pm}
\bea\label{twoloop}
{\cal L}_{\rm two-loop} &=&
2e^{-2\Phi}\left[ R_{KLMN} R^{KLMN} -{1\over 2} R_{KLMN}H_P^{~KL} H^{PMN}
\right. \cr
&~&\left.~~~ -{1\over 8} H_K^{~MN} H^{KPQ}H_{PQ}^{~~~L} + {1\over 24} H_{KLM}H_{PQ}^{~~~K}
H_R^{~~LP} H^{RMQ}\right]~,
\eea
derived from the condition of conformal invariance for the string theory $\sigma$-model
at two-loop order.\cite{Metsaev:1987zx}   This action corresponds to the bosonic
four derivative terms of an on-shell supersymmetric action.  That is to say,
the action truncated at four derivatives is only supersymmetric up to five derivative
terms, and the algebra only closes on-shell.  We therefore expect this action
to give the right correction to first subleading order in the charges, but not
beyond.   Indeed, it has been established\cite{Sahoo:2006pm} that (\ref{twoloop})
agrees with (\ref{Scardy}) to first subleading order, for both BPS and non-BPS
extremal black holes.

To close this discussion, let us note that besides these 5D black strings / 4D black
holes, we have also obtained results for the entropies of other black objects, such
as 5D spinning black holes.  In general, these black holes do not have an AdS$_3$
factor.  Consequently, we expect that our entropy formulas are valid to first subleading order in the charges, but we have no argument that they remain valid at higher orders.
At present, this issue is difficult to study, since we lack a microscopic description
of such black holes (we only have such a description for $N=4$ black holes,
in which case we can find an AdS$_3$ factor by making a duality rotation.)

\subsection{The Gauss-Bonnet action}
Yet another approach to higher derivative corrections is to consider the Gauss-Bonnet type
action
\ben\label{gaussbonnet}
{\cal L}_{\rm GB} = {1\over 192}c_{2I} M^I  \left( R_{\alpha\beta\gamma\delta}R^{\alpha\beta\gamma\delta}
-4R_{\alpha\beta}R^{\alpha\beta} + R^2\right)~.
\een
The overall coefficient is determined by the coefficient of the
$R^{\mu\nu\alpha\beta}R_{\mu\nu\alpha\beta}$ term obtained from the string S-matrix.
Our action (\ref{offfourd}) is the supersymmetric completion of that term, with the coefficient
of the protected gauge-gravity Chern-Simons term coming out the same as when it is determined
using anomalies. In (\ref{gaussbonnet}) the {\it Riemann squared} term has been
completed instead by forming the Gauss-Bonnet invariant.

Including only the Gauss-Bonnet term is not compatible with supersymmetry, and so {\it a priori},
there is no reason why this approach will give the correct black hole entropy.
Nevertheless, one can check that the Gauss-Bonnet action ({\ref{gaussbonnet}) has the
same on-shell value as the supersymmetric action (\ref{offfourd}) when evaluated on the magnetic attractor solution (discussed Section
\ref{sec:magattr}). This agreement extends to the non-rotating electric attractor (discussed in
Section \ref{sec:elecattr}), but not to rotating attractors (Section \ref{sec:rotatt}).
Combining with the diffeomorphism anomaly from the gauge-gravity
Chern-Simons term one can therefore recover both the left and right central charges of magnetic strings.
In four dimensions the corresponding term gives the correct result for
extremal BPS black holes,\cite{Maldacena:1997de} but it fails in the non-BPS
case.\cite{Sahoo:2006pm}

The successes of the Gauss-Bonnet  combination suggest that it may allow a supersymmetric completion. However, until that has been established clearly one must be cautious when
using  this term, because there is no
clear argument that explains the agreements that have been found, and also the
success is not universal.

\subsection{Higher dimensions}
In this review we have focussed on higher derivative corrections in five dimensions.
As we have discussed, many of the same issues have already been confronted in four
dimensions. In Section \ref{sec:fourdfived} we discussed the interrelation between four and five dimensions,
including some unresolved puzzles in that regard.

An interesting future direction is to consider dimensions six and higher. Ultimately we
would like to understand how higher derivative corrections in ten or eleven dimensions
modify various solutions. For example, many of the standard brane solutions are
singular in the lowest order approximation, but it could be that they are regular once
corrections are taken into account. Criteria for determining in which cases
this hypothesis is valid have not yet been established.   Since we have found a
smooth solution for the heterotic string in five dimensions, it seems natural to
look for similar smooth solutions for heterotic or type II strings in ten dimensions.
On general grounds, we would expect these to have a near horizon AdS$_3 \times S^7$
geometry.

A useful intermediate step would be to understand higher derivative corrections in six
dimensions. At one level, six dimensions is expected to be a relatively straightforward
lift of the five dimensional examples with AdS$_3\times S^2$ and
AdS$_2\times S^3$ near horizon geometries. On the other hand, in six dimensions
it is not generally possible to write Lorentz invariant actions, because of self-duality conditions
on tensor fields. Related to this, anomalies in six dimensions have a much richer
structure. These complications introduce new features which would be interesting
to develop.

\section*{Acknowledgments}

We would like to thank Kentaro Hanaki for useful discussions on the superconformal formalism, and Akhil Shah for useful discussions and collaboration on previous work reviewed in this article. The work of PK and JD is supported in part by NSF grant PHY-0456200. The work of
FL and AC is supported by DOE under grant DE-FG02-95ER40899.

\appendix

\section{Conventions}

We briefly summarize our conventions which are largely that of
Refs.~\refcite{Castro:2007hc} and \refcite{Castro:2007ci}. Latin
indices $a,b,\ldots$ denote tangent space indices and curved
space-time indices are denoted by Greek indices $\mu,\nu,\ldots$.
The metric signature is mostly minus, $\eta_{ab}= {\mathrm{
diag}}(+,-,-,-,-)$. Covariant derivatives of spinors are defined as
\ben {\cal D}_\mu =\partial_\mu + {1\over4} \omega_\mu^{~ab}
\gamma_{ab}~, \een
where $\omega^{ab}$ are the spin-connection one forms related to the
vielbein through the Cartan equation
\ben d e^a +  \omega^a_{~b} \wedge e^b  =0~. \een
Our convention for the two-form curvature is
\ben
R^a_{~b}=d\omega^a_{~b}+\omega^a_{~c}\wedge\omega^c_{~b}={1\over2}R^{a}_{~bcd}e^c\wedge
e^d~.\een
%
%
The scalar curvature is then, {\it e.g.}, $R={p(p-1) \over
\ell_A^2}-{q(q-1) \over \ell_S^2}$ for AdS$_p\times S^q$.  The Weyl
tensor is given by \ben
 C_{abcd}=R_{abcd}- {2 \over
3}(g_{a[c}R_{d]b}-g_{b[c}R_{d]a})+{1 \over 6}g_{a[c}g_{d]b}R~.
\een

Gamma-matrices satisfy the usual Clifford algebra
\ben \lt\{\g_a , \g_b \rt\}=2\eta_{ab}~. \een
Totally anti-symmetric products of $p$ - gamma-matrices are denoted
by
\ben
\g_{a_1\cdots a_p}~, \een
and is normalized such that $\gamma_{abcde} = \varepsilon_{abcde}$,
where $\varepsilon_{01234}=1$. The operation $\gamma\cdot \alpha$,
where $\alpha_{a_1\cdots a_p}$ is a $p$-form, is understood as
\ben \gamma\cdot \alpha= \gamma^{a_1\cdots a_p}\alpha_{a_1\cdots
a_p}~.\een

Finally, we take $G_5={\pi\over 4}$ and measure moduli in units of
$2\pi\ell_{11}$. In these units the charges are quantized (for
review see Ref.~\refcite{Larsen:2006xm}).


\end{document}